\documentclass[longauth]{aaEC}
\usepackage{lastpage}
\usepackage{graphicx}
\usepackage[labelfont=bf,labelsep=period]{caption}

\usepackage{subcaption}
\usepackage{txfonts}
\usepackage{float}
\usepackage{xcolor}
\usepackage{makecell}
\usepackage{esvect}
\usepackage{cellspace} %
\usepackage{stackengine}
\usepackage{amsmath}
\usepackage{pifont}
\usepackage{soul}
\usepackage{siunitx}
\usepackage{xfrac}
\usepackage{comment}
\usepackage{float}

\usepackage{ragged2e}

\usepackage[colorlinks=true,linkcolor=blue,citecolor=blue]{hyperref}

\usepackage{natbib}
\usepackage{scalerel}
\usepackage{euclid}

\makeatletter
\renewcommand*\aa@pageof{, page \thepage{} of \pageref*{LastPage}}
\makeatother
\usepackage[utf8]{inputenc}

\usepackage[switch, modulo]{lineno}
\nolinenumbers

\begin{document}

\title{\Euclid: Galaxy cluster detection through the weak lensing effect -- algorithm assessment and selection\thanks{This paper is published on behalf of the Euclid Consortium.}} 

\newcommand{\orcid}[1]{} %% if already defined in aa.cls: comment, or use renewcommand			   
\author{A.~Manj\'on-Garc\'ia\orcid{0000-0002-7413-8825}\inst{\ref{aff1}}
\and G.~Leroy\orcid{0009-0004-2523-4425}\thanks{\email{gavin.leroy@durham.ac.uk}}\inst{\ref{aff2},\ref{aff3}}
\and S.~Pires\orcid{0000-0002-0249-2104}\inst{\ref{aff4}}
\and J.~M.~Diego\orcid{0000-0001-9065-3926}\inst{\ref{aff5}}
\and C.~Giocoli\orcid{0000-0002-9590-7961}\inst{\ref{aff6},\ref{aff7}}
\and L.~Trobbiani\orcid{0009-0000-8058-834X}\inst{\ref{aff8},\ref{aff9}}
\and A.~D\'iaz-S\'anchez\orcid{0000-0003-0748-4768}\inst{\ref{aff1}}
\and M.~Maturi\orcid{0000-0002-3517-2422}\inst{\ref{aff10},\ref{aff11}}
\and M.~Oguri\orcid{0000-0003-3484-399X}\inst{\ref{aff12},\ref{aff13}}
\and G.~Panebianco\orcid{0000-0002-3410-8613}\inst{\ref{aff6}}
\and G.~W.~Pratt\inst{\ref{aff4}}
\and S.~Andreon\orcid{0000-0002-2041-8784}\inst{\ref{aff14}}
\and L.~Moscardini\orcid{0000-0002-3473-6716}\inst{\ref{aff15},\ref{aff6},\ref{aff7}}
\and M.~Sereno\orcid{0000-0003-0302-0325}\inst{\ref{aff6},\ref{aff7}}
\and C.~Carbone\orcid{0000-0003-0125-3563}\inst{\ref{aff16}}
\and A.~M.~C.~Le~Brun\orcid{0000-0002-0936-4594}\inst{\ref{aff17}}
\and M.~Lieu\inst{\ref{aff18}}
\and L.~Chappuis\orcid{0009-0005-9644-8858}\inst{\ref{aff19}}
\and B.~Altieri\orcid{0000-0003-3936-0284}\inst{\ref{aff20}}
\and T.~Castro\orcid{0000-0002-6292-3228}\inst{\ref{aff21},\ref{aff22},\ref{aff23},\ref{aff24}}
\and M.~Douspis\orcid{0000-0003-4203-3954}\inst{\ref{aff25}}
\and D.~Eckert\orcid{0000-0001-7917-3892}\inst{\ref{aff26}}
\and S.~Farrens\orcid{0000-0002-9594-9387}\inst{\ref{aff4}}
\and R.~Gavazzi\orcid{0000-0002-5540-6935}\inst{\ref{aff27},\ref{aff28}}
\and A.~H.~Gonzalez\orcid{0000-0002-0933-8601}\inst{\ref{aff29}}
\and L.~Ingoglia\orcid{0000-0002-7587-0997}\inst{\ref{aff9}}
\and E.~Jullo\orcid{0000-0002-9253-053X}\inst{\ref{aff27}}
\and S.~Maurogordato\inst{\ref{aff30}}
\and J.-B.~Melin\inst{\ref{aff31}}
\and N.~Okabe\orcid{0000-0003-2898-0728}\inst{\ref{aff32},\ref{aff33},\ref{aff34}}
\and F.~Pacaud\orcid{0000-0002-6622-4555}\inst{\ref{aff35}}
\and M.~Radovich\orcid{0000-0002-3585-866X}\inst{\ref{aff36}}
\and Z.~Sakr\orcid{0000-0002-4823-3757}\inst{\ref{aff37},\ref{aff38},\ref{aff39}}
\and L.~Salvati\inst{\ref{aff25}}
\and B.~Sartoris\orcid{0000-0003-1337-5269}\inst{\ref{aff40},\ref{aff21}}
\and S.~A.~Stanford\orcid{0000-0003-0122-0841}\inst{\ref{aff41}}
\and J.~E.~Taylor\orcid{0000-0002-6639-4183}\inst{\ref{aff42},\ref{aff43}}
\and J.~Weller\orcid{0000-0002-8282-2010}\inst{\ref{aff40},\ref{aff44}}
\and C.~Wood\orcid{0000-0002-3169-0126}\inst{\ref{aff45}}
\and C.~Baccigalupi\orcid{0000-0002-8211-1630}\inst{\ref{aff23},\ref{aff21},\ref{aff22},\ref{aff46}}
\and M.~Baldi\orcid{0000-0003-4145-1943}\inst{\ref{aff8},\ref{aff6},\ref{aff7}}
\and S.~Bardelli\orcid{0000-0002-8900-0298}\inst{\ref{aff6}}
\and P.~Battaglia\orcid{0000-0002-7337-5909}\inst{\ref{aff6}}
\and A.~Biviano\orcid{0000-0002-0857-0732}\inst{\ref{aff21},\ref{aff23}}
\and E.~Branchini\orcid{0000-0002-0808-6908}\inst{\ref{aff47},\ref{aff48},\ref{aff14}}
\and M.~Brescia\orcid{0000-0001-9506-5680}\inst{\ref{aff49},\ref{aff50}}
\and S.~Camera\orcid{0000-0003-3399-3574}\inst{\ref{aff51},\ref{aff52},\ref{aff53}}
\and V.~Capobianco\orcid{0000-0002-3309-7692}\inst{\ref{aff53}}
\and V.~F.~Cardone\inst{\ref{aff54},\ref{aff55}}
\and J.~Carretero\orcid{0000-0002-3130-0204}\inst{\ref{aff56},\ref{aff57}}
\and M.~Castellano\orcid{0000-0001-9875-8263}\inst{\ref{aff54}}
\and G.~Castignani\orcid{0000-0001-6831-0687}\inst{\ref{aff6}}
\and S.~Cavuoti\orcid{0000-0002-3787-4196}\inst{\ref{aff50},\ref{aff58}}
\and A.~Cimatti\inst{\ref{aff59}}
\and C.~Colodro-Conde\inst{\ref{aff60}}
\and G.~Congedo\orcid{0000-0003-2508-0046}\inst{\ref{aff61}}
\and L.~Conversi\orcid{0000-0002-6710-8476}\inst{\ref{aff62},\ref{aff20}}
\and Y.~Copin\orcid{0000-0002-5317-7518}\inst{\ref{aff63}}
\and F.~Courbin\orcid{0000-0003-0758-6510}\inst{\ref{aff64},\ref{aff65},\ref{aff66}}
\and H.~M.~Courtois\orcid{0000-0003-0509-1776}\inst{\ref{aff67}}
\and H.~Degaudenzi\orcid{0000-0002-5887-6799}\inst{\ref{aff26}}
\and S.~de~la~Torre\inst{\ref{aff27}}
\and G.~De~Lucia\orcid{0000-0002-6220-9104}\inst{\ref{aff21}}
\and H.~Dole\orcid{0000-0002-9767-3839}\inst{\ref{aff25}}
\and F.~Dubath\orcid{0000-0002-6533-2810}\inst{\ref{aff26}}
\and X.~Dupac\inst{\ref{aff20}}
\and S.~Escoffier\orcid{0000-0002-2847-7498}\inst{\ref{aff68}}
\and M.~Farina\orcid{0000-0002-3089-7846}\inst{\ref{aff69}}
\and R.~Farinelli\inst{\ref{aff6}}
\and S.~Ferriol\inst{\ref{aff63}}
\and F.~Finelli\orcid{0000-0002-6694-3269}\inst{\ref{aff6},\ref{aff70}}
\and P.~Fosalba\orcid{0000-0002-1510-5214}\inst{\ref{aff71},\ref{aff72}}
\and S.~Fotopoulou\orcid{0000-0002-9686-254X}\inst{\ref{aff45}}
\and M.~Frailis\orcid{0000-0002-7400-2135}\inst{\ref{aff21}}
\and E.~Franceschi\orcid{0000-0002-0585-6591}\inst{\ref{aff6}}
\and M.~Fumana\orcid{0000-0001-6787-5950}\inst{\ref{aff16}}
\and S.~Galeotta\orcid{0000-0002-3748-5115}\inst{\ref{aff21}}
\and K.~George\orcid{0000-0002-1734-8455}\inst{\ref{aff73}}
\and B.~Gillis\orcid{0000-0002-4478-1270}\inst{\ref{aff61}}
\and P.~G\'omez-Alvarez\orcid{0000-0002-8594-5358}\inst{\ref{aff74},\ref{aff20}}
\and J.~Gracia-Carpio\orcid{0000-0003-4689-3134}\inst{\ref{aff44}}
\and A.~Grazian\orcid{0000-0002-5688-0663}\inst{\ref{aff36}}
\and F.~Grupp\inst{\ref{aff44},\ref{aff40}}
\and S.~V.~H.~Haugan\orcid{0000-0001-9648-7260}\inst{\ref{aff75}}
\and H.~Hoekstra\orcid{0000-0002-0641-3231}\inst{\ref{aff76}}
\and M.~S.~Holliman\inst{\ref{aff61}}
\and W.~Holmes\inst{\ref{aff77}}
\and F.~Hormuth\inst{\ref{aff78}}
\and A.~Hornstrup\orcid{0000-0002-3363-0936}\inst{\ref{aff79},\ref{aff80}}
\and K.~Jahnke\orcid{0000-0003-3804-2137}\inst{\ref{aff81}}
\and M.~Jhabvala\inst{\ref{aff82}}
\and B.~Joachimi\orcid{0000-0001-7494-1303}\inst{\ref{aff83}}
\and S.~Kermiche\orcid{0000-0002-0302-5735}\inst{\ref{aff68}}
\and A.~Kiessling\orcid{0000-0002-2590-1273}\inst{\ref{aff77}}
\and B.~Kubik\orcid{0009-0006-5823-4880}\inst{\ref{aff63}}
\and M.~Kunz\orcid{0000-0002-3052-7394}\inst{\ref{aff84}}
\and H.~Kurki-Suonio\orcid{0000-0002-4618-3063}\inst{\ref{aff85},\ref{aff86}}
\and S.~Ligori\orcid{0000-0003-4172-4606}\inst{\ref{aff53}}
\and P.~B.~Lilje\orcid{0000-0003-4324-7794}\inst{\ref{aff75}}
\and V.~Lindholm\orcid{0000-0003-2317-5471}\inst{\ref{aff85},\ref{aff86}}
\and I.~Lloro\orcid{0000-0001-5966-1434}\inst{\ref{aff87}}
\and G.~Mainetti\orcid{0000-0003-2384-2377}\inst{\ref{aff88}}
\and O.~Mansutti\orcid{0000-0001-5758-4658}\inst{\ref{aff21}}
\and O.~Marggraf\orcid{0000-0001-7242-3852}\inst{\ref{aff35}}
\and M.~Martinelli\orcid{0000-0002-6943-7732}\inst{\ref{aff54},\ref{aff55}}
\and N.~Martinet\orcid{0000-0003-2786-7790}\inst{\ref{aff27}}
\and F.~Marulli\orcid{0000-0002-8850-0303}\inst{\ref{aff15},\ref{aff6},\ref{aff7}}
\and R.~J.~Massey\orcid{0000-0002-6085-3780}\inst{\ref{aff3}}
\and E.~Medinaceli\orcid{0000-0002-4040-7783}\inst{\ref{aff6}}
\and M.~Meneghetti\orcid{0000-0003-1225-7084}\inst{\ref{aff6},\ref{aff7}}
\and E.~Merlin\orcid{0000-0001-6870-8900}\inst{\ref{aff54}}
\and G.~Meylan\inst{\ref{aff89}}
\and A.~Mora\orcid{0000-0002-1922-8529}\inst{\ref{aff90}}
\and M.~Moresco\orcid{0000-0002-7616-7136}\inst{\ref{aff15},\ref{aff6}}
\and E.~Munari\orcid{0000-0002-1751-5946}\inst{\ref{aff21},\ref{aff23}}
\and R.~Nakajima\orcid{0009-0009-1213-7040}\inst{\ref{aff35}}
\and C.~Neissner\orcid{0000-0001-8524-4968}\inst{\ref{aff91},\ref{aff57}}
\and S.-M.~Niemi\orcid{0009-0005-0247-0086}\inst{\ref{aff92}}
\and J.~W.~Nightingale\orcid{0000-0002-8987-7401}\inst{\ref{aff93}}
\and C.~Padilla\orcid{0000-0001-7951-0166}\inst{\ref{aff91}}
\and S.~Paltani\orcid{0000-0002-8108-9179}\inst{\ref{aff26}}
\and F.~Pasian\orcid{0000-0002-4869-3227}\inst{\ref{aff21}}
\and K.~Pedersen\inst{\ref{aff94}}
\and V.~Pettorino\orcid{0000-0002-4203-9320}\inst{\ref{aff92}}
\and G.~Polenta\orcid{0000-0003-4067-9196}\inst{\ref{aff95}}
\and M.~Poncet\inst{\ref{aff96}}
\and L.~A.~Popa\inst{\ref{aff97}}
\and L.~Pozzetti\orcid{0000-0001-7085-0412}\inst{\ref{aff6}}
\and F.~Raison\orcid{0000-0002-7819-6918}\inst{\ref{aff44}}
\and A.~Renzi\orcid{0000-0001-9856-1970}\inst{\ref{aff98},\ref{aff99},\ref{aff6}}
\and J.~Rhodes\orcid{0000-0002-4485-8549}\inst{\ref{aff77}}
\and G.~Riccio\inst{\ref{aff50}}
\and E.~Romelli\orcid{0000-0003-3069-9222}\inst{\ref{aff21}}
\and M.~Roncarelli\orcid{0000-0001-9587-7822}\inst{\ref{aff6}}
\and C.~Rosset\orcid{0000-0003-0286-2192}\inst{\ref{aff100}}
\and R.~Saglia\orcid{0000-0003-0378-7032}\inst{\ref{aff40},\ref{aff44}}
\and A.~G.~S\'anchez\orcid{0000-0003-1198-831X}\inst{\ref{aff44}}
\and D.~Sapone\orcid{0000-0001-7089-4503}\inst{\ref{aff101}}
\and M.~Schirmer\orcid{0000-0003-2568-9994}\inst{\ref{aff81}}
\and P.~Schneider\orcid{0000-0001-8561-2679}\inst{\ref{aff35}}
\and A.~Secroun\orcid{0000-0003-0505-3710}\inst{\ref{aff68}}
\and G.~Seidel\orcid{0000-0003-2907-353X}\inst{\ref{aff81}}
\and E.~Sihvola\orcid{0000-0003-1804-7715}\inst{\ref{aff102}}
\and P.~Simon\inst{\ref{aff35}}
\and C.~Sirignano\orcid{0000-0002-0995-7146}\inst{\ref{aff98},\ref{aff99}}
\and G.~Sirri\orcid{0000-0003-2626-2853}\inst{\ref{aff7}}
\and L.~Stanco\orcid{0000-0002-9706-5104}\inst{\ref{aff99}}
\and P.~Tallada-Cresp\'{i}\orcid{0000-0002-1336-8328}\inst{\ref{aff56},\ref{aff57}}
\and A.~N.~Taylor\inst{\ref{aff61}}
\and I.~Tereno\orcid{0000-0002-4537-6218}\inst{\ref{aff103},\ref{aff104}}
\and N.~Tessore\orcid{0000-0002-9696-7931}\inst{\ref{aff105}}
\and S.~Toft\orcid{0000-0003-3631-7176}\inst{\ref{aff106},\ref{aff107}}
\and R.~Toledo-Moreo\orcid{0000-0002-2997-4859}\inst{\ref{aff108}}
\and F.~Torradeflot\orcid{0000-0003-1160-1517}\inst{\ref{aff57},\ref{aff56}}
\and I.~Tutusaus\orcid{0000-0002-3199-0399}\inst{\ref{aff72},\ref{aff71},\ref{aff38}}
\and J.~Valiviita\orcid{0000-0001-6225-3693}\inst{\ref{aff85},\ref{aff86}}
\and T.~Vassallo\orcid{0000-0001-6512-6358}\inst{\ref{aff21},\ref{aff73}}
\and G.~Verdoes~Kleijn\orcid{0000-0001-5803-2580}\inst{\ref{aff109}}
\and Y.~Wang\orcid{0000-0002-4749-2984}\inst{\ref{aff110}}
\and A.~Zacchei\orcid{0000-0003-0396-1192}\inst{\ref{aff21},\ref{aff23}}
\and G.~Zamorani\orcid{0000-0002-2318-301X}\inst{\ref{aff6}}
\and F.~M.~Zerbi\orcid{0000-0002-9996-973X}\inst{\ref{aff14}}
\and E.~Zucca\orcid{0000-0002-5845-8132}\inst{\ref{aff6}}
\and J.~Garc\'ia-Bellido\orcid{0000-0002-9370-8360}\inst{\ref{aff37}}
\and J.~Mart\'{i}n-Fleitas\orcid{0000-0002-8594-569X}\inst{\ref{aff111}}
\and V.~Scottez\orcid{0009-0008-3864-940X}\inst{\ref{aff112},\ref{aff113}}
\and M.~Viel\orcid{0000-0002-2642-5707}\inst{\ref{aff23},\ref{aff21},\ref{aff46},\ref{aff22},\ref{aff24}}}
										   
%%%% please do not edit the affiliation list -- contact ECEB Bureau for changes
\institute{Departamento F\'isica Aplicada, Universidad Polit\'ecnica de Cartagena, Campus Muralla del Mar, 30202 Cartagena, Murcia, Spain\label{aff1}
\and
Department of Physics, Centre for Extragalactic Astronomy, Durham University, South Road, Durham, DH1 3LE, UK\label{aff2}
\and
Department of Physics, Institute for Computational Cosmology, Durham University, South Road, Durham, DH1 3LE, UK\label{aff3}
\and
Universit\'e Paris-Saclay, Universit\'e Paris Cit\'e, CEA, CNRS, AIM, 91191, Gif-sur-Yvette, France\label{aff4}
\and
Instituto de F\'isica de Cantabria, Edificio Juan Jord\'a, Avenida de los Castros, 39005 Santander, Spain\label{aff5}
\and
INAF-Osservatorio di Astrofisica e Scienza dello Spazio di Bologna, Via Piero Gobetti 93/3, 40129 Bologna, Italy\label{aff6}
\and
INFN-Sezione di Bologna, Viale Berti Pichat 6/2, 40127 Bologna, Italy\label{aff7}
\and
Dipartimento di Fisica e Astronomia, Universit\`a di Bologna, Via Gobetti 93/2, 40129 Bologna, Italy\label{aff8}
\and
INAF, Istituto di Radioastronomia, Via Piero Gobetti 101, 40129 Bologna, Italy\label{aff9}
\and
Institut f\"ur Theoretische Physik, University of Heidelberg, Philosophenweg 16, 69120 Heidelberg, Germany\label{aff10}
\and
Zentrum f\"ur Astronomie, Universit\"at Heidelberg, Philosophenweg 12, 69120 Heidelberg, Germany\label{aff11}
\and
Center for Frontier Science, Chiba University, 1-33 Yayoi-cho, Inage-ku, Chiba 263-8522, Japan\label{aff12}
\and
Department of Physics, Graduate School of Science, Chiba University, 1-33 Yayoi-Cho, Inage-Ku, Chiba 263-8522, Japan\label{aff13}
\and
INAF-Osservatorio Astronomico di Brera, Via Brera 28, 20122 Milano, Italy\label{aff14}
\and
Dipartimento di Fisica e Astronomia "Augusto Righi" - Alma Mater Studiorum Universit\`a di Bologna, via Piero Gobetti 93/2, 40129 Bologna, Italy\label{aff15}
\and
INAF-IASF Milano, Via Alfonso Corti 12, 20133 Milano, Italy\label{aff16}
\and
Laboratoire d'etude de l'Univers et des phenomenes eXtremes, Observatoire de Paris, Universit\'e PSL, Sorbonne Universit\'e, CNRS, 92190 Meudon, France\label{aff17}
\and
School of Physics and Astronomy, University of Nottingham, University Park, Nottingham NG7 2RD, UK\label{aff18}
\and
Universit\'e Paris-Saclay, CEA, D\'epartement d'\'Electronique des D\'etecteurs et d'Informatique pour la Physique, 91191, Gif-sur-Yvette, France\label{aff19}
\and
ESAC/ESA, Camino Bajo del Castillo, s/n., Urb. Villafranca del Castillo, 28692 Villanueva de la Ca\~nada, Madrid, Spain\label{aff20}
\and
INAF-Osservatorio Astronomico di Trieste, Via G. B. Tiepolo 11, 34143 Trieste, Italy\label{aff21}
\and
INFN, Sezione di Trieste, Via Valerio 2, 34127 Trieste TS, Italy\label{aff22}
\and
IFPU, Institute for Fundamental Physics of the Universe, via Beirut 2, 34151 Trieste, Italy\label{aff23}
\and
ICSC - Centro Nazionale di Ricerca in High Performance Computing, Big Data e Quantum Computing, Via Magnanelli 2, Bologna, Italy\label{aff24}
\and
Universit\'e Paris-Saclay, CNRS, Institut d'astrophysique spatiale, 91405, Orsay, France\label{aff25}
\and
Department of Astronomy, University of Geneva, ch. d'Ecogia 16, 1290 Versoix, Switzerland\label{aff26}
\and
Aix-Marseille Universit\'e, CNRS, CNES, LAM, Marseille, France\label{aff27}
\and
Institut d'Astrophysique de Paris, UMR 7095, CNRS, and Sorbonne Universit\'e, 98 bis boulevard Arago, 75014 Paris, France\label{aff28}
\and
Department of Astronomy, University of Florida, Bryant Space Science Center, Gainesville, FL 32611, USA\label{aff29}
\and
Universit\'e C\^{o}te d'Azur, Observatoire de la C\^{o}te d'Azur, CNRS, Laboratoire Lagrange, Bd de l'Observatoire, CS 34229, 06304 Nice cedex 4, France\label{aff30}
\and
Universit\'e Paris-Saclay, CEA, D\'epartement de Physique des Particules, 91191, Gif-sur-Yvette, France\label{aff31}
\and
Physics Program, Graduate School of Advanced Science and Engineering, Hiroshima University, 1-3-1 Kagamiyama, Higashi-Hiroshima, Hiroshima 739-8526, Japan\label{aff32}
\and
Hiroshima Astrophysical Science Center, Hiroshima University, 1-3-1 Kagamiyama, Higashi-Hiroshima, Hiroshima 739-8526, Japan\label{aff33}
\and
Core Research for Energetic Universe, Hiroshima University, 1-3-1, Kagamiyama, Higashi-Hiroshima, Hiroshima 739-8526, Japan\label{aff34}
\and
Universit\"at Bonn, Argelander-Institut f\"ur Astronomie, Auf dem H\"ugel 71, 53121 Bonn, Germany\label{aff35}
\and
INAF-Osservatorio Astronomico di Padova, Via dell'Osservatorio 5, 35122 Padova, Italy\label{aff36}
\and
Instituto de F\'isica Te\'orica UAM-CSIC, Campus de Cantoblanco, 28049 Madrid, Spain\label{aff37}
\and
Institut de Recherche en Astrophysique et Plan\'etologie (IRAP), Universit\'e de Toulouse, CNRS, UPS, CNES, 14 Av. Edouard Belin, 31400 Toulouse, France\label{aff38}
\and
Universit\'e St Joseph; Faculty of Sciences, Beirut, Lebanon\label{aff39}
\and
Universit\"ats-Sternwarte M\"unchen, Fakult\"at f\"ur Physik, Ludwig-Maximilians-Universit\"at M\"unchen, Scheinerstr.~1, 81679 M\"unchen, Germany\label{aff40}
\and
Department of Physics and Astronomy, University of California, Davis, CA 95616, USA\label{aff41}
\and
Department of Physics and Astronomy, University of Waterloo, Waterloo, Ontario N2L 3G1, Canada\label{aff42}
\and
Waterloo Centre for Astrophysics, University of Waterloo, Waterloo, Ontario N2L 3G1, Canada\label{aff43}
\and
Max Planck Institute for Extraterrestrial Physics, Giessenbachstr. 1, 85748 Garching, Germany\label{aff44}
\and
School of Physics, HH Wills Physics Laboratory, University of Bristol, Tyndall Avenue, Bristol, BS8 1TL, UK\label{aff45}
\and
SISSA, International School for Advanced Studies, Via Bonomea 265, 34136 Trieste TS, Italy\label{aff46}
\and
Dipartimento di Fisica, Universit\`a di Genova, Via Dodecaneso 33, 16146, Genova, Italy\label{aff47}
\and
INFN-Sezione di Genova, Via Dodecaneso 33, 16146, Genova, Italy\label{aff48}
\and
Department of Physics "E. Pancini", University Federico II, Via Cinthia 6, 80126, Napoli, Italy\label{aff49}
\and
INAF-Osservatorio Astronomico di Capodimonte, Via Moiariello 16, 80131 Napoli, Italy\label{aff50}
\and
Dipartimento di Fisica, Universit\`a degli Studi di Torino, Via P. Giuria 1, 10125 Torino, Italy\label{aff51}
\and
INFN-Sezione di Torino, Via P. Giuria 1, 10125 Torino, Italy\label{aff52}
\and
INAF-Osservatorio Astrofisico di Torino, Via Osservatorio 20, 10025 Pino Torinese (TO), Italy\label{aff53}
\and
INAF-Osservatorio Astronomico di Roma, Via Frascati 33, 00078 Monteporzio Catone, Italy\label{aff54}
\and
INFN-Sezione di Roma, Piazzale Aldo Moro, 2 - c/o Dipartimento di Fisica, Edificio G. Marconi, 00185 Roma, Italy\label{aff55}
\and
Centro de Investigaciones Energ\'eticas, Medioambientales y Tecnol\'ogicas (CIEMAT), Avenida Complutense 40, 28040 Madrid, Spain\label{aff56}
\and
Port d'Informaci\'{o} Cient\'{i}fica, Campus UAB, C. Albareda s/n, 08193 Bellaterra (Barcelona), Spain\label{aff57}
\and
INFN section of Naples, Via Cinthia 6, 80126, Napoli, Italy\label{aff58}
\and
Dipartimento di Fisica e Astronomia "Augusto Righi" - Alma Mater Studiorum Universit\`a di Bologna, Viale Berti Pichat 6/2, 40127 Bologna, Italy\label{aff59}
\and
Instituto de Astrof\'{\i}sica de Canarias, E-38205 La Laguna, Tenerife, Spain\label{aff60}
\and
Institute for Astronomy, University of Edinburgh, Royal Observatory, Blackford Hill, Edinburgh EH9 3HJ, UK\label{aff61}
\and
European Space Agency/ESRIN, Largo Galileo Galilei 1, 00044 Frascati, Roma, Italy\label{aff62}
\and
Universit\'e Claude Bernard Lyon 1, CNRS/IN2P3, IP2I Lyon, UMR 5822, Villeurbanne, F-69100, France\label{aff63}
\and
Institut de Ci\`{e}ncies del Cosmos (ICCUB), Universitat de Barcelona (IEEC-UB), Mart\'{i} i Franqu\`{e}s 1, 08028 Barcelona, Spain\label{aff64}
\and
Instituci\'o Catalana de Recerca i Estudis Avan\c{c}ats (ICREA), Passeig de Llu\'{\i}s Companys 23, 08010 Barcelona, Spain\label{aff65}
\and
Institut de Ciencies de l'Espai (IEEC-CSIC), Campus UAB, Carrer de Can Magrans, s/n Cerdanyola del Vall\'es, 08193 Barcelona, Spain\label{aff66}
\and
UCB Lyon 1, CNRS/IN2P3, IUF, IP2I Lyon, 4 rue Enrico Fermi, 69622 Villeurbanne, France\label{aff67}
\and
Aix-Marseille Universit\'e, CNRS/IN2P3, CPPM, Marseille, France\label{aff68}
\and
INAF-Istituto di Astrofisica e Planetologia Spaziali, via del Fosso del Cavaliere, 100, 00100 Roma, Italy\label{aff69}
\and
INFN-Bologna, Via Irnerio 46, 40126 Bologna, Italy\label{aff70}
\and
Institut d'Estudis Espacials de Catalunya (IEEC),  Edifici RDIT, Campus UPC, 08860 Castelldefels, Barcelona, Spain\label{aff71}
\and
Institute of Space Sciences (ICE, CSIC), Campus UAB, Carrer de Can Magrans, s/n, 08193 Barcelona, Spain\label{aff72}
\and
University Observatory, LMU Faculty of Physics, Scheinerstr.~1, 81679 Munich, Germany\label{aff73}
\and
FRACTAL S.L.N.E., calle Tulip\'an 2, Portal 13 1A, 28231, Las Rozas de Madrid, Spain\label{aff74}
\and
Institute of Theoretical Astrophysics, University of Oslo, P.O. Box 1029 Blindern, 0315 Oslo, Norway\label{aff75}
\and
Leiden Observatory, Leiden University, Einsteinweg 55, 2333 CC Leiden, The Netherlands\label{aff76}
\and
Jet Propulsion Laboratory, California Institute of Technology, 4800 Oak Grove Drive, Pasadena, CA, 91109, USA\label{aff77}
\and
Felix Hormuth Engineering, Goethestr. 17, 69181 Leimen, Germany\label{aff78}
\and
Technical University of Denmark, Elektrovej 327, 2800 Kgs. Lyngby, Denmark\label{aff79}
\and
Cosmic Dawn Center (DAWN), Denmark\label{aff80}
\and
Max-Planck-Institut f\"ur Astronomie, K\"onigstuhl 17, 69117 Heidelberg, Germany\label{aff81}
\and
NASA Goddard Space Flight Center, Greenbelt, MD 20771, USA\label{aff82}
\and
Department of Physics and Astronomy, University College London, Gower Street, London WC1E 6BT, UK\label{aff83}
\and
Universit\'e de Gen\`eve, D\'epartement de Physique Th\'eorique and Centre for Astroparticle Physics, 24 quai Ernest-Ansermet, CH-1211 Gen\`eve 4, Switzerland\label{aff84}
\and
Department of Physics, P.O. Box 64, University of Helsinki, 00014 Helsinki, Finland\label{aff85}
\and
Helsinki Institute of Physics, Gustaf H{\"a}llstr{\"o}min katu 2, University of Helsinki, 00014 Helsinki, Finland\label{aff86}
\and
SKAO, Jodrell Bank, Lower Withington, Macclesfield SK11 9FT, UK\label{aff87}
\and
Centre de Calcul de l'IN2P3/CNRS, 21 avenue Pierre de Coubertin 69627 Villeurbanne Cedex, France\label{aff88}
\and
Institute of Physics, Laboratory of Astrophysics, Ecole Polytechnique F\'ed\'erale de Lausanne (EPFL), Observatoire de Sauverny, 1290 Versoix, Switzerland\label{aff89}
\and
Telespazio UK S.L. for European Space Agency (ESA), Camino bajo del Castillo, s/n, Urbanizacion Villafranca del Castillo, Villanueva de la Ca\~nada, 28692 Madrid, Spain\label{aff90}
\and
Institut de F\'{i}sica d'Altes Energies (IFAE), The Barcelona Institute of Science and Technology, Campus UAB, 08193 Bellaterra (Barcelona), Spain\label{aff91}
\and
European Space Agency/ESTEC, Keplerlaan 1, 2201 AZ Noordwijk, The Netherlands\label{aff92}
\and
School of Mathematics, Statistics and Physics, Newcastle University, Herschel Building, Newcastle-upon-Tyne, NE1 7RU, UK\label{aff93}
\and
DARK, Niels Bohr Institute, University of Copenhagen, Jagtvej 155, 2200 Copenhagen, Denmark\label{aff94}
\and
Space Science Data Center, Italian Space Agency, via del Politecnico snc, 00133 Roma, Italy\label{aff95}
\and
Centre National d'Etudes Spatiales -- Centre spatial de Toulouse, 18 avenue Edouard Belin, 31401 Toulouse Cedex 9, France\label{aff96}
\and
Institute of Space Science, Str. Atomistilor, nr. 409 M\u{a}gurele, Ilfov, 077125, Romania\label{aff97}
\and
Dipartimento di Fisica e Astronomia "G. Galilei", Universit\`a di Padova, Via Marzolo 8, 35131 Padova, Italy\label{aff98}
\and
INFN-Padova, Via Marzolo 8, 35131 Padova, Italy\label{aff99}
\and
Universit\'e Paris Cit\'e, CNRS, Astroparticule et Cosmologie, 75013 Paris, France\label{aff100}
\and
Departamento de F\'isica, FCFM, Universidad de Chile, Blanco Encalada 2008, Santiago, Chile\label{aff101}
\and
Department of Physics and Helsinki Institute of Physics, Gustaf H\"allstr\"omin katu 2, University of Helsinki, 00014 Helsinki, Finland\label{aff102}
\and
Departamento de F\'isica, Faculdade de Ci\^encias, Universidade de Lisboa, Edif\'icio C8, Campo Grande, PT1749-016 Lisboa, Portugal\label{aff103}
\and
Instituto de Astrof\'isica e Ci\^encias do Espa\c{c}o, Faculdade de Ci\^encias, Universidade de Lisboa, Tapada da Ajuda, 1349-018 Lisboa, Portugal\label{aff104}
\and
Mullard Space Science Laboratory, University College London, Holmbury St Mary, Dorking, Surrey RH5 6NT, UK\label{aff105}
\and
Cosmic Dawn Center (DAWN)\label{aff106}
\and
Niels Bohr Institute, University of Copenhagen, Jagtvej 128, 2200 Copenhagen, Denmark\label{aff107}
\and
Universidad Polit\'ecnica de Cartagena, Departamento de Electr\'onica y Tecnolog\'ia de Computadoras,  Plaza del Hospital 1, 30202 Cartagena, Spain\label{aff108}
\and
Kapteyn Astronomical Institute, University of Groningen, PO Box 800, 9700 AV Groningen, The Netherlands\label{aff109}
\and
Caltech/IPAC, 1200 E. California Blvd., Pasadena, CA 91125, USA\label{aff110}
\and
Aurora Technology for European Space Agency (ESA), Camino bajo del Castillo, s/n, Urbanizacion Villafranca del Castillo, Villanueva de la Ca\~nada, 28692 Madrid, Spain\label{aff111}
\and
Institut d'Astrophysique de Paris, 98bis Boulevard Arago, 75014, Paris, France\label{aff112}
\and
ICL, Junia, Universit\'e Catholique de Lille, LITL, 59000 Lille, France\label{aff113}}    

\authorrunning{Manj\'on-Garc\'ia et al.}
\titlerunning{Weak lensing selected cluster}

\date{Received XXX; Accepted XXX}

\abstract{Weak gravitational lensing offers a powerful way to detect galaxy clusters by directly tracing their total matter content. Its effectiveness increases with the density of background galaxies, a requirement that is now being met by the high sensitivity and wide coverage of current and upcoming large-area surveys. 
A prime example is the \Euclid telescope, which will map approximately 14\,000\,deg$^2$ of the sky, measuring the shapes of billions of galaxies and opening a meaningful window for detecting galaxy clusters uniquely through their weak lensing signal. We present the results of nine galaxy cluster detection algorithms in a blind challenge, using 1200\,deg$^2$ of synthetic \Euclid-like weak lensing observations from the DEMNUni-Cov simulations. The performance of these methods was assessed by matching their detections to known synthetic clusters with signal-to-noise ratios greater than 2, satisfying the selection cut $\logten(\Mvircr/\si{\solarmass}) > 13.3 + 1.049\,z + 0.489\,z^{2}$, and adopting two different matching procedures. 
The purpose of the challenge was to identify and improve strategies for galaxy cluster detection via weak lensing in preparation of the upcoming \Euclid data releases. We pre-selected four methods based on their individual performance and complementary nature. Each pre-selected method adopts a distinct approach: AMICO-WL uses an optimal filtering technique, DoG employs Gaussian filtering, O21 relies on aperture-mass filtering, and W234 applies multi-scale wavelet filtering. 
Together, the results of these methods can be merged to enhance the processing of \Euclid data.
Individually, these algorithms reach approximately 10\% completeness for a mean purity of 90\%, while their combination leads to roughly a two-fold improvement in overall performance, exceeding 70\% completeness for low-redshift, high-mass clusters.  When extrapolating the results of this work to \Euclid Data Release 1, we expect to detect approximately 2500 galaxy clusters via weak gravitational lensing.}


\maketitle
\nolinenumbers
%\def\thefootnote{\equalcontribution}\footnotetext{These authors contributed equally to this work.}

%===================================================================
\section{Introduction}
\label{section_1}
%===================================================================

Cosmic structure formation and evolution are central issues in cosmology. Modern cosmological models show that the present-day cosmic structure is the result of a gradual, hierarchical merging process driven by gravitational collapse \citep{White_Rees_1978, Davis_1985, Ciardi_Ferrara_2005}. As the largest gravitationally bound structures known in the Universe, clusters of galaxies play a pivotal role in the formation and growth of the cosmic structure over time, acting as a key probe for measuring the properties of the entire Universe \citep{Castillo-Morales_2003, Voit_2005, Allen_2011}. Galaxy clusters are predominantly composed of dark matter ($\sim$85\%), which is distributed in a halo. The remaining baryonic matter is primarily in the form of ionised hot gas in the intracluster medium (ICM), which makes up around 12\% of the total cluster mass, whereas the stellar and gaseous components of the galaxies themselves account for only 3\% of the total.

Wide-area X-ray and Sunyaev–Zeldovich (SZ) surveys have produced catalogues of thousands of clusters, identified either through the X-ray emission of their ICM \citep[e.g.,][]{Bohringer_2001, mcxc, Pacaud_2016} or via the SZ effect at millimetre wavelengths \citep[e.g.,][]{Bleem_2015, Planck_2016, Hilton_2021_ACT_SZ_cluster_catalogue}.
In parallel, clusters have also been detected in the optical and near-infrared, either as overdensities of galaxies \citep[e.g.,][]{Kepner_1999, Werner_2023} or through the characteristic luminosity and colour distributions of their members \citep[e.g.,][]{ Koester_2007, Andreon_2009_JKCS, Rykoff_2014}.
However, these approaches rely on baryonic tracers (X-ray luminosity, SZ flux, optical richness), which only account for a small fraction of the total cluster mass, and do not directly probe the dominant dark matter component \citep[e.g.,][]{Debackere_2021}.
Such detections introduce selection effects, for instance towards cool-core systems with centrally-peaked emission \citep[e.g.,][]{Eckert_2011}, biasing samples towards brighter clusters and potentially overlooking fainter ones.
A robust understanding of structure formation therefore requires well-characterised samples anchored to their underlying dark matter haloes.
These limitations point to the need for approaches that can directly trace clusters through their full matter content, rather than relying solely on baryonic observables.

Galaxy clusters, containing most of their mass in dark matter together with baryons, act as natural lenses: their gravity bends light rays from background galaxies, distorting and magnifying their images. This effect, known as gravitational lensing \citep{Schneider_1992}, provides a direct link to the total matter distribution in clusters.
With the exception of a few rare cases, this lensing effect is so weak that it requires statistical methods to be detected; for this reason, it is known as weak gravitational lensing \citep[WL;][]{Bartelmann_Schneider_2001}. The subtle image distortions in the shapes of distant galaxies are induced by WL, and WL caused by large-scale structure (LSS) is referred to as cosmic shear. WL is a sensitive probe for detecting galaxy clusters through their total projected matter density along the line of sight. Unlike other methods, its selection is largely independent of baryonic physics and of the cluster's dynamical state, relying primarily on the cluster's total mass and redshift. 
While WL signals are measured statistically, the high sensitivity and wide coverage of current deep, wide-field surveys provide the necessary background galaxy densities, enabling robust, mass-selected samples of galaxy clusters. Thus, these WL-selected samples offer the potential to enhance cluster cosmology, as they enable a more precise calibration of the mass-observable relation and provide an independent probe of cluster abundance with a selection function that differs from baryonic methods.
As leading examples of this, the \Euclid space telescope \citep{Laureijs11, EuclidSkyOverview} and the Large Synoptic Survey Telescope \citep[LSST;][]{LSST_2009} will open a unique window for detecting galaxy clusters solely through their WL signal. 
Launched successfully in July 2023, \Euclid aims, among its primary science goals, to measure cosmic shear through WL and to probe LSS via galaxy clustering.
The Euclid Wide Survey \citep[EWS;][]{Scaramella-EP1} will cover around $14\,000\,\rm{deg}^2$ of the extragalactic sky, having already delivered its first Early Release Observation (ERO) data \citep{EROData, EROLensData}, covering a small area of $\sim$10\,deg$^2$, and its first Quick Data Release \citep[Q1;][]{Q1-TP001}, surveying around 63\,deg$^2$. With its outstanding depth, high spatial resolution, and wide angular coverage \Euclid's imaging instruments are particularly well suited to find galaxy clusters based on their WL signal. Within \Euclid, the OU-LE3 Organisational Unit (OU) of the Science Ground Segment (SGS) is responsible for the optical cluster detection and the production of the corresponding catalogues, although WL information is not used for the detection process.

As a result, the upcoming \Euclid data necessitate the development and optimisation of efficient algorithms for detecting galaxy clusters using WL information (see \citealt{2025PJAB..101..129O} for a review). Early WL techniques in optical surveys involved applying a Gaussian filter and a subsequent thresholding of the convergence maps \citep{White_2002,Hamana_2004,Gavazzi_Soucail_2007}. Nowadays, one of the most widely used filtering approaches is the aperture-mass (AM) technique \citep{Schneider1996}, which involves convolving the lensing maps with a filter function of a specific scale (i.e., the aperture radius). The filter function must be designed to maximise the signal-to-noise ratio (S/N) at a chosen scale larger than the scale on which the noise is dominant to boost the lensing signal. Numerous single-scale filters have been developed based on the AM method, in an effort to maximise its performance \citep{Schneider_1998, Schirmer_2004, Hetterscheidt_2005, Dietrich_Hartlap_2010, Hamana_2012, Lin_2015, Miyazaki_2018}. It is worth noting that an AM filter function at a specific scale is mathematically equivalent to a wavelet transform at that same scale. In contrast to typical AM filter functions, wavelet functions (such as Mexican hat and Morlet) are localised both in real and Fourier space (\citealp{Leonard2012}; see \citealp{Starck1998} for a review). Another approach involves optimal filters, which are kernels optimised to incorporate prior information on the shape of the halo profile,
while decreasing the shape noise and the LSS contribution \citep{Hennawi_Spergel_2005, maturi2005optimal, Wittman_2006}. A noteworthy example of this approach is the AMICO (Adaptive Matched Identifier of Clustered Objects) algorithm, which, following its successful application to optical surveys such as the Kilo-Degree Survey \citep[KiDS;][]{maturi2019amico,bellagamba2019amico,giocoli2021amico}, the Javalambre-Physics of the Accelerated Universe Astrophysical Survey \citep[J-PAS;][]{maturi2023minijpas}, COSMOS \citep{Toni_2024}, and COSMOS-WEB \citep{Toni_2025}, has been adapted for WL detections \citep{Trobbiani2025}. AMICO is one of the two detection algorithms implemented in the \Euclid SGS optical cluster-detection pipeline. Nevertheless, given that no single scale is sufficient to detect all clusters, multi-scale techniques have also been adopted with promising results \citep{Starck2006, Leonard2015, Lanusse2016, Leroy2023}.

The observations from the \Euclid mission will make it possible to detect numerous galaxy clusters over a vast area by identifying the regions with high concentration of galaxies \citep[][Euclid Collaboration: Melin et al. in prep]{Adam-EP3}. In particular, \cite{Q1-SP050} already detected 426 clusters in the Q1 survey from the identification of their galaxy members. Additionally, \Euclid %, designed to measure the WL distortions in billions of galaxies, 
enables to trace the clusters' total matter components, including dark matter, avoiding the bias inherent in methods that only trace the baryonic content or ICM emission \citep{2022A&A...661A..14R}. Notably, recent advances in galaxy cluster detection through WL, driven by the Subaru Hyper Suprime-Cam (HSC) survey \citep{2018PASJ...70S...4A} have already led to the discovery of hundreds of WL-detected clusters \citep{Miyazaki_2018,Hamana2020,Oguri2021}. Therefore, by combining these new samples with existing data sets, \Euclid will contribute to a more representative characterisation of the overall cluster population.

Motivated by these prospects, a challenge for detecting galaxy clusters using WL data was carried out within the Euclid Consortium. 
The main goal of this challenge, the results of which are presented in this paper, is %to define the limits on what can realistically be achieved 
to determine the capabilities of WL for galaxy cluster finding, 
and to better understand the strengths and weaknesses of different detection methods by comparing their relative performance. Here, the WL detection methods were tested on 1200\,deg$^2$ of \Euclid-like synthetic observations extracted from the DEMNUni-Cov simulations \citep{2016JCAP...07..034C, 2022JCAP...11..041P}. 
Additional validation tests will follow with increasingly realistic simulations to mimic Euclid Data Release 1 (DR1).

The structure of this paper is as follows. Section~\ref{section_2} summarises the theoretical framework behind WL. In Sect.~\ref{section_3}, we describe the simulations and present the challenge. Section~\ref{section_4} introduces the different WL detection methods compared in this work. We describe the two matching approaches used to associate detections with haloes in Sect.~\ref{section_5} and present the performance of the detection methods in Sect.~\ref{section_6}, comparing the results of an `integrated approach' (considering all the galaxies across the full redshift range) with a `masked approach' that takes into account the existence of masked regions. Finally, our conclusions are detailed in Sect.~\ref{section_8}. We provide a deeper discussion of the complementarity between the detection methods in Appendix~\ref{app:complementarity}, while Appendix~\ref{app:tomography_results} presents the results of various `tomographic approaches', that split galaxies into different redshift bins.

Throughout this paper, we assume a flat $\Lambda$CDM cosmological model, consistent with that adopted for the synthetic data: $\Omm=0.31$, $\OmLa=0.69$, and $H_{0}=67\,\kmsMpc$.

%===================================================================
\section{Weak lensing formalism} \label{section_2}
%===================================================================

The trajectory of light from distant galaxies is deflected by the gravitational field of the matter it encounters, leading to the observed shapes of these galaxies  appearing distorted. Known as (reduced) shear, these distortions trace the underlying distribution of matter on large scales, such as filaments, galaxy clusters, or individual galaxies. Here, we briefly introduce the formalism of gravitational lensing by galaxy clusters (see, e.g., \citealt{Bartelmann_Schneider_2001}, and \citealt{Umetsu2020} for reviews).
%\cite[see e.g.][for reviews]{Bartelmann_Schneider_2001,Umetsu2020}.

We can consider a galaxy cluster as a lens, denoted by its Cartesian angular position $\vec{\theta}$. %at distance \Dd\ from the observer.
Its three-dimensional mass density, $\rho(\vec{\theta},s)$, is linked to its surface mass density $\Sigma(\vec{\theta})$, by the integral
\begin{equation}
    %\Sigma(\vec{\theta}) = \int \diff s \;\rho(\vec{\theta},s)\;, 
    \Sigma(\vec{\theta}) = \int_{-S}^{+S} \diff s \;\rho(\vec{\theta},s)\;, 
\end{equation}
where the integration is performed along the line of sight $s$ and $S$ represents the half-extent of the cluster along this direction, chosen such that the integration interval $[-S,+S]$ encompasses the full mass of the cluster.

The effective two-dimensional deflection potential of the cluster, $\psidp$, describes its gravitational deflection field. This potential depends on the cluster's surface mass density, its position $\vec{\theta}$, and the angular-diameter distances from the lens to the background sources $\Dds$, from the observer to the source $\Ds$, and from the observer to the lens $\Dd$. It is expressed as 
\begin{equation}
    \psidp(\vec{\theta}) = \frac{4 \GN}{c^2}\frac{\Dd \Dds}{\Ds} \int_{\mathbb{R}^2} \diff^2\theta' \;\Sigma(\vec{\theta}') \ln{(\abs{\vec{\theta}-\vec{\theta}'}})\;,
    \label{Eqn:psi}
\end{equation}
where $c$ is the vacuum speed of light and $\GN$ is the gravitational constant. 

The distortion and magnification induced by the lens can be quantified using the Jacobian matrix,
\begin{equation}
    \AJ \equiv \frac{\partial \vec{\beta}}{\partial \vec{\theta}} = \left(\delta^{\rm K}_{ij} - \frac{\partial^2 \psidp(\vec{\theta})}{\partial \theta_i \partial \theta_j} \right) = 
    \begin{pmatrix} 
    1 - \kappa - \gamma_1 & - \gamma_2 \\
    - \gamma_2 & 1 - \kappa + \gamma_1
    \end{pmatrix}\;\textcolor{orange}{,}
\end{equation}
which describes the linear mapping between the lensed, $\vec{\theta}$, and the true, $\vec{\beta}$, coordinates. For brevity, we will use the shorthand notation
\begin{equation}
  \frac{\partial^2}{\partial \theta_i \partial \theta_j} \equiv \partial^2_{ij}\;,
\end{equation}
and define both components of the complex shear:
\begin{equation}
\gamma=\gamma_1+\rm{i} \gamma_2 = \frac{1}{2} ( \partial_{11}^2 - \partial_{22}^2)\, \psidp + i \partial_{12}\,\psidp\;.
\label{Eqn:gamma}
\end{equation}These two components describe the shape distortion of the observed image, while the convergence
\begin{equation}
    \kappa= \frac{1}{2} ( \partial_1^2 + \partial_2^2)\, \psidp\;
    \label{Eqn:kappa}
\end{equation}
characterises its  contraction or dilation.

From Eqs.~(\ref{Eqn:psi}) and~(\ref{Eqn:kappa}), the convergence can be rewritten as
\begin{equation}
    \kappa(\vec{\theta}) = \frac{\Sigma(\vec{\theta})}{\Sigcr}\;,
\end{equation}
where $\Sigcr$ is the critical value of the surface mass density defined as
\begin{equation}
    \Sigcr \equiv \frac{c^2}{4\, \pi\, \GN} \frac{\Ds}{\Dd \Dds}\;.
\end{equation} In practice, the measurement of the ellipticities of the background galaxies only provides a measure of the reduced shear $g$, defined as
\begin{equation}
g \equiv \frac{\gamma}{1 - \kappa}\;.
\end{equation}
However, when $\kappa \ll 1$, the reduced shear $g$ is approximated by the shear $\gamma$, such that $g \approx \gamma$. Although $\kappa$ can reach values of a few tenths at the centre of massive clusters, in this paper we adopted this approximation in the simulations, as it is expected to affect all methods in the same way. Moreover, since $\kappa$ increases towards the cluster centre, $g$ is correspondingly enhanced in these regions. Therefore, neglecting the reduced-shear correction provides a conservative assumption, as including it would enhance the WL signal near the $\kappa$ peaks and, consequently, lead to more efficient cluster detection. In addition to shear from individual clusters, uncorrelated LSS along the line of sight contributes an extra shear component that introduces
correlated noise in the reconstructed mass maps \citep{Hoekstra2001, Hoekstra2003}.

Since WL observations constrain the integrated matter distribution along the line of sight, analyses are typically conducted in terms of the projected mass. 
A widely used approach to recover this distribution -- proportional to the convergence field -- from the shear field is the inversion method proposed by \cite{Kaiser_Squires_1993}, which relates shear and convergence in Fourier space. This approach, based on smoothing with a fixed kernel, helps to regularise the noise, preventing the resulting convergence map from being dominated by white noise arising from galaxies' intrinsic shape noise.

In complex notation, the convergence field, $\kappa_{\mathbb{C}}$ is represented as $\kappa_{\mathbb{C}} = \kappa_{E} + \mathrm{i}\,\kappa_{B}$, where the real part, $\kappa_E$, and the imaginary part, $\kappa_B$, are referred to as the \textit{E}- and \textit{B}-modes, respectively. 
Because WL arises from a scalar potential, it produces only \textit{E}-modes. In contrast, intrinsic alignments and imperfect PSF corrections generate both \textit{E}- and \textit{B}-modes, making the presence of \textit{B}-modes a useful diagnostic of residual systematics in WL surveys.
By taking the Fourier transform of Eqs.~\eqref{Eqn:gamma} and \eqref{Eqn:kappa}, we obtain
\begin{equation}
  \hat{\gamma} = \hat{P}\,\hat{\kappa}\,,
\end{equation}
with

\begin{equation}
\hat{P} = \hat{P}_1 + \mathrm{i}\,\hat{P}_2 = \frac{k_1^2 - k_2^2}{k_1^2 + k_2^2} + \mathrm{i} \,\frac{2\,k_1 k_2}{k_1^2 + k_2^2}\,,
\end{equation}where the hat symbol denotes the Fourier transform, and $k_i$ is the wave number. Considering the conjugate $\hat{P}^{*} = \hat{P}_1 - \mathrm{i} \hat{P}_2$, $\kappa_E$ and $\kappa_B$ can be reconstructed from the complex shear $\gamma$ as:

\begin{equation}
    \hat{\kappa}_E = \hat{P}_1 \hat{\gamma}_1  + \hat{P}_2 \hat{\gamma}_2
    \quad \text{and} \quad
    \hat{\kappa}_B = -\hat{P}_2 \hat{\gamma}_1  + \hat{P}_1 \hat{\gamma}_2. 
\end{equation}This method offers an efficient framework for reconstructing mass maps from WL data and is adopted in this work by the following methods: DoG, J04, S96, TANH, MRLens, and W234 (see Sect.~\ref{section_4}).

%===================================================================
\section{Synthetic \Euclid weak lensing observations}  \label{section_3}
%===================================================================

In order to evaluate and compare the performance of detection algorithms for the future exploitation of \Euclid data, it is essential to test them on a common data set. In this work, we use \Euclid-like synthetic observations. This section first introduces the simulations on which the galaxy cluster finders were run and then describes the organisation and execution of the challenge.

\subsection{The light-cone simulations} \label{subsection_3_1}

The simulations used in this work are two realisations of the Dark Energy and Massive Neutrino Universe \citep[DEMNUni-Cov,][]{2016JCAP...07..034C, 2022JCAP...11..041P}\, from each of which six light-cones are constructed. This is a set of cosmological $N$-body simulations designed to study the evolution of the LSS with and without massive neutrinos. The specific realisations considered here do not include neutrinos. These simulations are particularly well-suited for our study due to their large volume and high resolution, making them ideal for cluster WL studies and the analysis of galaxy clustering and other cosmological observables.

The DEMNUni-Cov simulations have been designed to follow the gravitational evolution of $1024^3$ cold dark matter (CDM) particles with mass resolution $m_{\rm p} \approx 8 \times 10^{10} \, \si{\h^{-1}} \si{\solarmass}$ in a box of comoving size equal to $1 \, h^{-1} \rm Gpc$ on a side. Initial conditions were generated at $z=99$, using a theoretical linear power spectrum calculated using \texttt{CAMB} \citep{camb}, and standard cosmological parameters were set in agreement with the \cite{Planck_2014_results} values stated in the introduction. 
A total of 63 snapshots were recorded during the simulations, spanning from redshift $z = 99 $ to $ z = 0 $. This extensive range enabled the construction of continuous past light-cones out to high redshifts.

Since the simulations used in this study are based on only two DEMNUni-Cov realisations, the light-cones originating from the same parent realisation are not fully independent on the largest scales, limiting the precision of absolute quantities such as cluster abundances and completeness forecasts. However, this has limited impact on our analysis, which focuses on WL measurements on scales up to $\sim 10\,\mathrm{Mpc}$. Similarly, the finite simulation volume mainly affects the largest-scale modes, which are not directly relevant to the scales considered here. The only practical limitation is that the rarest and most massive clusters are not fully sampled, but these objects would be readily detected by \Euclid owing to their large lensing signal.

\subsection{The halo catalogues} \label{subsection_3_2}

The halo catalogues were produced by identifying collapsed dark matter haloes in every simulation by means of a Friends-of-Friends algorithm \citep[FoF,][]{Davis_1985}. It was run using a linking length $\lambda$ = 0.2\;$d$, where $d$ is the mean separation distance between dark matter particles. The SUBFIND algorithm \citep{Springel_2001} was later applied to detect gravitationally bound structures and to estimate the standard quantities of each FoF-identified halo. These quantities include the angular position (i.e., RA and Dec), the redshift, the virial mass $\Mvircr$, and the virial radius $\rvircr$. The virial mass and radius are defined by a spherical volume on a  halo enclosing 200 times the critical density of the Universe, $\rho_{\mathrm{cr}}$, at that specific redshift:
% Definition of M200cr
\begin{equation}
    M_{200\mathrm{,cr}} = \frac{4}{3} \, \pi \, r_{200\mathrm{,cr}}^3 \, 
    200 \, \rho_{\mathrm{cr}}(z) ,
\end{equation}
with the critical density given by
\begin{equation}
    \rho_{\mathrm{cr}}(z) = \frac{3 \,H^2(z)}{8\, \pi\, \GN} \,.
\end{equation}

\subsection{The galaxy catalogues} \label{subsection_3_3}

In line with the approach adopted in \cite{Ajani-EP29}, WL past light-cones from $z = 0$ to $z = 4$ were built from the simulation using the \texttt{MapSim} routine \citep{Giocoli_2014}. This reconstruction is based on the slicing of a set of comoving particle snapshots. The pipeline extracts the positions of the particles stored in 40 different snapshots and projects them onto 43 lens planes to recompose the projected matter density distribution. Several different light-cone realisations can be obtained by randomising the input comoving cosmological boxes. Light rays are traced through these lens planes from various source redshifts, located at the upper bound of each lens plane, down to the observer placed at the vertex of the pyramid. The convergence and shear maps are computed with the \texttt{MOKA} library pipeline \citep{giocoli12a} and resolved with 4096 pixels, which corresponds to a pixel scale of $\ang{;;8.8}$. This resolution is adequate for WL cluster studies. Under the assumption of the Born approximation \citep{Bartelmann_Schneider_2001}, light rays are approximated as straight radial lines and the deflections arising from lens-lens couplings are neglected. %This approximation has proven to be a reliable estimation for weak cosmic shear. 
Although negligible for shear two-point statistics, post-Born corrections can affect higher-order observables and peak counts. Recent full-sky studies generally find small differences, but ray tracing is required for percent-level precision \citep{Ferlito2024}. We therefore consider the Born approximation adequate for the present algorithm pre-selection, while noting its possible impact on the absolute S/N of the highest peaks.

The resulting galaxy catalogues are composed of the angular position of the galaxies (RA and Dec), the two components of the shear, and their corresponding redshift. Galaxies are uniformly distributed on the sky, so source clustering is not accounted for in this work. The redshift distributions $n(z)$ of these galaxy catalogues are derived from the COSMOS2015 photometric redshift data \citep{Laigle_2016}, with a cut at $\IE \leq 24.5$ mag, to mimic the expected \Euclid $n(z)$. A full characterisation of these redshift distributions can be found in section 2.3 of \cite{Ajani-EP29}.

\subsection{Incorporating statistical uncertainties} \label{subsection_3_4}

We model shape and measurement noise as a Gaussian distribution with mean $\mu = 0$ and standard deviation $\sigma_{\epsilon} = 0.26$ \citep{Leauthaud_2007, Schrabback_2018, Martinet-EP4}. For each galaxy, realistic shape noise was included by adding Gaussian noise to the two shear components.
Photometric redshift Gaussian errors were included with a standard deviation of $\sigma_{z}=0.05\;(1+z)$. In addition, 10\% of the galaxies were randomly selected and assigned redshifts that differ significantly from their original values, either higher or lower, to mimic catastrophic outliers. 
Real catastrophic photometric-redshift outliers may exhibit spatial or redshift-dependent structure rather than being randomly distributed. This simplification may therefore affect the preliminary tomographic tests, but not the integrated analysis on which our main conclusions are based. The galaxy catalogues also include \Euclid-like masks, that correspond to about 22\% of missing data. The masks are derived from the \Euclid Data Challenge 2 catalogues produced by the Euclid Collaboration using the code FLASK \citep{flask:xavier16}. 
Although these masks are slightly conservative compared to the current expectations for the final EWS -- we expect a lower percentage of missing data from DR1 -- they are sufficient to study the impact of masking on the detection methods. In the future, the masks will be directly derived from the true data. 

The noisy galaxy catalogues serve as inputs for the detection algorithms described in Sect.~\ref{section_4}. Each one of the simulated fields used in this work contains around 10.8 million galaxies randomly distributed in a field of view of $10\degree\times10\degree$, with a galaxy distribution extending up to $z = 3$ \citep{Blanchard-EP7}. This translates into a galaxy density of about $n_{\rm g}$ = $30\,\rm{arcmin^{-2}}$.

It should be noted that a number of systematic effects were not considered in this study,  including reduced shear, source clustering \citep[e.g.,][]{2024MNRAS.527L.115G}, intrinsic alignments of galaxies \citep[e.g.,][]{2026ApJ...996...36L}, and baryonic feedback \citep[e.g.,][]{2024PhRvD.110j3539G}. The detection methods are sensitive to these effects, which become particularly important on small scales. In future work, we plan to account for all these systematic effects by characterising the selection function through the injection of simulated cluster signals into real observational data.

\subsection{Inputs of the challenge} \label{subsection_3_5}

The detection algorithms participating in the WL galaxy cluster detection challenge were applied blindly to 12 simulated fields of $10\degree\times10\degree$ each. These fields, totalling 1200 deg$^2$, were built using two realisations of the DEMNUni-Cov simulations (six random light-cones each). Each simulated field consists of a galaxy catalogue and a halo catalogue, and can hereafter be referred to interchangeably as either `field' or `realisation'. These realisations should not be confused now with the DEMNUni-Cov realisations from which the simulated fields were derived. 

The challenge was blind, meaning that the positions of the synthetic clusters were unknown to the participants. However, a non-blind preparatory phase with a single field ($100\,\mathrm{deg^{2}}$) was conducted first. This decision was made for several reasons: (i) to give the participants the chance to set up the implementation of their algorithms so as to improve their performance on this specific field; (ii) to help the participants to familiarise with the data format; (iii) to calibrate the matching procedure and the metrics of the challenge; and (iv) to facilitate the development of codes and tools needed for comparing as well as assessing the performance of the different methods. 

\subsection{Outputs of the challenge} \label{subsection_3_6}
Each detection method was required to provide 12 catalogues of detections, one for each realisation. These catalogues represent the `integrated approach', using all the galaxies in the galaxy catalogues. Each detection catalogue had to include the RA and Dec of the candidate clusters, a size proxy (or filter radius) in arcminutes, and the S/N of the detections.

In addition, there were two optional outputs. The first of these, called `masked approach', involved accounting for the masks available in the galaxy catalogue. The second, called `tomographic approach', consisted of splitting the galaxies in the shear catalogue into redshift bins, with suggested bins: $z\in[0, 0.65]$, $z\in[0.65, 1.05]$ and $z\in[1.05, 3]$. Some of the methods used different binning schemes. 

The results of the integrated approach without masks are provided in Sect.~\ref{integrated_wo_mask}, while those with masks are presented in Sect.~\ref{integrated_with_mask}. 
The tomographic analysis has been moved to the Appendix~\ref{app:tomography_results} as the results are preliminary and the use of different tomographic splitting strategies prevented a fair comparison between the methods.

We notice that the different detection approaches compared in this work do not provide catalogues of detections down to the same detection limit. In fact, each participating method produces a catalogue of detections adopting its own S/N definition, which follows the theoretical definition

\begin{equation}
  \mathrm{S/N}\;(\vec{\theta}) = \frac{A_{0}(\vec{\theta})}{\sigma_{\rm n}}\,,
\end{equation}
where $A_{0}(\vec{\theta})$ is the amplitude of the detection in the filtered mass map and the dispersion of the noise is described as
\begin{equation} 
  \sigma^2_{\rm n} = \frac{1}{2\pi} \iint_{0}^{2\pi} \diff^2 \ell \; \abs{\Psi(\ell)}^2\,P_{\rm n}(\ell)\,,
\end{equation}
where $\Psi(\ell)$ is the filter and $P_{\rm n}(\ell)$ the noise power spectrum.

Overall, the catalogues produced here are internal WL detection-candidate catalogues rather than final cosmology-ready products. Cluster redshift assignment, including optical cross-matching and the impact of method-dependent localisation accuracy, will be assessed in future work.

%===================================================================
\section{Galaxy cluster detection algorithms} \label{section_4}
%===================================================================
In this section, we introduce the nine different detection methods that have participated in the WL galaxy cluster detection challenge. Table~\ref{tab:summary_methods_challenge} summarises their names, primary detection principles, and corresponding references for more detailed information. All the methods share the same input catalogues and provide their results in the same output format, but followed different approaches. To clarify their distinct features,  their common scheme can be split into three steps: (A) processing the input data to adapt their format to the specific need of the algorithm, (B) filtering the data and extracting the signal from the clusters, and (C) estimating the noise and selecting signal peaks corresponding to potential clusters.

The filtering step aims to maximise the S/N of cluster detections. Each method is based on a different filter, which can be applied either to the shear or to the convergence. The filter parameters were chosen empirically to achieve optimal average S/N for clusters. For shear-based approaches, some methods apply the filter directly to the shear at the galaxy positions, while others bin the shear into a regular grid before applying the filter. In the latter case, the implementation can be done either in Fourier space or directly in real space. The noise estimation procedure also varies depending on the method.

The following subsections provide an overview of the methodology, implementation, and assumptions used by each algorithm. To avoid redundancy, algorithms based on the same principle are grouped together under a single description.

\begin{table*}
  \centering
  \captionsetup{font=small}
  \caption{Summary of the detection algorithms that participated in the WL cluster-detection challenge, grouped by approach. The pre-selected methods are highlighted in bold.}
  \smallskip
  \label{tab:summary_methods_challenge} 
  \smallskip
  \small
  \begin{tabular}{lll} % four columns, alignment for each
    \hline \hline
    \noalign{\vskip 3pt}
    Method name & Detection principle & Main reference \\
    \noalign{\vskip 2pt}
    \hline
    \noalign{\vskip 3pt}
     \textbf{AMICO-WL} & Optimal matched filter & \cite{maturi2005optimal}, \cite{Trobbiani2025} \\
     Damped filter & Optimal filter & \cite{Diego_Herranz_2008} \\
     TANH & Aperture-mass filter & \cite{Schirmer_2004} \\
     \hline
     \noalign{\vskip 1pt}
     \textbf{DoG} & Gaussian filter & Manjón-García et al. (in prep.) \\
     \hline
     \noalign{\vskip 1pt}
     J04 & Aperture-mass filter & \cite{Jarvis_2004} \\
     S96 & Aperture-mass filter & \cite{Schneider1996} \\
     
     \textbf{O21} & Aperture-mass filter & \cite{Oguri2021} \\
     \hline
     \noalign{\vskip 1pt}
     MRLens & Bayesian Multi-scale wavelet filter & \cite{Starck2006} \\
     \textbf{W234} & Multi-scale wavelet filter & \cite{Leroy2023} \\
     
     \hline
    \end{tabular}
\end{table*}

\subsection{Gaussian filter} 
\label{subsection_Gaussian}

The simple but fruitful procedure of applying a Gaussian filter to WL convergence maps, followed by a thresholding, has proven to be effective for detecting galaxy clusters \citep[e.g.,][]{Miyazaki2002, Miyazaki2007, Fan_2010, Shan_2012, Shan_2018}. 

\subsubsection*{Double Gaussian filter (DoG)} \label{subsection_DoG}
%===================================================================

Here we introduce a WL detection method based on the implementation of a two-stage Gaussian filtering. In the first instance, a filtering is applied to the shear in real space (as described in \citealt{Schneider_Seitz_1995}) at the position of every galaxy using a Gaussian kernel, $G_{\rm k}$, with $\sigma = 1 \arcmin$, according to:

\begin{equation}
  \gamma_{\rm s}(\vec{\theta}) =   \int_{\mathbb{R}^2} \diff^2 \theta' \; \gamma(\vec{\theta}') \; G_{\rm k}(\vec{\theta}-\vec{\theta}')\,.
\end{equation}

This smoothed shear signal, $\gamma_{\rm s}$, is then binned into a regular grid to build shear maps of a given pixel size. The corresponding convergence map is derived using the \cite{Kaiser_Squires_1993} mass-inversion technique on $750 \times 750$ pixel shear maps. To further enhance the signal from embedded haloes, a second Gaussian filter with $\sigma = \ang{;0.8;}$ is applied to the convergence map. Detection peaks are identified as significant enhancements measured in a \textbf{$3 \times 3$-pixel} window more than $3\sigma$ above the background (excluding masked regions if they exist). The detections resulting from this process make up the integrated approach detections catalogue. The procedure followed in the masked approach is the same, but excluding the masked regions in both the input shear data and the peak selection.

The filter parameters, namely the scales of the two Gaussian smoothings, were chosen based on trials carried out during the preparatory phase and informed by earlier work \citep[e.g.,][]{Gavazzi_Soucail_2007}. Performing a Gaussian smoothing of the shear signal of the galaxies before downsampling it to a regular pixelised grid of a given pixel size further improves the results. Even if the Gaussian filtering is a commutative process, the subsequent pixelisation step is not. Applying the Gaussian filtering before the pixelisation therefore is crucial and helps to manage issues such as missing data (Manjón-García et al. in prep.).

For the tomographic approach, the galaxies were split into the three non-overlapping redshift bins suggested: $z\in[0, 0.65]$, $z\in[0.65, 1.05]$, and $z\in[1.05, 3]$. The results for the tomographic approach broadly followed the same procedure explained for the integrated approach. However, here the shear data smoothing, the building of the convergence maps, and the second Gaussian filtering were performed separately for each redshift slice. For each redshift bin the shear smoothing was consistent, but the convergence maps were filtered with different kernel sizes: $\sigma = \ang{;6.4;}$, $\ang{;2.4;}$, and $\ang{;0.8;}$, respectively for bins with increasing redshifts. Finally, the three resulting detection catalogues are merged into a single catalogue, by performing angular cross-matching and selecting the maximum S/N for common detections.

%===================================================================
\subsection{Aperture-mass filters}
\label{subsection_Aperture_Mass_Filters}

The aperture-mass \citep[AM,][]{Schneider1996} formalism, introduced in Sect.~\ref{section_1}, provides a powerful filtering technique to extract WL signals. It measures the projected matter density within a circular aperture by convolving the convergence or, equivalently, the tangential shear with a compensated filter function. This compensation ensures that the statistic is insensitive to the mass-sheet degeneracy and suppresses large-scale noise contributions. By optimising the choice of the filter, AM methods enhance the detectability of cluster-scale structures while providing a direct connection between shear measurements and the underlying matter distribution.

To obtain the AM map at position $\boldsymbol \theta_{\rm o}$, the tangential shear is convolved with a filter 
$Q$,
\begin{equation}
    M_{\rm ap}(\boldsymbol{\theta}_{\rm o})= \int_{\mathbb{R}^2} \diff^2 {\rm \theta}\;\gamma_{\rm t}(\boldsymbol{\rm \theta})\, Q(\lvert\, \boldsymbol{\rm \theta} \, -\, \boldsymbol{\rm \theta}_{\rm o}\rvert)\,,
\label{Eqn:aperture_mass_shear}    
\end{equation}
or on the convergence with a filter $U$,
% Aperture mass definition with convergence
\begin{equation}
    M_{\rm ap}(\vec{\theta}_{\rm o}) = \int_{\mathbb{R}^2} \diff^2 {\rm \theta} \; \kappa(\vec{\theta}) \;
    U(|\vec{\theta} - \vec{\theta}_{\rm o}|) \,.
\end{equation}

For a compensated, radially symmetric filter $U(\theta)$ with finite support $\theta \le \theta_{\rm ap}$,
this definition is equivalent to the shear-based expression involving the filter
$Q(\theta)$. The two filters are related by
\begin{equation}
    U(\theta)
    =
    2 \int_{\theta}^{\theta_{\rm ap}}
    \mathrm{d}\theta'\,
    \frac{Q(\theta')}{\theta'}
    - Q(\theta),
    \label{Eqn:QtoU}
\end{equation}
with the inverse relation
\begin{equation}
    Q(\theta)
    =
    \frac{2}{\theta^2}
    \int_0^{\theta}
    \mathrm{d}\theta'\,
    \theta' U(\theta')
    - U(\theta).
\end{equation}
The filter $U$ is required to be compensated, that is,
\begin{equation}
    \int_0^{\theta_{\rm ap}}
    \mathrm{d}\theta\;
    \theta\,U(\theta) = 0.
\end{equation}

\subsubsection{J04, S96, and TANH}\label{subsection_J04_S96_TANH}
%===================================================================

This approach consists of convolving the WL convergence maps with a well-designed filter following the methodology described in \cite{Leroy2023}. 
For reasons of computational efficiency, we adopt an implementation based on convergence maps instead of applying the filter at the positions of the galaxies. On real data, where the galaxy distribution is non-uniform, the codes will be adapted to work with the shear maps, since this enables the computation of the noise per pixel within the aperture \citep{Schneider1996}.

In this procedure, we reconstructed the shear maps ($\gamma_1$ and $\gamma_2$) by summing the simulated shear values within each pixel, as this strategy yields similar results to those obtained with the real-space approach in terms of the number and distribution of the detections \citep{Leroy2023}. In practice, the simulated shear catalogues are binned into a grid of $1024\times1024$ pixels, corresponding to a pixel scale of \ang{;0.59;}. The associated convergence maps are derived through the \cite{Kaiser_Squires_1993} inversion, which decomposes them into \textit{E}-modes, corresponding to gravitational lensing by mass, and \textit{B}-modes, which are not generated by lensing but serve as a useful diagnostic of systematic errors. In this challenge, we applied three AM filters optimised for detecting the galaxy cluster signal to the convergence maps: 
\begin{enumerate}
    \item AM filter J04: following the design by \cite{Jarvis_2004}, this is defined as the second derivative of a Gaussian filter with a standard deviation $\sigma=4\arcmin$ and a truncation radius of $R = 20\arcmin$.
    \item AM filter S96: following the design of the filter in \cite{Schneider1996} we adopt the parametrisation defined in equation (24) of \cite{Leroy2023}, with the following values: $\nu_1 = 0.1$, $\nu_2 = 0.9$, $R = 9\arcmin$, $\alpha = 0.8531$, $b = -329.8$, and $c= 0.2415$.
    \item AM TANH filter: this filter is numerically computed from the design by \cite{Schirmer_2007}, using a fixed scale radius of $x_{\rm c} = 0.1$ and a truncation radius $R = 7\arcmin$. In practice, this filter is defined following the projected Navarro--Frenk--White (NFW) profile and could be classified among the special apertures (see Sect.~\ref{subsection_Optimal_filters}). However, since its practical implementation closely resembles that of J04 and S96, we describe its application in this section.
\end{enumerate}  
Note that these three filters were originally constructed for an application on the shear catalogues directly. While the implementation of the AM on pixelised maps provide results of lower fidelity than the baseline AM formalism introduced in \cite{Schneider1996}, the trade-off between accuracy and speed is essential in view of the large data volume expected from the EWS. For the sake of simplicity, we shall refer to the original filter papers as the reference for this variant of the AM implementation, albeit with a slight abuse of terminology.

The detections are identified as local maxima with respect to their eight neighbouring pixels in the filtered {\textit{E}-mode convergence map, having values above the noise level, estimated as the average maximum of the filtered \textit{B}-mode convergence maps. 
The result is a catalogue, produced for each field and each filter, containing the detection positions and their measured S/N. This procedure defines the integrated approach.

%mask 
In parallel, for each field, we also produce a detection catalogue after applying the simulated masks on the shear catalogues. For such a data set, the previous procedure was adapted to discard the detections obtained at a distance equal to the filter radius from the borders of missing data. 

% tomographic approach
Additionally, we produce detection catalogues using a tomographic approach by applying the same procedure on three separate, but overlapping redshift selection cuts on the shear catalogues. These selections were obtained by considering three minimal redshift cuts ($z_{\mathrm{min}}$ = 0, 0.65, and 1.05). 
Subsequently, for each field and each filter, the detections were regrouped by matching the positions within a radius corresponding to half the filter radius.

%===================================================================
\subsubsection{O21} \label{subsection_O21}
%===================================================================
This follows the methodology developed in \citet{Oguri2021}, in which the spatial filter of \citet{Schneider1996} is used to compute the AM map. More specifically we use the `TI05' set-up, which detected the largest number of clusters among the three set-ups considered in \citet{Oguri2021}. 
This spatial filter is truncated at the inner radius of $0\farcm5$, and extends out to approximately \ang{;20;}, and the parameters of the filter are carefully chosen to maximise the S/N for massive clusters, while minimising the impact of the LSS noise \citep[see Appendix of ][]{Oguri2021}. The AM map is created from a shear map on a two-dimensional regular grid with a pixel size of \ang{;0.25;} using the fast Fourier transform. The boundary of the survey region as well as the masked regions of source galaxy samples are handled using the number density map. More specifically, a smoothed number density map of source galaxies is created using a $\sigma=\ang{;8;}$ Gaussian kernel. Any region that has smoothed number densities less than 0.5 times the average value is then masked to ensure only areas with sufficient galaxy data contribute. The shot noise is estimated by randomly rotating the orientation of galaxies, creating 500 randomised AM maps. Any region with noise values estimated from the randomised maps 1.5 times higher than the average value is also masked. Interested readers are referred to \citet{Oguri2021} for more details, in which hundreds of WL shear-selected clusters from the Subaru HSC survey \citep{2018PASJ...70S...4A} are presented. Cosmological constraints from the Subaru HSC WL shear-selected cluster sample constructed with this O21 cluster detection algorithm are presented in \citet{2024OJAp....7E..90C} and \citet{2025OJAp....8E...2C}.

The efficiency of selecting massive clusters from AM maps can be increased by using multiple source galaxies with different redshift cuts \citep{Hamana2020,Oguri2021}. Hence, in addition to the integrated and masked cases, we also consider peak catalogues built splitting source galaxies into the three non-overlapping suggested redshift bins, as well as the four overlapping redshift bin approach adopted in \citet{Oguri2021}.

\subsection{Special apertures}
\label{subsection_Optimal_filters}

Optimal matched filtering is a simple but yet powerful technique to identify objects in noisy data sets. This method relies on a model describing the observational features of the objects to be found and the power spectrum of the noise, which can be directly measured from the data. This approach exploits a constrained minimisation  to determine the optimal filter used to convolve the data, providing an estimate of the detection signal, which is unbiased and of minimum variance, meaning that the filter maximises the S/N of the detections.

\subsubsection{Adaptive Matched Identifier of Clustered Objects (AMICO-WL)}

\label{subsection_AMICO}
%===================================================================

The first development of this filter for WL applications was presented by \cite{maturi2005optimal} and it has since been used on real as well as on synthetic data \citep{maturi2007searching,pace2007testing}. For WL applications, the filter template relies on a radial profile of the halo shear pattern, which we assume to be produced by a NFW dark matter halo with a mass of $10^{15}\,\si{\h^{-1}}\,\si{\solarmass}$. 
This value sets the angular scale of the filter rather than acting as a mass threshold: the amplitude estimator is unbiased by construction, irrespective of the true mass of the detected halo, while a mismatch between the assumed and true halo masses only degrades the S/N, without excluding lower-mass haloes from the sample. Indeed, \citet{Trobbiani2025} showed that, using this same fixed template, the matching procedure recovers haloes down to $5\times10^{13}\,\si{\h^{-1}}\,\si{\solarmass}$, more than an order of magnitude below the template mass, with completeness decreasing smoothly rather than dropping abruptly (see their figures~10 and 14).
The noise power spectrum comprises three distinct contributions: the shear produced by the LSS, the galaxy shot noise, and the intrinsic ellipticity. This idea is implemented in AMICO-WL \citep{Trobbiani2025}, a branch of the AMICO algorithm \citep{bellagamba2018amico, maturi2019amico}, which on top of the standard filtering technique, implements an iterative approach for deblending objects which might be hidden by nearby ones. To set the lowest S/N to be used for sample selection, we use a criterion based on the study of the cumulative distribution of the S/N of both positive (representing the true signal contribution) and negative peaks (representing the statistics of spurious peaks produced by observational noise and contributions from the LSS) in the filtered WL maps. The ratio between such cumulative distributions provides an estimate of the sample purity as a function of the S/N. %While \texttt{SExtractor} is initially used to study the distribution of the peaks in the map and compute a threshold for the detection procedure, the final detections are obtained with AMICO-WL functions using standard peak detection (i.e. local maxima with respect to eight neighbours).
While \texttt{SExtractor} is initially employed for its computational efficiency to rapidly study the distribution of the positive and negative peaks down to low S/N and compute the threshold for the detection procedure, the final detections are obtained with AMICO-WL functions using standard, more robust but computationally intensive peak detection (i.e., local maxima with respect to the eight neighbouring pixels) and deblending techniques. We required a purity of 70\% completeness to determine the S/N thresholds and build the catalogues. To further refine our processing strategy, we exclusively use galaxies with redshifts greater than $z>0.6$. This selection of the background sources reduces the foreground contribution which would dilute the lensing signal. The strategy described above was adopted to produce the detection catalogues of the integrated and masked approaches.

In the absence of the masks that \Euclid will provide, a conservative approach was employed, completely disregarding possible detections falling in the masked areas. We generated binary maps with the same pixelisation as the signal and variance maps to identify the regions without galaxies, and then masked the empty pixels.
The detection algorithm subsequently ignored the masked regions entirely. Although the AMICO algorithm can actively account for masks by setting internal correlation factors, we did not implement this feature in AMICO-WL. While this approach does not fully exploit the filter formalism, it is justified by the compact size of the optimal filter, which reduces the impact of masked regions on the final results.

Currently, AMICO-WL does not employ a tomographic approach to fully exploit the photo-$z$s. However, we created and applied three filters, each optimised for one of the three suggested redshift bins ($z\in[0, 0.65]$, $z\in[0.65, 1.05]$, and $z\in[1.05, 3]$) by using the corresponding number density of galaxies as well as the LSS power spectrum. After determining the S/N thresholds for each catalogue, we then combine the results through angular matching within a radius of \ang{;0.6;}. Whenever multiple detections were made, we kept the detection with the highest S/N.

%===================================================================
\subsubsection{Damped filter} \label{subsection_Damped_filter}
%===================================================================

This algorithm directly applies an optimal filter to the shear measurements in real space, as opposed to the more traditional \cite{Kaiser_Squires_1993} convolution in Fourier space. Following \cite{Diego_Herranz_2008}, the filter functions are defined as

\begin{equation}
F_{\gamma_1}(i,j) = \frac{\delta x_{ij}^2-\delta y_{ij}^2}{R_{ij}} 
\quad \text{and} \quad
F_{\gamma_2}(i,j) = \frac{2\,\delta x_{ij} \,\delta y_{ij}}{R_{ij}}\;,
\end{equation}
where indices $i$ and $j$ refer to the pixel position, $\delta x_{ij}$ and $\delta y_{ij}$ are the projected distances to the origin in the $x$- and $y$-axis, respectively, the term $R_{ij}$ is defined as $R_{ij}^2=\delta x_{ij}^2+\delta y_{ij}^2+S_{\rm o}^2$, and $S_{\rm o}$ is a predetermined smoothing length. The role of $S_{\rm o}$ is two-fold: to damp the filter at its maximum (when both $\delta x$ and $\delta y$ are approximately 0), and to set the filter scale. Physically, $S_{\rm o}$ is related to the angular scale of the lens. We explored a range of scales between $1\arcminute$ and $20\arcminute$. This revealed that small values of $S_{\rm o}$ are better suited for small or high-redshift clusters, while large values of $S_{\rm o}$ are better for larger clusters and/or very small redshifts. For this work we chose \mbox{$S_{\rm o}=2\arcmin$}. 
Ignoring $S_{\rm o}$, the shape of the filter resembles the classic \cite{Kaiser_Squires_1993} filter (although not exactly the same), but defined in real space \citep[see also the similar definition in][]{Schneider1996}. 

The filter is used in the standard way, as a special case of Eq.~\eqref{Eqn:aperture_mass_shear}, but applied to the individual $\gamma_1$ and $\gamma_2$ components and combined afterwards. For the integrated approach, all galaxies are included in a single redshift bin, with equal weight assigned to all shear measurements. In contrast, for the tomographic approach, the shear measurements are first divided into 100 equally spaced linear redshift slices from $z=0$ to $z=1.6$. Within each $1' \times 1'$ pixel of a given slice, the 
shear measurements ($\gamma_1$ and $\gamma_2$) are stacked and then convolved by the corresponding filter ($F_{\gamma_1}$ for $\gamma_1$ and $F_{\gamma_2}$ for $\gamma_2$), and summed to produce an estimate of the convergence for that specific redshift slice, $\kappa_z$. Finally, we sum the individual $\kappa_z$ maps to create a final convergence map. Since each $\kappa_z$ map is derived from a very narrow redshift slice and is therefore noisy, this combination is performed as a weighted sum to improve the S/N. The weighting is given by the lensing factor $\Dds\,(\Dd\,\Ds)^{-1}$, assuming a mean redshift for the lens of $z_{\rm d}=0.1$ to avoid penalising the detection of low-redshift clusters.
Structures at higher redshift would have different weights, but only the relative weight between redshift slices is relevant. Pixels in masked regions contain no galaxies and therefore no shear data, so they do not contribute to the final stacked maps.

\subsubsection{TANH} 
\label{subsection_TANH}

Lastly, the AM filter TANH (\citealt{Schirmer_2007}), introduced in Sect.~\ref{subsection_J04_S96_TANH}, can also be regarded as a special aperture filter, as its overall shape was optimally designed to closely approximate the tangential shear profile of a projected NFW halo.
Through the inclusion of a hyperbolic tangent term, it rises linearly at small radii while regularising the singular behaviour at the centre and limiting contributions from the aperture edges. This design enhances cluster detection by maximising S/N for NFW–like shear profiles and by suppressing noise arising from both intrinsic ellipticity scatter and LSS projections. Mathematically, the TANH filter is defined as:
\begin{equation}
Q_{\mathrm{TANH}}(x)
= E(x)\,
\frac{\tanh\left(x/x_{\mathrm{c}}\right)}{x/x_{\mathrm{c}}},
\end{equation}
where $E(x)$ is a smooth truncation function ensuring finite support at a truncation radius of $R = 7\arcmin$, and $x_{\mathrm{c}} = 0.1$ controls the core size of the filter. In the context of this challenge, its application follows exactly that of J04 and S96. In practice, the filter is applied to the convergence maps, and the corresponding filter function $U$ for the AM implementation is derived from Eq.~\eqref{Eqn:QtoU}.

%===================================================================

%===================================================================

\subsection{Multi-scale wavelet filters}
\label{subsection_Multi-scale wavelet filters}

The following methods, MRLens and W234, rely on the multi-scale principle, which exploits the fact that galaxy clusters imprint characteristic signals across different spatial scales in the convergence field.
By decomposing the data into these scales, noise can be more efficiently suppressed while enhancing cluster-like structures.

\subsubsection{Multi-Resolution method for gravitational Lensing (MRLens)} \label{subsection_MRLens}

%===================================================================

The MRLens detection method relies on the MRLens filter, which has been shown to be efficient for denoising convergence maps \citep{Starck2006, Massey2007}. However, we expect this detection method to suffer from spurious detections coming from LSS because the MRLens filter does not filter them.

In practice, the MRLens method consists of binning the shear into a regular grid to build the shear maps. The binned shear is obtained by summing the signal from the galaxies that fall in each pixel as described in \cite{Leroy2023}. Some pixels remain empty as they are devoid of galaxies. The number of empty pixels depends on the size of the pixel and on the mask applied to the data. For the challenge, the fields have been downsampled to $1024  \times 1024$ pixels, which correspond to a pixel size of about \mbox{\ang{;0.59;}}. 
We used the non-linear mass mapping method proposed by \cite{mm:Pires2020} to reconstruct the noisy convergence maps from these incomplete shear maps. This mass inversion method includes corrections for mass mapping systematic effects (e.g., missing data, border effects, etc.). 

After that, the noisy convergence maps are filtered using the MRLens filter. This filter follows a Bayesian approach using a multi-scale entropy prior based on the wavelet transform. The thresholds at each wavelet scale are computed from the corresponding convergence \textit{E}-mode map obtained from the mass inversion, assuming a Gaussian distribution for the noise \citep{Benjamini1995}, which can be non-stationary in the case of real data. Finally, the different wavelet scales are recombined to build the filtered convergence maps, and peak detection (i.e., local maxima with respect to the eight neighbours) is performed on them. 

For the tomographic catalogues, we followed the approach proposed by \cite{Hamana2020} and \cite{Oguri2021}, using three overlapping redshift bins, with minimum redshifts of $z_{\mathrm{min}}$ = 0, 0.65, and 1.05. We then cross-matched the detections made in each redshift bin to build a single catalogue.

%===================================================================
\subsubsection{W234} \label{subsection_W234}
%===================================================================

This method is based on the multi-scale algorithm presented in \cite{Leroy2023}, which uses the starlet transform \citep{Starck1998,Starck2006} applied to the convergence maps.

In practice, for each field we started by binning the noisy shear catalogues by summing the shear signal in each pixel, considering a pixel size of \ang{;0.59;}, as detailed in \cite{Leroy2023}. With the resulting $1024 \times 1024$ pixels shear maps ($\gamma_1$ and $\gamma_2$), we performed the \cite{Kaiser_Squires_1993} mass inversion to recover the \textit{E}- and \textit{B}-modes convergence maps. 
Regarding the starlet transform of the multi-scale procedure, we considered the three complementary scales called `W2', `W3', and `W4' used in \cite{Leroy2023} as they are designed to extract the cluster signal. 
The resulting filtered \textit{E}-mode convergence maps are individually thresholded by the noise level estimated from the average maximum value of the filtered \textit{B}-mode convergence maps, and the peak detection is made on each scale independently. For real data, the noise level will be estimated on a pixel-by-pixel basis to take into account the non-uniform distribution of the galaxies. This will be done by building noise realisations through random rotations of the galaxy orientations in the shear catalogue. The detections are taken as the remaining positive local maxima with respect to eight neighbours in these thresholded and filtered \textit{E}-mode convergence maps. Lastly, we regrouped the detections from the three scales based on their positions and the scale at which they were obtained. In practice, detections were merged as one when their separation distance was smaller than the radius of the wavelet that detected them. %closer than the radius of the wavelet that detected them were merged.
Following this implementation, we created the detection catalogues of the integrated approach.

Additionally, we provided the catalogues corresponding to the two variants of the implementation: the masked and the tomographic approaches. 
In the first variant, the mask is added at the level of the noisy shear catalogue before the binning step. 
Then, we applied the previous multi-scale procedure, but we discarded detections obtained less than five pixels away from the edges of any masked regions to mitigate potential border effects. 
For the tomographic approach, we split the shear catalogues using the same redshift binning as MRLens. Then, we repeated the multi-scale procedure on each of these selected shear catalogues and regrouped the detections obtained based on their positions, as defined previously. Based on \cite{Hamana2020}, this minimum redshift selection is expected to improve both our precision and the number of detections, as it helps mitigate the dilution effect caused by the foreground galaxies.

Note that the filters used for W234 are shared with the MRLens method, but the two methods largely differ in how they build the filtered maps, estimate noise, and ultimately the resulting detections.

%===================================================================
\subsection{Shape of the filters} \label{Sect_2_filters ditribution }
%===================================================================
\begin{figure}
\centering
  \includegraphics[width=\linewidth]{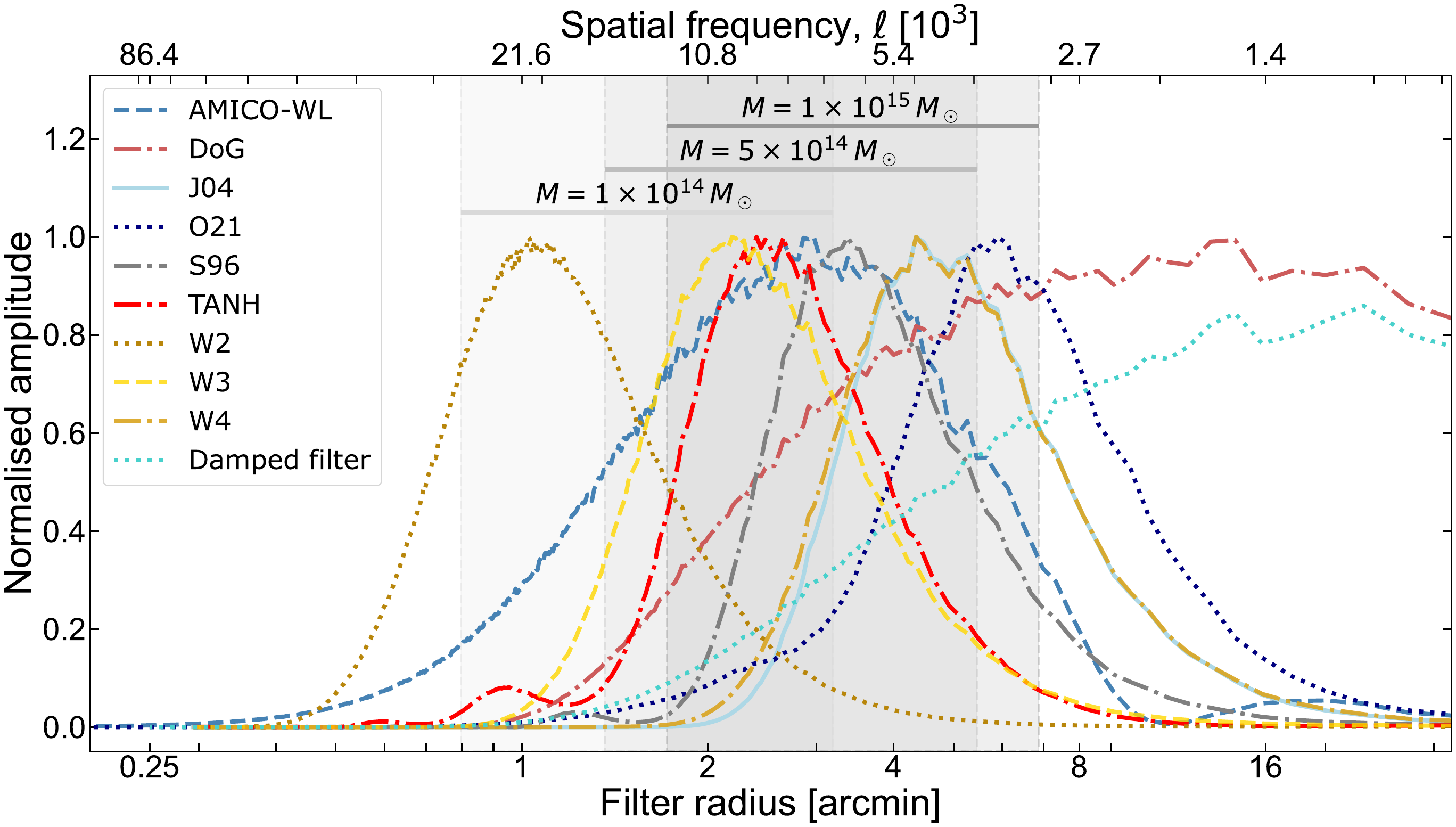}
\captionsetup{font=small}  
\caption{Shape of the average radial power spectra in Fourier space for the filters used by the participating methods in the integrated and the masked approaches. The shaded regions indicate the frequency ranges corresponding to clusters with different NFW profiles. The spatial frequencies, $\ell$, appear at the top of the figure, while at the bottom they are translated into arcminutes to highlight the selectivity of each filter in angular units. Note that the filter distribution for W2, W3, and W4 are common between MRLens and W234.}
\label{fig:distribution of the filters - Fourier}
\end{figure}

For this challenge, different methods have been defined and applied to the shear data or at the level of the convergence maps or on both. To get a better understanding of the difference between each filter, we highlight their shape in Fourier space in Fig.~\ref{fig:distribution of the filters - Fourier}. This figure shows the average radial power spectrum for each of the filters which participated in the challenge as a function of the spatial frequency translated into scales. To obtain these spectra, we created 12 white Gaussian noise catalogues (i.e., with flat power spectra). We then applied the different detection methods that participated in the challenge to produce  filtered noise maps and observe their frequency response, averaging the results over the 12 catalogues. To do so, the filtered maps were processed using a Fourier transform, and the frequencies were translated into arcminutes by scaling each filter according to its resolution. Note that MRLens is not included in the figure as the filters W2, W3, and W4 used are common with W234.

Representing the filter in the frequency domain offers a clear advantage: it allows us to analyse the filter behaviour in terms of spatial frequencies. When translated to angular scales (arcminutes), the spectrum highlights the selectivity of the filter in terms of cluster signal. In particular, we can see that most of these filters are band-pass filters, allowing a careful selection of the cluster signal while denoising at small angular scales and avoiding the contamination from the LSS at large scales. These spectra show that the AM filters TANH, J04, and S96, as well as the wavelet ones W2, W3, and W4, yield results consistent with previous studies \citep{Leonard2012,Leroy2023}. The scale targeted by the filter O21 is larger than for the other compensated filters. This is because the shape of the filter was designed to remove the contribution from the central region of the clusters to avoid signal dilution. The complex definition of the optimal filter in AMICO-WL leads to a band-pass filter which covers a wider range of frequencies. The damped and DoG filters are consistent with low-pass filters, in agreement with their descriptions. By construction, they should be more impacted by the LSS signal, compared to all the other filters. By design, the different methods participating in the challenge cover a large range of filter families and approaches. The result is an almost exhaustive comparison and analysis of the state of the art in WL cluster detection.
%===================================================================

%===================================================================
\section{Challenge analysis}
\label{section_5}
%===================================================================

Once the detections of an algorithm are definitive, it is essential to decide the criteria to associate the candidate detections with the original synthetic clusters. 
The chosen matching procedure is a key tool to assess the performance of a detection algorithm, as well as to compare it with others. If not optimal, it could introduce many false associations. 
Moreover, if the matching favours a certain type of algorithm or behaviour, the comparison will be unreliable.
In this section, we first define the selection cut applied to the synthetic halo catalogues, and then we present and explain the methodologies of the two matching procedures adopted for the analysis: the ranked geometrical matching (GM) and the maximum matching distance (MMD) procedures. Finally, we describe the metrics used to assess the performance of the methods.

\subsection{Mass-redshift cut}

In order to select the haloes that are more likely detectable by the algorithms and to decrease the number of potential false associations, the following cut was applied:
\begin{equation}
    \logten(\Mvircr/\si{\solarmass}) > 13.3 + 1.049\,z + 0.489\,z^{2}\;.
    \label{eq:diagonal_cut}
\end{equation} 
This selection, which is illustrated more clearly in the redshift-mass plane ($z$--$M$ plane hereafter) shown in Fig.~\ref{fig:diagonal_cut_zM_plane}, is based on equation~(12) of \cite{Andreon_Berge_2012}, which was designed to select clusters based on their theoretical S/N. In that theoretical definition, the noise includes all sources of non-cluster contributions, namely shape noise, LSS signal, projection systematics, etc. While that equation, adapted to the EWS, selects clusters with $\mathrm{S/N} > 5$, we adopt a $z$--$M$ cut with a lower threshold of $\mathrm{S/N} > 2$. Because the theoretical S/N assumes an idealised filter and exhibits substantial scatter with respect to the measured S/N, changing the halo selection defined by Eq.~\eqref{eq:diagonal_cut} primarily affects the absolute completeness and number of matches, while the relative comparison remains based on the same selection for all methods. This choice provides a good balance between maximising sample completeness, limiting the false association rate, and taking into account the scatter between theoretical and measured S/N. It ensures that the matching procedures are applied to clusters exhibiting significant WL signal, thereby substantially mitigating potential false positives arising from line-of-sight alignments.
The synthetic halo catalogues considered in this work contain a total of 24\,198 haloes, with an average of about 2000 haloes per catalogue. The masking results in a loss of about 22\% of those haloes. These halo catalogues are the truth tables used in the matching procedure to quantify the completeness and purity of each detection method. 

\begin{figure}
  \includegraphics[width=\linewidth]{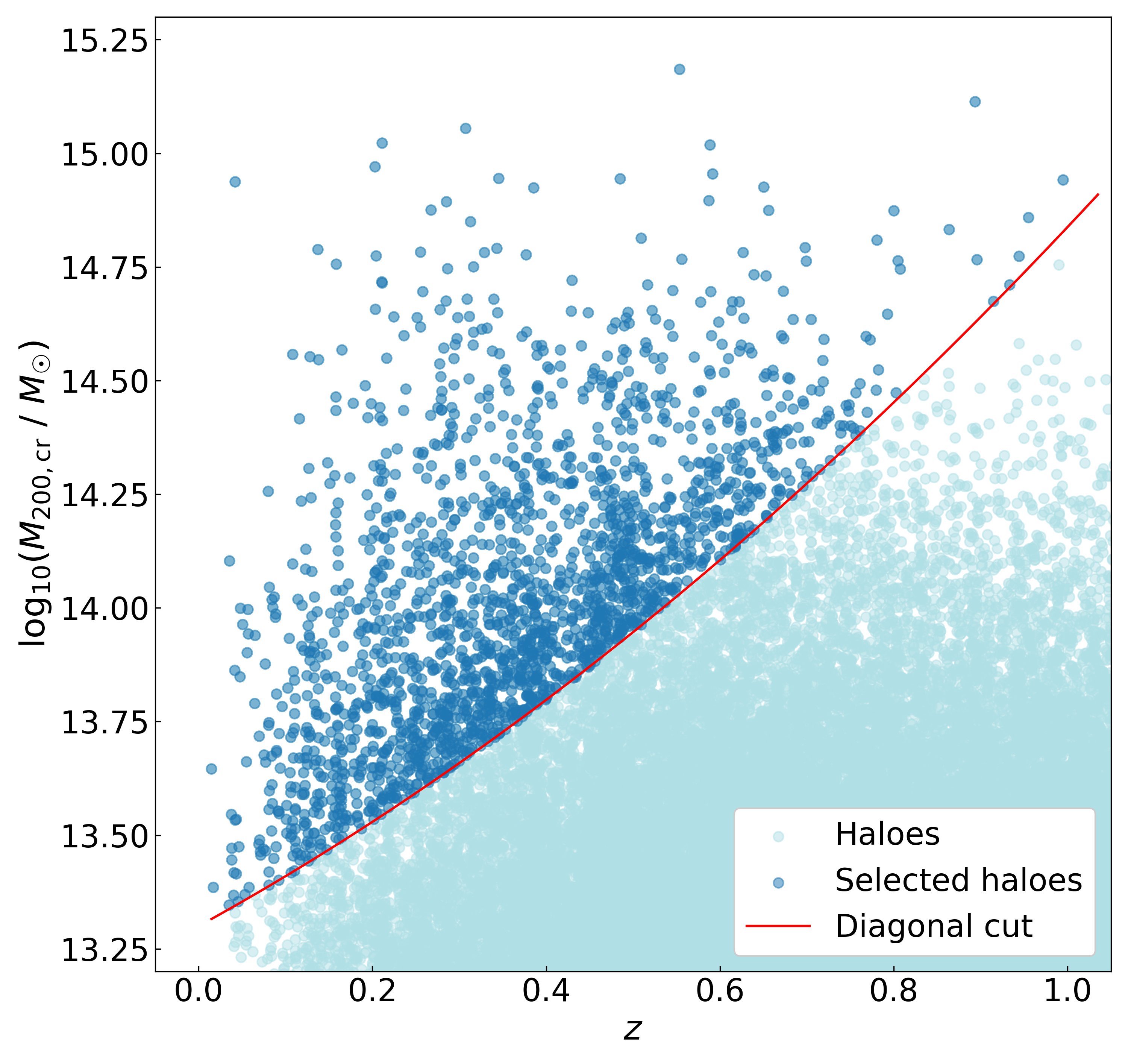}
\captionsetup{font=small}
\caption{Example of the halo selection in the $z$--$M$ plane using the cut (red line) from Eq.~(\ref{eq:diagonal_cut}) for one of the $\ang{10;;}\times\ang{10;;}$ fields. The light blue dots correspond to all the haloes in the simulated field, while the dark blue dots are the selected haloes.} 
\label{fig:diagonal_cut_zM_plane}
\end{figure}

\subsection{Matching procedures}

\subsubsection{Ranked geometrical matching (GM)}

The ranked geometrical matching is performed starting from the synthetic clusters and searching for associated detections within a defined area around each cluster. In the GM procedure, the matching distance depends on the projected halo radius, as the detected centre may be offset from the true centre due to shape noise, cluster triaxiality, or other effects such as pixel resolution. In this work, the GM procedure was implemented on pixelised maps with a pixel size of $\ang{;0.6;}$, chosen to ensure consistency with the resolutions of the shear and convergence maps adopted by the methods being compared and with the corresponding filter sizes. Additional tests comparing this implementation with a pixel-free version yielded consistent results within the associated uncertainties for all methods, indicating that the adopted pixelisation has only a minor impact on the results presented here. The GM is implemented as follows:
\begin{enumerate}
\item The synthetic clusters are sorted by decreasing mass, as provided from the halo catalogues.

\item For each synthetic cluster, we define a circular region of matching radius equal to $2\,\theta_{\rm 200,cr}/3$ centred on the cluster position. Here $\theta_{\rm 200,cr}$ is the projected halo radius. The detections are treated as point-like objects, such that only their positions are used to determine whether they fall within the matching region. We search for the detection counterparts that fall within this region. This matching radius was chosen to limit the fraction of false associations, $\tau_{\rm FAR}$, to about 15\% (see Sect.~\ref{subsection_performance_metrics} for details). Its average value for the synthetic clusters considered here is $2\,\bar{\theta}_{\rm 200,cr}/3 = \ang{;2.9;} \pm \ang{;1.2;}$.

\item The detections within that region are inspected. If multiple counterparts are found, the synthetic cluster is associated with the detection having the highest S/N. If two or more detections have the same highest S/N, the one nearest to the synthetic cluster centre is selected and removed to prevent it from being associated with another cluster.

\item The second and third steps are repeated for all the objects in the halo catalogue.
\end{enumerate}

This GM procedure only takes into account the characteristic $\theta_{\rm 200,cr}$ of the haloes, without paying attention to the filter scale of the detection method. Since detections and synthetic clusters are removed from their respective catalogues as they are matched to one another, this matching procedure is bijective by construction and therefore corresponds to a two-way matching. This matching procedure follows the same approach as the \texttt{ranking} matching adopted in \cite{Adam-EP3}.

\subsubsection{Maximum matching distance (MMD)}
\label{subsect MMD procedure}

The second matching procedure used to compare the different methods is the optimal matching approach developed by \cite{Leroy2023}. 
The principle of this procedure is to derive the maximum matching distance (MMD) corresponding to the empirical limiting distance that differentiates between the correct and incorrect match depending on the filter radius. This distance is computed from both the halo properties and the filter radius. Using the filter radius for the association is important, as the error in the position of the detection is directly related to the size of the filter.

In practice, this matching procedure is a two-step process in which the same procedure is applied twice consecutively: first to compute the MMD, and then to match the detections based on the resulting MMD.
This core procedure can be summarised in four key points: 

\begin{enumerate}
\item We compute the projected angular separations, $D$, between the detection peaks and each halo in the catalogue. These distances are then normalised by the projected halo radius $\theta_{\rm 200,cr}$.
\item We search for the smallest normalised distance among all halo-detection peak pairs. By doing so, we favour the best detection candidate while taking the halo properties into account.
\item Once this pair is identified, we remove the detection and halo pair from the catalogues to ensure uniqueness of the association. 
\item We repeat steps 2 and 3. 
\end{enumerate}

First, in order to compute the MMD value for each algorithm, this core procedure is repeated iteratively for all detections, regardless of the smallest distance value. Following this approach, we can distinguish between correct and false halo–detection pairs, as shown in the results of \citet{Planck2014}. Through a Gaussian fit, we can extract the standard deviation, $\sigma_{\rm MMD}$, and, the mean, $\mu_{\rm MMD}$, corresponding to the distribution of the matches that are correct given the filter and halo radii. Following \cite{Leroy2023}, we define the MMD as $5\sigma_{\rm MMD}$ + $\mu_{\rm MMD}$, whose values for every participating method are shown in Table~\ref{tab:MMD_values}. Due to a significant number of multiple detections, the MMD could not be measured for the damped filter, which is therefore excluded from the MMD analysis in this work.

Second, we repeat the core procedure, keeping only the matched pairs for which the corresponding projected angular separation, $D$, is smaller than the measured MMD. This approach ensures that each match between a detection and a halo is unique, while optimising the analysis for each filter. 
Compared to the GM, this approach has the advantage of taking into account both halo and filter properties, as well as offering an alternative point of view on the performance analysis.
However, the MMD calculation relies on an empirical and statistical approach, in which accuracy scales with the number of detections for a given method. 
In particular, the number of detections required to achieve the same confidence in the MMD value varies with the filter scale.
Thus, this uncertainty is more significant for filters with large radii, such as W4, J04, and O21 (as shown in Fig.~\ref{fig:distribution of the filters - Fourier}).

\begin{table}
  \centering 
  \captionsetup{font=small}
  \caption{MMD values measured for the participating methods, following the procedure described in \cite{Leroy2023}.}
  \label{tab:MMD_values}
  \resizebox{0.65\linewidth}{!}
  {\begin{tabular}{lc}
    \hline\hline
    \noalign{\vskip 2pt}
    \multicolumn{1}{c}{Method} & MMD value [arcmin] \\
    \hline
    \noalign{\vskip 3pt}   
    AMICO-WL & $2.1$  \\
    Damped filter & $\dots$  \\
    DoG & $3.8$  \\
    J04 & $5.5$  \\
    S96 & $4.3$  \\
    TANH & $2.7$  \\
    O21 & $3.1$  \\
    MRLens & $2.9$ \\  
    W2 & $2.0$\\
    W3 & $3.1$ \\
    W4 & $5.2$\\
    \hline
  \end{tabular}} 
  \end{table}

By construction, this matching approach relies entirely on the detection catalogue to estimate the properties of the method. Consequently, issues with the catalogues or the presence of multiple detections affect the ability to compute the MMD (e.g., the damped filter).

Another important point to consider with the MMD procedure is the false association rate ($\tau_{\rm FAR}$, see Sect.~\ref{subsection_performance_metrics}). 
Indeed, including the filter radius in the association causes the maximum matching distance used by the MMD procedure to vary between different detection methods. As a result, the fraction of false associations is not constant across methods, unlike in the GM procedure (see Sect.~\ref{subsection_performance_metrics}). Therefore, it is essential to consider both the GM and MMD matching procedures when interpreting the results.

\subsection{Performance metrics}
\label{subsection_performance_metrics}

To evaluate the performance of the different methods and to
compare them, we use five different metrics: the completeness $C$, the purity $P$, the masking loss $\Delta_{\rm ML}$, the tomographic gain $\Delta_{\rm TG}$, and the false association rate $\tau_{\rm FAR}$. 
For each S/N threshold, the metrics are computed from the counts summed over all 12 fields, rather than by averaging the corresponding field-level ratios.

The completeness is defined as
\begin{equation}
    C = \frac{N_{\rm match}}{N_{\rm haloes}}\;,
    \label{eq:completeness}
\end{equation}
where $N_{\rm haloes}$ is the number of clusters in the halo catalogue and $N_{\rm match}$ is the number of those clusters that are matched by a given method. The completeness does not only depend on the number of matched detections, but also on the number of objects present in the halo catalogue. Therefore, moving up or down the $z$--$M$ cut shown in Fig.~\ref{fig:diagonal_cut_zM_plane} will have an impact on the completeness resulting from the matching between clusters and detections. It might seem that increasing the number of haloes boosts the probability of matching a detection, but it also increases the number of false detections incorrectly matched. 
In fact, if many low-mass and distant clusters, which are not detectable with WL, are added to the target halo catalogue, true matches will likely be outnumbered by false matches, resulting in a completeness dominated by random associations. Although minimising this effect is important, the comparison between the methods would still be fair, since all of them would be affected to the same extent.

While the completeness measures the capability of a detection method at finding all the existing haloes, the purity is an indicator of the reliability of its detections. 
The purity is defined as
\begin{equation}
    P = \frac{N_{\rm match}}{N_{\rm det}}\;,
    \label{eq:purity}
\end{equation}
with $N_{\rm det}$ being the number of objects in the detection catalogue. This quantity measures the fraction of matched detections among all the detections made, and can be used to evaluate the false detection rate.

In this work, we are comparing the performance of detection methods, both ignoring and using the masks provided in the galaxy catalogues. 
For this reason, we quantify the `masking loss' for each detection method. This is a measure of the relative decrease in the number of matched haloes due to masking. We define it as
\begin{equation}
\Delta_{\rm ML} = 100\,\left( \frac{N_{\rm match}^{\rm mask}}{N_{\rm match}^{\rm no\;mask}} - 1 \right)\;,
\label{eq:percentage_masking_loss}
\end{equation}
where $N_{\rm match}^{\rm mask}$ is the number of matched haloes achieved by a detection method when considering the masking from the galaxy catalogue and $N_{\rm match}^{\rm no \;mask}$ is the same when the masks are ignored.

Furthermore, we evaluate the performance of the tomographic approaches proposed by the different detection methods with respect to their corresponding integrated versions. For this, we compute the tomographic gain of every detection method as
\begin{equation}
\Delta_{\rm TG} = 100\,\left(\frac{N_{\rm match}^{\rm tomo}}{N_{\rm match}^{\rm int}} - 1 \right)\:,
\label{eq:percentage_tomographic_gain}
\end{equation} where $N_{\rm match}^{\rm tomo}$ is the number of matched by the tomographic approach and $N_{\rm match}^{\rm int}$ is the number of haloes matched by the integrated approach.

Lastly, we evaluate the false association rate, $\tau_{\rm FAR}$, for each detection method using both matching procedures. This is computed by matching the detections with random positions of the synthetic haloes. This random matching process is repeated numerous times, and the final measured values are obtained by averaging the outcomes of these repetitions. 
Note that $\tau_{\rm FAR}$ is determined by the association distance. In the GM procedure, this distance depends only on the halo sizes; consequently, $\tau_{\rm FAR}$ is nearly identical across all methods. In contrast, for the MMD procedure, $\tau_{\rm FAR}$ depends on the adopted filter radius.
Consequently, it represents a key tool to evaluate the  performance of the different filters. It is important to note that if the false detections are not randomly distributed, which could be the case due to the LSS, the estimated $\tau_{\rm FAR}$ could be underestimated. We expect the detection methods targeting the largest scales to be more affected by this, and plan to study this issue in more depth in future work. Through these two association approaches and the various metrics, we aim to fully assess the performance of the participating methods.

%===================================================================
\section{Results} 
\label{section_6}
%===================================================================

In this section we present the results and compare the performance of the integrated approaches of the nine detection methods that participated in the challenge (see Table~\ref{tab:summary_methods_challenge}). Section~\ref{integrated_wo_mask} addresses the results obtained in the absence of masks, while Sect.~\ref{integrated_with_mask} considers the impact of the masks on the performance of the methods. Finally, in Sect.~\ref{complementarity_integrated} we explore the complementarity among the different methods.

\subsection{Integrated approach without masks} 
\label{integrated_wo_mask}
After performing the matching between candidate clusters and synthetic clusters (as described in Sect. \ref{section_5}) for all detection methods following the integrated approach, our analysis focuses on the results obtained without applying any masks.

\subsubsection{Comparison with varying purity and completeness} 

First, we examine completeness as a function of purity, as these quantities are the main metrics for evaluating the performance of detection methods. % \Sandrine{and 
A higher number of detections typically increases completeness but may compromise purity, thus, they are intrinsically related. As a result, two detection methods cannot be fairly compared using a single completeness-purity value. A robust comparison requires an analysis of the evolution of both metrics together to understand how one changes with the other. 

To do this, we computed the purity and completeness, as defined in Sect.~\ref{subsection_performance_metrics}, for each detection method by applying different S/N thresholds to their corresponding detection catalogues. Only detections with an S/N value above the chosen threshold were considered in the matching procedure. The resulting evolution curves of completeness with respect to purity, averaged over the 12 realisations, are shown in Fig.~\ref{fig:completeness_purity_integrated} for all detection methods. In the left panel, we display the results when using the GM procedure, while the right panel shows the results from the MMD procedure. Focusing on the GM results, AMICO-WL outperforms the other detection methods at the high-purity end, but below around 88\% purity, DoG takes the lead. Overall, AMICO-WL, DoG, O21, and W234 are the best-performing methods. For the MMD procedure, the best performance by far is achieved by W234 and J04, which evolve similarly, and outperform the other methods across the entire purity range. DoG and S96 rank third and fourth, respectively. These results collectively highlight the strong performance of DoG, W234, O21, and AMICO-WL.

\begin{figure*}[!ht]
\centering
\includegraphics[angle=0,width=1.0\hsize]{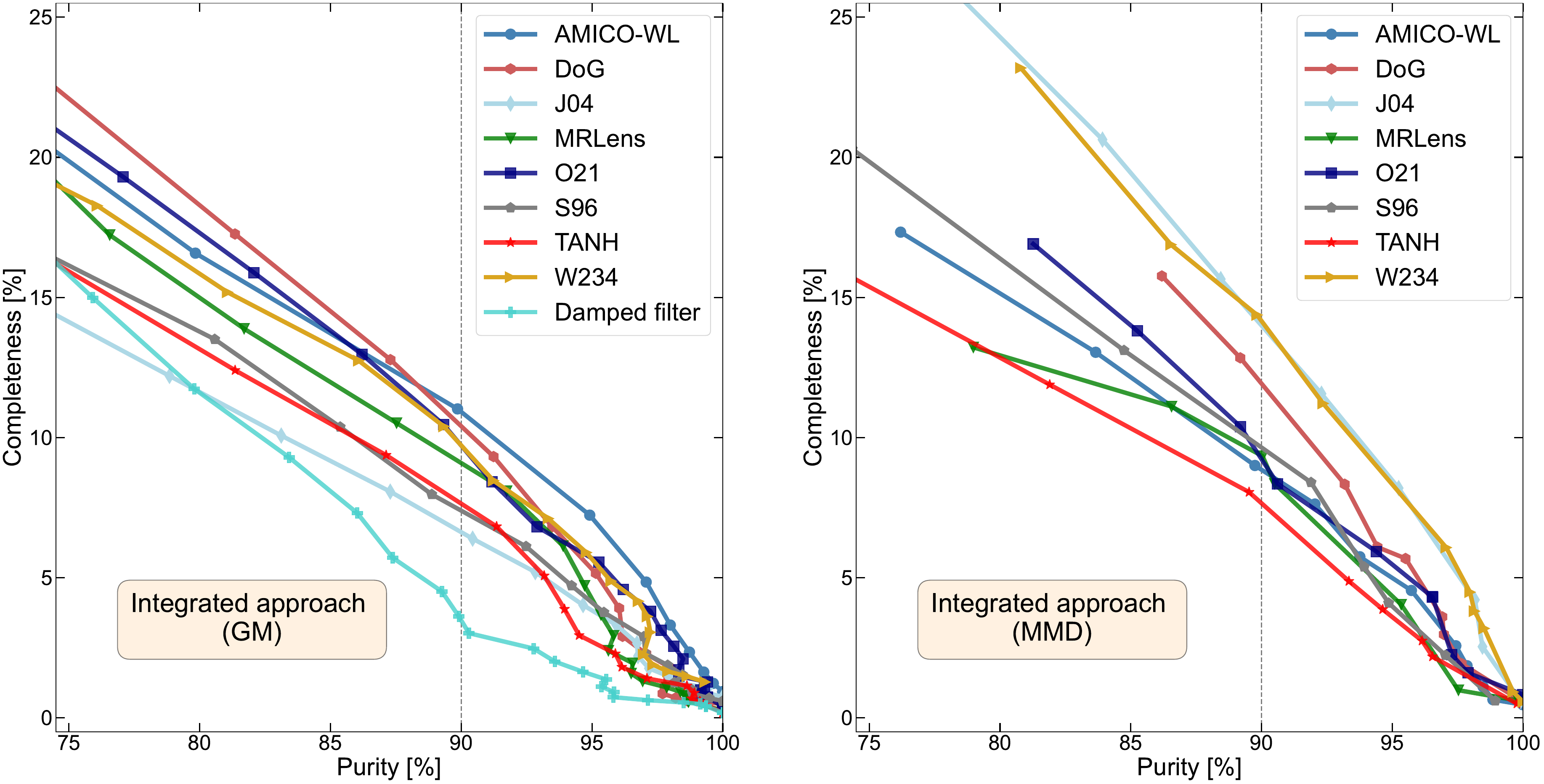}
\captionsetup{font=small}
\caption{Completeness and purity computed by varying the S/N thresholds of the detection methods following an integrated approach, when matching the detections and the synthetic clusters with the GM procedure (\textit{left}) and the MMD procedure (\textit{right}). The markers represent where the purity and completeness are computed, averaging over the 12 realisations. Error bars are consistent with a maximum deviation of $\pm 0.3$ percentage points for all the methods and matching approaches, except for the damped filter, for which the uncertainty is larger.}
\label{fig:completeness_purity_integrated}
\end{figure*}

\subsubsection{Comparison at 90\% purity}
\label{sec:integrated_90_purity}

Based on the completeness-purity behaviours seen in Fig.~\ref{fig:completeness_purity_integrated}, we chose a cut at 90\% purity to compare in more detail the performance of the methods in terms of matched haloes and completeness. This purity cut is maintained throughout this work. 
 
Table~\ref{tab:integrated_approach_90_percent_purity} shows the average performance of the detection methods at around 90\% purity, estimated over the 12 realisations on which they were run. These results represent an average estimation of the performance that each method would have on a $10\degree\times10\degree$ synthetic field simulated under similar conditions. From this table, we can identify which methods are more efficient in terms of completeness at 90\% purity. Considering the GM procedure, AMICO-WL (10.9\%), DoG (10.5\%), O21 (9.8\%), and W234 (9.7\%) are the most complete detection methods. Henceforth, any percentage in parentheses following the name of a method denotes its completeness at 90\% purity, unless stated otherwise. 
With the MMD procedure, W234 (14.0\%), J04 (13.8\%), and DoG (12.2\%) lead in terms of completeness, followed by  S96 (9.9\%), MRLens (9.3\%), and O21 (9.3\%). 
While the GM procedure results in all methods having similar $\tau_{\rm FAR}$ values (around 15\%), the MMD procedure, exhibits a much larger variation, ranging from 7.1\% (AMICO-WL) to 36.2\% (J04). 
In this context, we observe that methods with low $\tau_{\rm FAR}$ values (due to a smaller optimal matching radius) tend to have lower completeness. This explains the differences in method rankings observed between the MMD and GM procedures. It is also worth noting that the three most complete methods under the MMD procedure all have $\tau_{\rm FAR}$ values greater than 20\%.

The uncertainties reported in Tables~\ref{tab:integrated_approach_90_percent_purity} and~\ref{tab:integrated_mask_approach_90_percent_purity} are standard errors on the mean completeness across the 12 fields. To further quantify the uncertainty, for the four pre-selected methods, the field-to-field standard deviation ranges from 0.44 to 0.63 percentage points for GM and from 0.57 to 0.74 percentage points for MMD, while the bootstrap standard error of the global completeness ranges from 0.13 to 0.18 and from 0.16 to 0.20 percentage points, respectively. Paired permutation tests show that, under GM, AMICO-WL and DoG are statistically consistent ($p=0.056$), as are O21 and W234 ($p=0.31$), whereas DoG has significantly higher completeness than O21 ($p=0.002$). Under MMD, all adjacent differences in the ordering W234 > DoG > O21 > AMICO-WL are statistically significant ($p=4.9\times10^{-4}$).

\begin{table*}
  \centering 
  \captionsetup{font=small}
  \caption{Performance at 90\% purity of the detection methods following an integrated approach. $N_{\mathrm{det}}$ and $N_{\mathrm{match}}$ are the mean numbers of detections and matches per $\ang{10;;}\times\ang{10;;}$ field. $C$ and $\tau_{\rm FAR}$ are ratios of the corresponding counts summed over all 12 fields, measured as described in Sect.~\ref{subsection_performance_metrics}.}

  % are measured as described in  Sect.~\ref{subsection_performance_metrics} and are ratios of the corresponding counts summed over all 12 fields.}}  
  \smallskip
  \label{tab:integrated_approach_90_percent_purity}
  \smallskip
  \resizebox{\textwidth}{!}{\begin{tabular}{lcccccccccc}
    \hline\hline
    \noalign{\vskip 2pt}
    & \multicolumn{5}{c}{GM} & \multicolumn{5}{c}{MMD} \\
    \hline
    \noalign{\vskip 3pt}
    \multicolumn{1}{c}{Method} & S/N & $N_{\mathrm{det}}$ & $N_{\mathrm{match}}$ & $C$ (\%) & $\tau_{\rm FAR}$ (\%) & S/N & $N_{\mathrm{det}}$ & $N_{\mathrm{match}}$ & $C$ (\%) & $\tau_{\rm FAR}$ (\%) \\
    \noalign{\vskip 3pt}
    \hline
    \noalign{\vskip 3pt}
    
    AMICO-WL & 3.54 & $245 \pm \phantom{0}4$ & $221 \pm \phantom{0}3$ & $10.9 \pm 0.2$ & $14.9 \pm 0.7$ & 3.67 & $201\pm3$ & $180\pm3$ & $\phantom{0}9.1\pm0.2$ & $\phantom{0}7.1\pm0.4$ \\

    Damped filter & 5.54 & $\phantom{0}85 \pm 16$ & $\phantom{0}75 \pm 12$ & $\phantom{0}3.7 \pm 0.6$ & $14.7 \pm 1.1$ & \dots & \dots & \dots & \dots & \dots \\
  
    DoG & 3.37 & $236 \pm \phantom{0}3$ & $212 \pm \phantom{0}3$ & $10.5 \pm 0.1$ & $14.6 \pm0.6$ & 3.24 & $273\pm5$ & $246\pm2$ & $12.2\pm0.2$ & $20.3\pm0.6$ \\

    J04 & 4.33 & $149 \pm \phantom{0}3$ & $134 \pm \phantom{0}3$ & $\phantom{0}6.7 \pm 0.1$ & $15.5 \pm 0.7$ & 3.78 & $307 \pm 4$ & $276 \pm 3$ & $13.8 \pm 0.2$ & $36.2 \pm 0.9$ \\

    S96 & 4.51 & $168 \pm \phantom{0}4$ & $151 \pm \phantom{0}3$ & $\phantom{0}7.5 \pm 0.1$ & $14.7 \pm 0.8$ & 4.23 & $222 \pm 4$ & $200 \pm 4$ & $\phantom{0}9.9 \pm 0.2$ & $24.6 \pm 0.8$ \\
        
    TANH & 4.61 & $178 \pm \phantom{0}4$ & $160 \pm \phantom{0}4$ & $\phantom{0}7.9 \pm 0.2$ & $14.7 \pm 0.7$ & 4.62 & $176 \pm 4$ & $158 \pm 4$ & $\phantom{0}7.8 \pm 0.2$ & $11.3 \pm 0.8$ \\

    O21 & 4.77 & $219 \pm \phantom{0}5$ & $197 \pm \phantom{0}4$ & $\phantom{0}9.8 \pm 0.2$ & $15.1 \pm 0.7$ & 4.84 & $206 \pm 5$ & $187 \pm 5$ & $\phantom{0}9.3 \pm 0.2$ & $14.3 \pm 0.6$ \\
        
    MRLens & 4.85 & $209 \pm \phantom{0}4$ & $188 \pm \phantom{0}4$ & $\phantom{0}9.3 \pm 0.2$ & $14.7 \pm 0.6$ & 4.85 & $209 \pm 4$ & $187 \pm 5$ & $\phantom{0}9.3 \pm 0.2$ & $12.9 \pm 0.7$ \\
    
    W234 & 4.29 & $216 \pm \phantom{0}4$ & $195 \pm \phantom{0}4$ & $\phantom{0}9.7 \pm 0.2$ & $15.1 \pm 0.4$ & 3.66 & $311 \pm 5$ & $279 \pm 4$ & $14.0 \pm 0.2$ & $23.4 \pm 0.9$ \\

    \hline
  \end{tabular}} \\
  \vspace{0.1cm}
  \begin{justify}
  {\small \textbf{Notes.} For each detection method, we estimate the S/N cut that provides an average purity of approximately 90\% on all simulated fields and apply it separately to each realisation. These results are obtained by averaging over the 12 realisations considered ($\ang{10;;}\times\ang{10;;}$ fields), for both the ranked geometrical (GM) and the maximum matching distance (MMD) matching approaches adopted. S/N thresholds are not recalibrated per field, as this would require realisation-specific truth information and would artificially impose the same purity on every field. Uncertainties in S/N are $\leq 0.01$.}
  \end{justify}
\end{table*}

\subsection{Integrated approach with masks}
\label{integrated_with_mask}
The aim of this challenge is to better understand the optimal strategies for enhancing cluster detection using real \Euclid WL data. To this end, we also study the current performance of the methods taking into account masks that mimic the expected EWS mask pattern.

\subsubsection{Strategies for handling masks}

The nine WL detection algorithms presented here used different strategies to account for the masks provided in the galaxy catalogues. All methods discarded the masked regions in the noisy shear data, so galaxies in those regions did not contribute to the shear maps. However, methods such as AMICO-WL and DoG applied their filters directly at the position of the galaxies, mitigating the impact of missing data. MRLens used a non-linear mass mapping method \citep{mm:Pires2020} to fill in the gaps (masks) in the shear maps via interpolation. Cluster detection was then performed on complete convergence maps derived from these interpolated shear maps. Other methods applied masks to low-density areas. For instance, AMICO-WL generated custom masks from the input shear catalogues by building binary maps with the same pixel grid as the signal and variance maps, where $0$ ($1$) denotes the absence (presence) of galaxies.
O21 followed a similar strategy, but masking regions with a galaxy number density below 0.5 times and above 1.5 times the average value. For their part, DoG, J04, S96, and TANH excluded detections located at a distance from the edges of masked regions less} than or equal to the filter radius. W234 followed the same strategy but used a fixed exclusion zone of about $3\arcmin$.

\subsubsection{Comparison of the masked approaches}

When we carried out the matching between the synthetic clusters and the detections for the approaches that used masks, we realised that some methods (J04, O21, S96, TANH, and the damped filter) still had detections falling within masked regions. 
These detections harmed the performance of those methods. 
Therefore, in the analysis carried out using the GM and MMD procedures, we removed these invalid detections. The criterion for removal being if any of the eight pixels surrounding a detection is fully or partially masked, that detection is removed from the catalogue. We then performed the matching process again for those specific methods.

\begin{figure*}
\centering
\includegraphics[angle=0,width=1.0\hsize]{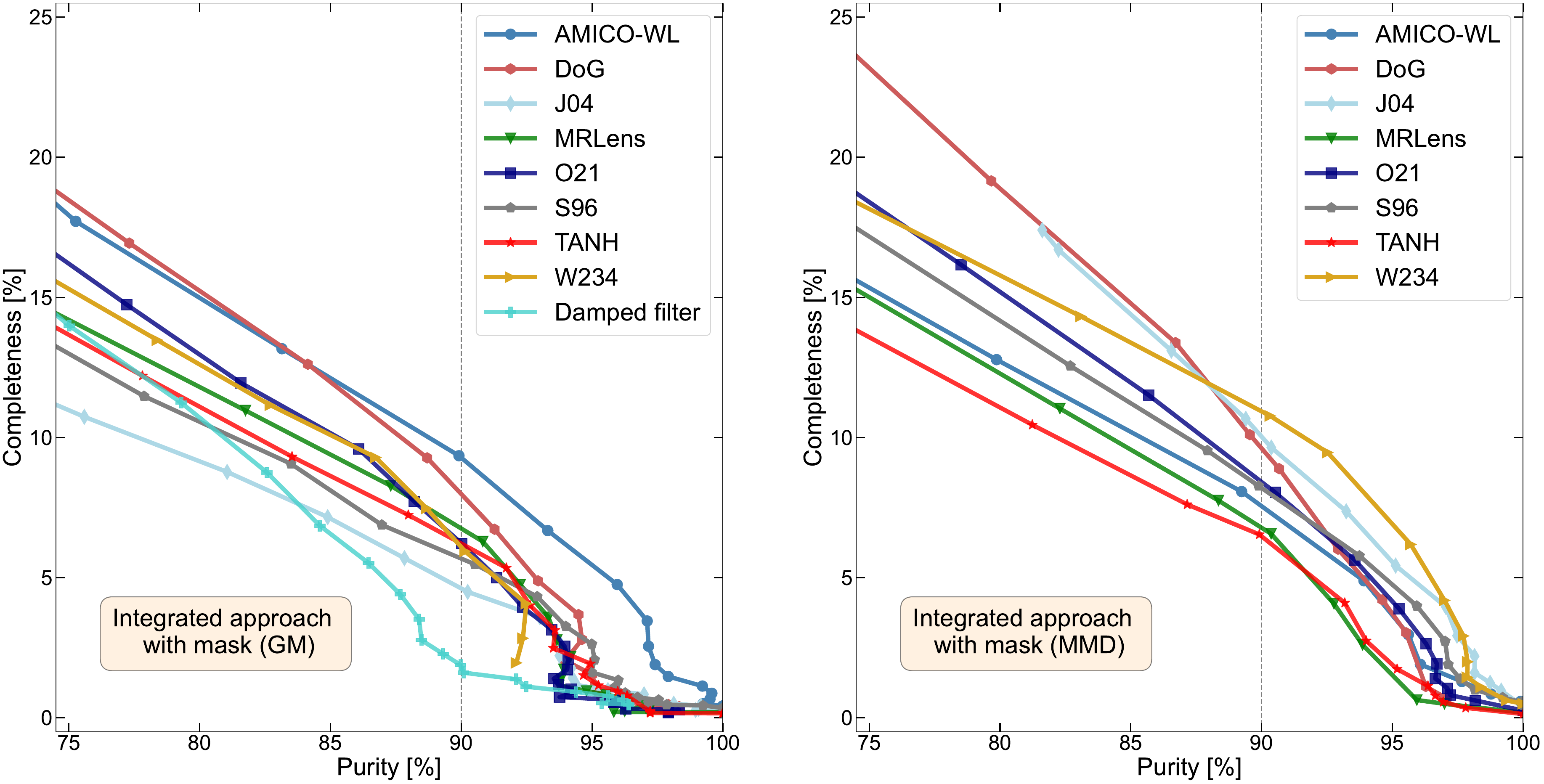}
\captionsetup{font=small}
\captionsetup{justification=raggedright,singlelinecheck=false}
\caption{As Fig.~\ref{fig:completeness_purity_integrated}, but for the masked integrated approach.}
\label{fig:completeness_purity_integrated_masks}
\end{figure*}

Figure~\ref{fig:completeness_purity_integrated_masks} shows the completeness as a function of purity for the same methods displayed in Fig.~\ref{fig:completeness_purity_integrated}, but this time taking into account the presence of masks in the galaxy catalogues. Here, completeness was computed excluding masked-out galaxy clusters. As before, the results obtained using the GM procedure are shown in the left panel, while those from the MMD procedure are displayed in the right panel. Table~\ref{tab:integrated_mask_approach_90_percent_purity} reports the average performance over the 12 realisations considered at 90\% purity, including also the percentage relative loss of matched haloes due to the masking, $\Delta_{\rm ML}$, as defined in Eq.~\eqref{eq:percentage_masking_loss}. By comparing Fig.~\ref{fig:completeness_purity_integrated_masks} with Fig.~\ref{fig:completeness_purity_integrated} and Table~\ref{tab:integrated_mask_approach_90_percent_purity} with Table~\ref{tab:integrated_approach_90_percent_purity}, we can see that the completeness of all methods decreases, as expected. 
However, the method ranking changes. Turning first to the GM results, we see that AMICO-WL (9.4\%) remains the most complete method, ahead of the other methods at higher purities and expanding its dominance to lower purities compared to the case without masking. It is now clear that DoG (7.8\%) is the second best-performing method. MRLens (6.9\%) stands out at 90\% purity, but at lower purities, it is surpassed by O21 (6.4\%) and W234 (6.1\%). On the other hand, the MMD results show that W234 (11.0\%) and J04 (9.9\%) consistently outperform other methods in the high-purity regime. % between 90\% and 97\%. 
When all galaxies are used to build the convergence maps, W234 follows an increasing and undisturbed trend as purity decreases (see Fig.~\ref{fig:completeness_purity_integrated}). However, when part of the galaxies are masked (see Fig.~\ref{fig:completeness_purity_integrated_masks}), W234 shows a more gradual decline compared to the other methods, and is overtaken by J04 and DoG (9.5\%) below 88\% purity. More work will be carried out in the future to fully understand this behaviour.

\begin{table*}
  \centering
\captionsetup{font=small}  
\captionsetup{justification=raggedright,singlelinecheck=false}
  \caption{As Table~\ref{tab:integrated_approach_90_percent_purity}, but for the masked integrated approach. The masking loss, $\Delta_{\rm ML}$, is defined in Eq.~(\ref{eq:percentage_masking_loss}).}
  \smallskip
  \label{tab:integrated_mask_approach_90_percent_purity}
  \smallskip
  \resizebox{\textwidth}{!}{\begin{tabular}{lcccccccccc}
  \hline\hline
  \noalign{\vskip 2pt}
   & \multicolumn{5}{c}{GM}  & \multicolumn{5}{c}{MMD} \\
   \hline
   \noalign{\vskip 3pt}
   \multicolumn{1}{c}{Method} & $N_{\mathrm{det}}$ & $N_{\mathrm{match}}$ & $C$ (\%) & $\Delta_{\rm ML}$ (\%) & $\tau_{\rm FAR}$ (\%) & $N_{\mathrm{det}}$ & $N_{\mathrm{match}}$ & $C$ (\%) & $\Delta_{\rm ML}$ (\%) & $\tau_{\rm FAR}$ (\%) \\
   \noalign{\vskip 3pt}
   \hline
   \noalign{\vskip 3pt}
    
    AMICO-WL & $163 \pm 5$ & $147 \pm 4$ & $9.4 \pm 0.2$ & $-33 \pm \phantom{0}3$ & $12.0 \pm 0.8$ & $ 136 \pm 3$ & $121\pm3$ & $\phantom{0}7.7\pm0.1$ & $-33 \pm 3$ & $\phantom{0}3.7\pm0.7$\\

    Damped filter & $\phantom{0}38 \pm 6$ & $\phantom{0}34 \pm 5$ & $2.1 \pm 0.3$ & $-55 \pm 14$ & $12.2 \pm 1.9$ & \dots & \dots & \dots & \dots & \dots \\

    DoG & $136 \pm 3$ & $122 \pm 2$ & $7.8 \pm 0.1$ & $-42 \pm \phantom{0}2$ & $12.3 \pm 0.9$ & $165\pm3$ & $148\pm3$ & $\phantom{0}9.5\pm0.2$ & $-40 \pm 2$ & $11.4\pm1.3$\\

    J04 & $\phantom{0}79 \pm 3$ & $\phantom{0}71 \pm 2$ & $4.5 \pm 0.1$ & $-47 \pm \phantom{0}3$ & $12.4 \pm 1.1$ & $172\pm3$ & $155\pm3$ & $\phantom{0}9.9\pm0.2$ & $-44 \pm 2$ & $27.7\pm1.8$ \\

    S96 & $\phantom{0}98 \pm 3$ & $\phantom{0}88 \pm 2$ & $5.6 \pm 0.1$ & $-42 \pm \phantom{0}2$ & $11.9 \pm 1.1$ & $143\pm4$ & $128\pm3$ & $\phantom{0}8.2\pm0.1$ & $-36 \pm 3$ & $19.0\pm1.7$ \\

    TANH & $112 \pm 4$ & $101 \pm 3$ & $6.4 \pm 0.2$ & $-37 \pm \phantom{0}3$ & $11.6 \pm 0.9$ & $117\pm4$ & $105\pm4$ & $\phantom{0}6.7\pm0.2$ & $-34 \pm 4$ & $\phantom{0}7.6\pm0.9$ \\

    O21 & $112 \pm 4$ & $100 \pm 3$ & $6.4 \pm 0.2$ & $-49 \pm \phantom{0}3$ & $12.1 \pm 0.9$ & $145\pm2$ & $130\pm2$ & $\phantom{0}8.3\pm0.1$ & $-30 \pm 3$ & $11.4\pm1.1$ \\
        
    MRLens & $121 \pm 4$ & $109 \pm 4$ & $6.9 \pm 0.2$ & $-42 \pm \phantom{0}3$ & $12.1 \pm 1.0$ & $120\pm3$ & $108\pm2$ & $\phantom{0}6.8\pm0.1$ & $-42 \pm 3$ & $\phantom{0}8.7\pm0.9$\\
   
    W234 & $106 \pm 3$ & $\phantom{0}96 \pm 3$ & $6.1 \pm 0.1$ & $-51 \pm \phantom{0}3$ & $12.0 \pm 0.5$ & $192\pm6$ & $172\pm4$ & $11.0\pm0.3$ & $-38 \pm 2$ & $12.9\pm1.5$ \\
    
    \hline
  \end{tabular}} \\
  \vspace{0.1cm}
  \begin{justify}
  {\small \textbf{Notes.} The J04, O21, S96, TANH, and the damped filter methods had detections within masked regions, which were removed before performing the matching. Their completeness values would decrease, by 0.2, 2.6, 0.1, 0.1, and 0.5 percentage points, respectively, if those detections were taken into account for the GM procedure.}
  \end{justify}
\end{table*}

An important point to note is that although only 22\% of the galaxy sources were masked, Table~\ref{tab:integrated_mask_approach_90_percent_purity} shows that $\Delta_{\rm ML}$ results are roughly a factor of two higher for several detection methods. The average $\Delta_{\rm ML}$ due to masking across all methods, when considering the GM procedure, is approximately 44.2\%. AMICO-WL is the least affected by masking, with $\Delta_{\rm ML} \approx 33\%$. The only other method with $\Delta_{\rm ML}<40\%$ is TANH. At the opposite end, O21 ($\Delta_{\rm ML} \approx 49\%$), W234 ($\Delta_{\rm ML} \approx 51\%$) and the damped filder ($\Delta_{\rm ML} \approx 55\%$) suffer larger performance drops due to masking. Under the MMD procedure, where the damped filter could not be evaluated, the average $\Delta_{\rm ML}$ is approximately 37.1\%. In this case, J04 and MRLens are the only two methods with $\Delta_{\rm ML}>40\%$. J04 suffers the greatest decrease at 90\% purity ($\Delta_{\rm ML} \approx 44\%$), yet it remains the second most complete method after masking is applied. The methods that are least impacted by masking under the MMD procedure are O21 ($\Delta_{\rm ML} \approx 30\%$) and AMICO-WL ($\Delta_{\rm ML} \approx 33\%$). 

There are several effects that can explain why the average $\Delta_{\rm ML}$ is larger than the actual fraction of masked galaxy sources. First, clusters are not point sources: they have spatial extent, which increases the fraction of clusters that are affected by the masks. This effect should impact all methods similarly. The second effect is due to the size of the filter: the larger the filter, the more the filtered map is influenced by masked areas. This may explain why AMICO-WL and TANH are the least impacted by the masks. Finally, another reason is due to the impact of masking on purity. Achieving a 90\% purity level requires increasing the S/N threshold used to detect the clusters, which in turn reduces the number of detections.

In summary, introducing masked regions into the data has a significant impact on the performance of the methods, resulting in a loss of around 40\% of matched detections at 90\% purity compared to the performance without masks. Among all methods, AMICO-WL exhibited the smallest loss with both matching procedures, suggesting that a small filter size combined with masking low-density regions helps preserve purity. Although currently not very efficient, the use of in-painting needs to be investigated further for larger filter sizes because it brings an improvement on MRLens compared to W234, where no in-painting was applied. Further work is needed for most methods to better handle missing data and optimise their performance under realistic \Euclid conditions.

\subsection{Complementarity between the integrated approaches}
\label{complementarity_integrated}

In order to study the complementarity between the integrated approaches of the participating detection methods, we created several combinations of the haloes detected using the various methods and matched with both the GM and the MMD procedures. We present here the main results, but further discussion and details on the complementarity between detection methods can be found in Appendix~\ref{app:complementarity}.

Figure~\ref{fig:completeness_matrices_z-M_plane_integrated_combining_methods_GM} shows the completeness distribution in the $z$--$M$ plane obtained with the GM approach for four of these combinations, ordered from lowest to highest overall completeness. Here, we note that although each detection method’s catalogue is obtained at 90\% purity, their combinations yield purities below 90\%.
As expected, more massive and less distant clusters, which are fewer, tend to be easily detected, whereas less massive and more distant clusters, which outnumber the former, are harder to find. 
Any detection observed close to the detection limit (bottom-right region of the $z$--$M$ plane) should be treated with caution. The low number of matches near the diagonal in the $z$--$M$ planes supports the validity of our cut, defined in Eq.~(\ref{eq:diagonal_cut}). To gain a more detailed understanding of the matching biases induced by the matching approaches, we divide each $z$--$M$ plane into four subareas, defined by two grey dashed lines: one at $z = 0.6$ and another one at $\Mvircr = 3 \times 10^{14} \si{\solarmass}$. We analysed the completeness in these areas using different combinations of detection methods.
\begin{itemize}
    \item High confidence merged catalogue (top left): we merged the halo detections achieved by at least 2/3 of the nine detection methods. This yielded a low but confident completeness of 6.1\% across the entire $z$--$M$ plane (see Table~\ref{tab:summary_performance_methods_challenge_90_percent_purity}), and 49.3\% in the upper left subarea.  
    \item Best performing combinations of three and four methods (top right and bottom left): we combined all haloes detected by the three (AMICO-WL, DoG, and W234) and four (AMICO-WL, DoG, O21, and W234) methods that together find the largest number of unique haloes according to both the GM and MMD procedures. These also happen to be the four most complete methods individually under the GM approach. These combinations  yield overall completeness values of 16.4\% and 17.2\%, respectively.
    \item All methods merged (bottom right): we merged the halo detections achieved by all the methods, that is all matched haloes are added provided that they are detected by, at least, one of the methods. This represents the maximum achievable completeness with the nine detection methods following an integrated approach and adopting the GM procedure: 18.5\%.
\end{itemize}

\begin{figure*}[tp]
% ------
\centering
\includegraphics[width=0.85\linewidth]{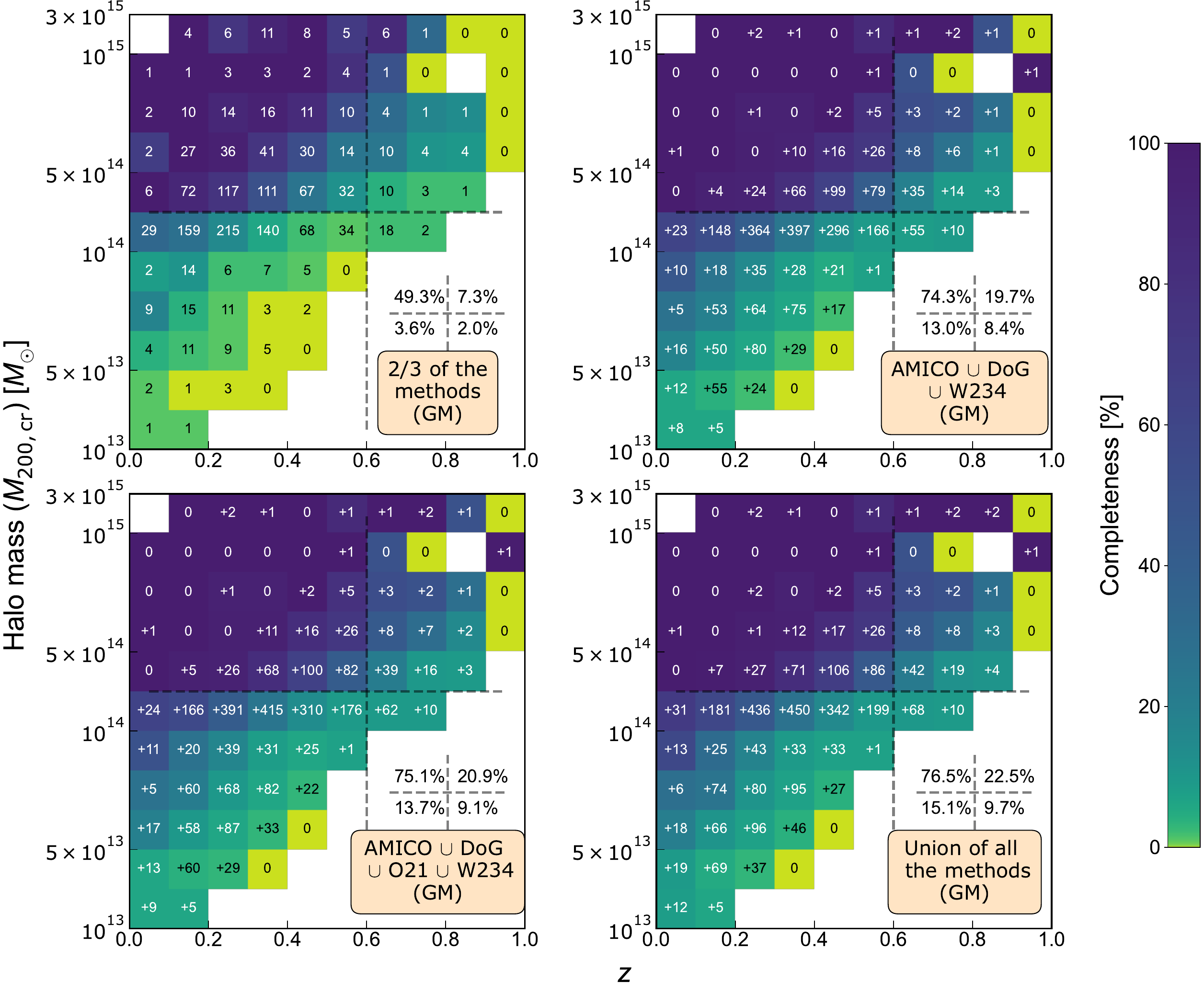}
\captionsetup{font=small}    
\caption{Completeness distribution in the $z$--$M$ plane for several combinations of the detection methods listed in Table~\ref{tab:summary_methods_challenge}, following an integrated approach without masks and with the GM matching procedure. Individual catalogues have 90\% purity, but their combinations lower the resulting purity. The panels show: haloes detected by two-thirds of the methods (upper left); the combined detections from the three most complete methods at 90\% purity (upper right); the same for the top four methods (lower left); and for the combined detections obtained from all methods (lower right). Cell colours give completeness per redshift–mass bin (percent; see colour bar). Numbers in the upper-left panel are the detected-halo counts, while the other panels show additional detections relative to the upper-left case. Grey dashed lines at $z=0.5$ and $\Mvircr = 10^{14}\si{\solarmass}$ split the plane into four regions, and the completeness in each region is reported.}
\label{fig:completeness_matrices_z-M_plane_integrated_combining_methods_GM}
\end{figure*}

The former overall completeness values, obtained following the GM and MMD procedures for these four combinations, are reported in Table~\ref{tab:summary_performance_methods_challenge_90_percent_purity}. In addition, Figs.~\ref{fig:matrix_comparing_int_methods_GM} and~\ref{fig:matrix_comparing_int_methods_MMD} detail the detections in common between the different methods for both matching procedures. As a result of these comparisons, we observe that almost all detections made by J04 are also recovered by W234, as expected given the similarity between the J04 filter and the W234 method’s W4 scale.

When comparing the results from the GM approach (Fig.~\ref{fig:completeness_matrices_z-M_plane_integrated_combining_methods_GM}) with those from the MMD approach (Fig.~\ref{fig:completeness_matrices_z-M_plane_integrated_combining_methods_MMD}), we see that they are consistent but not identical. The differences reflect the false associations made by the two matching approaches. The GM procedure tends to consistently have more haloes matched in the upper left region. In contrast, the MMD procedure favours low-redshift clusters and is therefore able to match them more accurately than the GM approach. This is because the MMD radius can be larger than the GM radius, so a low-redshift detection made by a larger filter has a higher chance of being matched with MMD. As a result, MMD exhibits notable differences in completeness in the bottom-left region compared to GM. These differences can be further amplified by the uncertainty in the MMD values (see Sect.~\ref{subsect MMD procedure}). We also observed these matching biases in the $z$--$M$ planes of individual methods. This is particularly evident with AMICO-WL, which was designed to target high-mass clusters. When applying the MMD procedure, the low MMD value measured for AMICO-WL may undermine this design and reduce the algorithm's performance in this region as well as in the overall field, explaining its poorer performance reported in Table~\ref{tab:integrated_approach_90_percent_purity}. 

Finally, regarding forecasts for the number of detectable clusters in \Euclid, assuming noise characteristics similar to those of the simulations used in this work and without data loss due to masking, we expect to detect at least two galaxy clusters per square degree down to 10$^{13} \si{\solarmass}$ using just one of the best-performing detection methods presented here. Since \Euclid data will include masked regions, we also analysed the performance of the methods in the presence of \Euclid-like masks (22\% of missing data). Based on these results, we estimate that approximately one galaxy cluster per square degree can be detected when using only a single method. These predictions improve when combining the detection capabilities of multiple methods. Using the four best-performing methods together, we could detect up to four galaxy clusters per square degree in the absence of masking, and between two and three when accounting for masking. Detailed predictions from both matching procedures for individual methods and their combinations (as shown in Figs.~\ref{fig:completeness_matrices_z-M_plane_integrated_combining_methods_GM} and \ref{fig:completeness_matrices_z-M_plane_integrated_combining_methods_MMD}) are reported in Table~\ref{tab:summary_forecasts_methods_challenge_90_percent_purity}.

%===================================================================
\section{Conclusions and perspectives}
\label{section_8}
%===================================================================

In this work, we have presented the methodology and the results of a WL galaxy cluster detection challenge carried out on \Euclid-like synthetic observations built using the DEMNUni-Cov simulations \citep{2016JCAP...07..034C, 2022JCAP...11..041P}. The nine detection algorithms that participated in this challenge were applied blindly to 12 simulated fields covering a total area of 1200\,deg$^2$. The purity and completeness of the resulting detection catalogues were compared in order to better understand the strengths and weaknesses of the different detection methods, and hence identify factors that can improve the galaxy cluster finding proficiency on WL data from future \Euclid data releases.

How detections from a given method are matched to the true synthetic clusters is crucial for evaluating its performance and comparing it to others. For this reason, we adopted two different matching approaches. The first one (GM) matches each synthetic cluster to the highest S/N detection located within two-thirds of the projected halo radius. The second approach (MMD) performs a spatial matching while computing, for each method, a maximum matching distance that takes into account both the halo properties and the filter size. The results obtained with these two approaches show clear differences. According to the GM results, the two most efficient methods are AMICO-WL and DoG, whereas W234 and J04 perform best under the MMD approach. 

We observe that the best-performing methods using the MMD approach achieve higher completeness than the top methods using the GM procedure  
This stems from the distinct matching biases induced by these two procedures. Another effect of these biases is that the GM procedure tends to pair detections to high-redshift, low-mass clusters, while the MMD approach seems prone to match low-redshift clusters. While the rate of false associations is restricted to around 15\% for integrated approaches using the GM method, it can vary substantially under the MMD procedure. In fact, the three most complete methods for the MMD approach have false association rates exceeding 20\%. Both association approaches help assess the performance of the methods and provide complementary insights.

Given their performance and complementarity, we pre-selected the best method from each methodology, resulting in four methods to run on the upcoming \Euclid DR1 data: AMICO-WL, DoG, O21, and W234.
AMICO-WL implements a linear optimal matched filter to maximise the S/N in WL data for galaxy cluster detection. The filter is derived by using a prior for the expected cluster signal and measuring the noise directly from the data. DoG implements a two-step Gaussian filtering, with the first one directly applied to the shear at  galaxy positions; O21 uses an AM filter whose parameters are set to minimise LSS noise and maximise the S/N of massive clusters. 
Finally, W234 performs a multi-scale approach that combines three complementary wavelet scales to extract the halo signal from the WL data. This pre-selection does not imply that we will produce a catalogue for each of the methods. Instead, the challenge aimed to identify the most effective detection techniques and explore their complementarity, to learn how to combine their key elements into the most complete and pure WL cluster catalogue from \Euclid data. AMICO-WL and W234 have the best performance with the GM and MMD approaches, respectively, while DoG and O21 yield competitive results with both matching approaches. In particular, the combination of these four methods exceeds 70\% completeness for low-redshift, high-mass clusters. We discarded J04 due to its high degree of overlap with W234, as 96\% of J04 associations are also detected by W234 when adopting the GM approach. A dedicated follow-up study will quantify, for each method, the relation between detection S/N and true halo mass, including its scatter and bias, and derive the corresponding selection function.

The general performance of the methods, when accounting for the masks, reveals that detection losses tend to increase with filter radius, exceeding 40\% for many methods. The way the mask is handled requires further investigation. However, our analysis suggests strategies that can substantially reduce these losses. For instance, applying the filter to the shear at the galaxy positions enables mitigation of the mask impact, and the use of in-painting is also a promising tool for filters with larger scales. Therefore, if the noise level and mask coverage in \Euclid are similar to those assumed in this study, our conservative forecast suggests we will detect at least 1.2 galaxy clusters (with masses above 10$^{13}$ $\si{\solarmass}$) per square degree. This would imply detecting more than 2400 clusters in DR1 ($\sim2000\,{\rm deg}^2$) and more than $16\,000$ in the final EWS ($\sim14\,000\,{\rm deg}^2$). In a more optimistic scenario, a detection campaign combining the best-performing methods would result in the detection of up to 2.5 galaxy clusters per square degree, which translates into about 5000 and 35\,000 clusters in DR1 and the full EWS, respectively. For comparison, optical galaxy detections in \cite{Q1-SP050} reach about 6.8 clusters per square degree. 

%Conclusions on the tomography
The tomographic tests in Appendix~\ref{app:tomography_results} suggest that overlapping redshift bins perform best, with four bins outperforming three. However, tomography brings little improvement for the pre-selected methods, mainly benefiting lower-completeness methods without surpassing them. Further work is needed to optimally combine redshift information for \Euclid.

Through this milestone analysis, the detection methods have been adapted to process this extensive data set and provided a performance forecast for the pre-selected methods. This preparatory work paves the way for the construction of the WL-selected galaxy cluster catalogue based on the \Euclid DR1 data.

Overall, the present results should be interpreted as a controlled blind comparison of WL cluster-detection algorithms on a common set of idealised \Euclid-like mocks. The relative performance of the methods is informative for pre-selecting promising and complementary approaches, but the absolute completeness, purity, and signal-to-noise calibration remain subject to additional observational and astrophysical systematics not included here. These effects will need to be incorporated into future end-to-end validation before constructing a final cosmology-ready \Euclid cluster catalogue.

\begin{acknowledgements}
\AckEC
A.M.G. and A.D.S. acknowledge the support of project PID2022-141915NB-C22 funded by MCIU/AEI/10.13039/501100011033 and FEDER/UE. G.L. acknowledges support from STFC via
grant ST/X001075/1, and the UK Space Agency via grant ST/W002612/1. J.M.D. acknowledges the support of projects PID2022-138896NB-C51 (MCIU/AEI/MINECO/FEDER, UE) Ministerio de Ciencia, Investigaci\'on y Universidades and SA101P24. M.O. acknowledges this work was supported by JSPS KAKENHI Grant Numbers JP22K21349, JP24K00684, JP25H00662, JP25H00672. S.A. acknowledges PRIN-MIUR grant 20228B938N “Mass and selection biases of galaxy clusters: a multi-probe approach” funded by the European Union Next generation EU, Mission 4 Component 1 CUP C53D2300092 0006. L.M. acknowledges the financial contribution from the PRIN-MUR 2022 20227RNLY3 grant “The concordance cosmological model: stress-tests with galaxy clusters” supported by Next Generation EU and from the grant A.S.I. n. 2024-10-HH.0 “Attività scientifiche per la missione \Euclid – fase E”. During part of this work, A.M.C.L.B. was supported by a Paris Observatory-PSL University Fellowship, hosted at the Paris Observatory.
\end{acknowledgements}

%%%%%%%%%%%%%%%%%%%%%%%%%%%%%%%%%%%%%%%%%%%%%%%%%%
%\bibliographystyle{aa}
\bibliography{manjon, pires, gavin}
%%%%%%%%%%%%%%%%%%%%%%%%%%%%%%%%%%%%%%%%%%%%%%%%%%

%%%%%%%%%%%%%%%%%%%%%%%%%%%%%%%%%%%%%%%%%%%%%%%%%%
%%%%%%%%%%%%%%%%% APPENDICES %%%%%%%%%%%%%%%%%%%%%

\newpage
\appendix
\onecolumn

% ---------------------------------------------------------
% Part of the Appendix on Complementarity between methods
% ---------------------------------------------------------
\section{Complementarity between the integrated approaches}
\label{app:complementarity}

Table~\ref{tab:summary_performance_methods_challenge_90_percent_purity} presents the results from the nine detection methods that participated in the challenge under both the integrated and masked approaches, along with results derived from selected combinations of methods. Here, we note that although each method’s catalogue is obtained at 90\% purity, their combinations yield purities below 90\%. In this table, in addition to presenting the individual completeness achieved by each detection method, we show that, when using the GM procedure, the combination of AMICO-WL and DoG yields a completeness of 14.8\%. These two methods together detect the largest number of different haloes. The best combination of three different detection methods is obtained with AMICO-WL, DoG, and W234, reaching a combined completeness of 16.4\%, while adding O21 results in the best four-method combination with a completeness of 17.2\%. When considering the MMD procedure, the best two-method combination is between W234 and DoG, with a completeness of 17.1\%. Adding AMICO-WL detections to this pair produces the best three-method combination, with 19.0\% completeness, and adding O21 further increases it to 19.4\%. To conclude, both matching procedures identify the same most complete four-method combination: AMICO-WL, DoG, O21, and W234.

\begin{table}[!ht]
  \centering 
  \captionsetup{font=small}
  \caption{Summary of the performance at 90\% purity of the WL detection algorithms. The values of $N_{\mathrm{match}}$ and $C$ are the number of matched haloes and the completeness, respectively, achieved over all the $\ang{10;;}\times\ang{10;;}$ fields. The completeness is computed as defined in Eq.~(\ref{eq:completeness}), but the number of haloes, $N_{\mathrm{haloes}}$, varies depending on whether the masks are taken into account or not.}
  \smallskip
  \label{tab:summary_performance_methods_challenge_90_percent_purity}
  \renewcommand{\arraystretch}{1.2}
  \begin{tabular}{lcccccccc}
    \hline\hline
    \noalign{\vskip 2pt}
    & \multicolumn{4}{c}{Integrated approach} & \multicolumn{4}{c}{Masked approach} \\
    \cline{2-9}
    \noalign{\vskip 2pt}
    & \multicolumn{2}{c}{GM} & \multicolumn{2}{c}{MMD} & \multicolumn{2}{c}{GM} & \multicolumn{2}{c}{MMD} \\
    \noalign{\vskip 2pt}
    \hline
    \noalign{\vskip 2pt}
    \multicolumn{1}{c}{Method} & $N_{\mathrm{match}}$ & $C$ (\%) & $N_{\mathrm{match}}$ & $C$ (\%) & $N_{\mathrm{match}}$ & $C$ (\%) & $N_{\mathrm{match}}$ & $C$ (\%) \\
    \noalign{\vskip 2pt}
    \hline
    \noalign{\vskip 2pt}
    AMICO-WL & $2647$ & $10.9\%$ & $2162$ & $\phantom{0}8.9\%$ & $1762$ & $9.4\%$ & $1449$ & $\phantom{0}7.7\%$ \\

    Damped filter & $\phantom{0}896$ & $\phantom{0}3.7\%$ & \dots & \dots & $\phantom{0}402$ & $2.1\%$ & \dots & \dots \\
    
    DoG & $2546$ & $10.5\%$ & $2954$ & $12.2\%$ & $1467$ & $7.8\%$ & $1781$ & $\phantom{0}9.5\%$ \\
    
    J04 & $1611$ & $\phantom{0}6.7\%$ & $3316$ & $13.7\%$ & $\phantom{0}853$ & $4.5\%$ & $1862$ & $\phantom{0}9.9\%$ \\

    S96 & $1812$ & $\phantom{0}7.5\%$ & $2404$ & $\phantom{0}9.9\%$ & $1051$ & $5.6\%$ & $1539$ & $\phantom{0}8.2\%$ \\
        
    TANH & $1922$ & $\phantom{0}7.9\%$ & $1895$ & $\phantom{0}7.8\%$ & $1209$ & $6.4\%$ & $1253$ & $\phantom{0}6.7\%$ \\

    O21 & $2362$ & $\phantom{0}9.8\%$ & $2239$ & $\phantom{0}9.3\%$ & $1205$ & $6.4\%$ & $1564$ & $\phantom{0}8.3\%$ \\
        
    MRLens & $2251$ & $\phantom{0}9.3\%$ & $2244$ & $\phantom{0}9.3\%$ & $1307$ & $6.9\%$ & $1295$ & $\phantom{0}6.9\%$\\
        
    W234 & $2336$ & $\phantom{0}9.7\%$ & $3357$ & $13.9\%$ & $1146$ & $6.1\%$ & $2060$ & $10.9\%$ \\
    
    \hline
    \noalign{\vskip 2pt}
    AMICO-WL $\cup$ DoG & $3588$ & $14.8\%$ & $3671$ & $15.2\%$ & $2222$ & $11.8\%$ & $2314$ & $12.3\%$ \\

    AMICO-WL $\cup$ O21 & $3495$ & $14.4\%$ & $3136$ & $13.0\%$ & $2124$ & $11.3\%$ & $2165$ & $11.5\%$ \\

    AMICO-WL $\cup$ W234 & $3491$ & $14.4\%$ & $4008$ & $16.6\%$ & $2090$ & $11.1\%$ & $2515$ & $13.4\%$ \\

    DoG $\cup$ W234 & $3110$ & $12.9\%$ & $4141$ & $17.1\%$ & $1668$ & $\phantom{0}8.9\%$ & $2529$ & $13.4\%$ \\

    AMICO-WL $\cup$ DoG $\cup$ O21 & $3882$ & $16.0\%$ & $3893$ & $16.1\%$ & $2340$ & $12.4\%$ & $2550$ & $13.5\%$ \\

    AMICO-WL $\cup$ DoG $\cup$ W234 & $3969$ & $16.4\%$ & $4588$ & $19.0\%$ & $2350$ & $12.5\%$ & $2851$ & $15.1\%$ \\
    
    AMICO-WL $\cup$ O21 $\cup$ W234 & $3864$ & $16.0\%$ & $4259$ & $17.6\%$ & $2280$ & $12.1\%$ & $2779$ & $14.8\%$ \\
    
    AMICO-WL $\cup$ DoG $\cup$ O21 $\cup$ W234 & $4151$ & $17.2\%$ & $4687$ & $19.4\%$ & $2445$ & $13.0\%$ & $2993$ & $15.9\%$ \\

    AMICO-WL $\cup$ DoG $\cup$ TANH $\cup$ W234 & $4093$ & $16.9\%$ & $4672$ & $19.3\%$ & $2448$ & $13.0\%$ & $2925$ & $15.5\%$ \\

    Union of all the methods & $4467$ & $18.5\%$ & $5556$ & $23.0\%$ & $2697$ & $14.3\%$ & $3541$ & $18.8\%$ \\
    
    \hline
    \noalign{\vskip 2pt}

    AMICO-WL $\cap$ DoG & $1605$ & $6.6\%$ & $1452$ & $6.0\%$ & $1007$ & $5.3\%$ & $940$ & $5.0\%$ \\
        
    AMICO-WL $\cap$ DoG $\cap$ MRLens & $1335$ & $5.5\%$ & $1032$ & $4.3\%$ & $\phantom{0}800$ & $4.2\%$ & $725$ & $3.9\%$ \\
    
    AMICO-WL $\cap$ DoG $\cap$ O21 & $1378$ & $5.7\%$ & $\phantom{0}997$ & $4.1\%$ & $\phantom{0}792$ & $4.2\%$ & $794$ & $4.2\%$ \\

    AMICO-WL $\cap$ DoG $\cap$ MRLens $\cap$ W234 & $1212$ & $5.0\%$ & $1107$ & $4.6\%$ & $\phantom{0}687$ & $3.6\%$ & $672$ & $3.6\%$ \\

    Intersection of $\sfrac{2}{3}$ of the methods & $1488$ & $6.1\%$ & $1527$ & $6.3\%$ & $\phantom{0}831$ & $4.4\%$ & $952$ & $5.1\%$ \\
    
    \hline
  \end{tabular} \\
  \vspace{0.1cm}
  \begin{justify}
  {\small \textbf{Notes.} For each detection method, we show the results obtained with the integrated and masked approaches, adopting both the ranked geometrical (GM) and the maximum matching distance (MMD) matching procedures. The results of combining the individual catalogues have purities below 90\%. The symbol $\cup$ (union) denotes that all haloes detected by the combination of the specified methods are counted, while the symbol $\cap$ (intersection) means that only haloes detected by all of the given methods are counted. In addition, we considered both the union of all methods and the intersection of two-thirds of the methods.} 
  \end{justify}
\end{table}

\begin{figure}[!ht]
    % \begin{subfigure}{0.5\linewidth}
    \centering
    \includegraphics[width=0.9\linewidth]{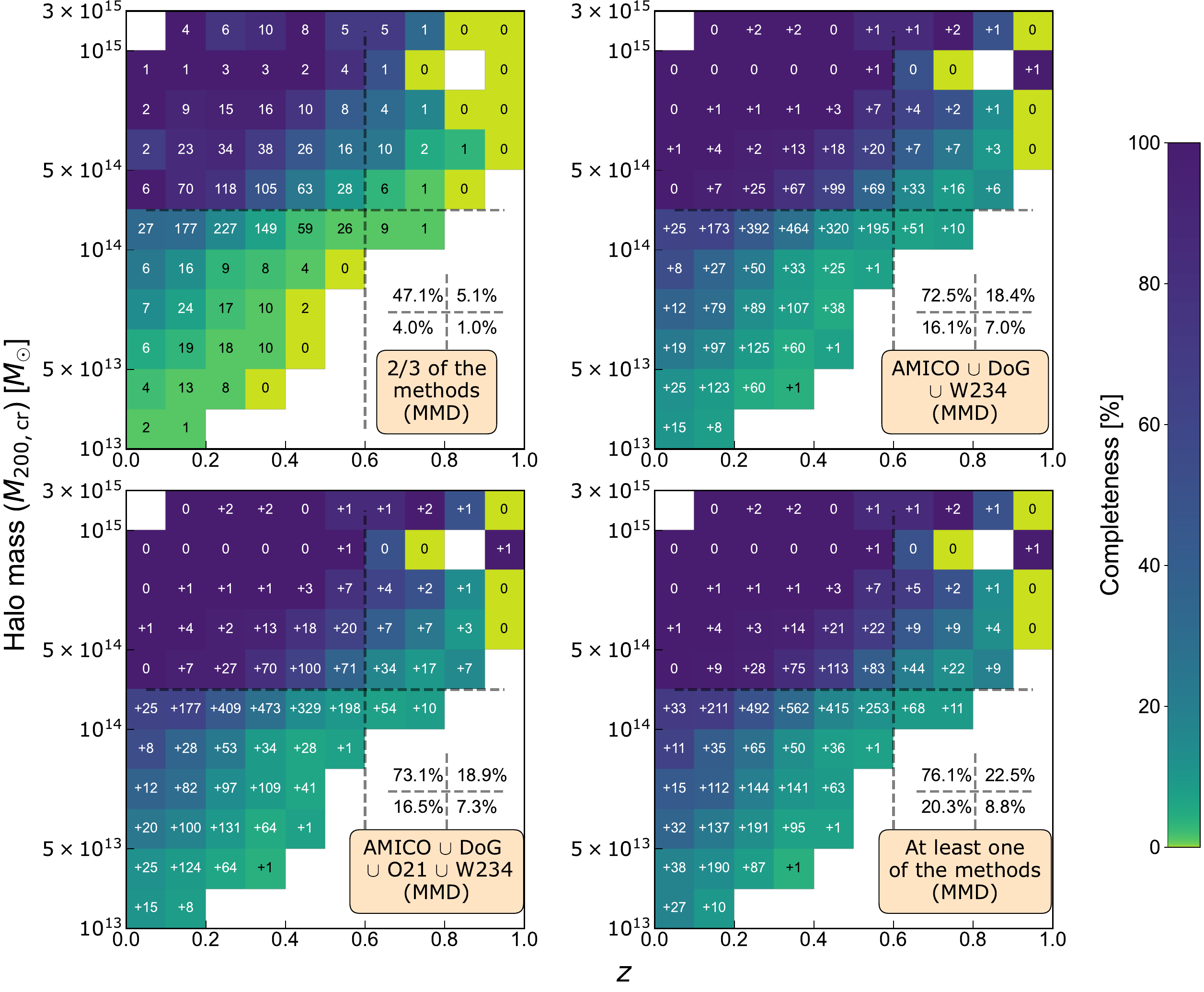} 
\captionsetup{justification=raggedright,singlelinecheck=false}
  \captionsetup{font=small}
 \caption{Same as Fig.~\ref{fig:completeness_matrices_z-M_plane_integrated_combining_methods_GM}, but adopting the MMD procedure.}
\label{fig:completeness_matrices_z-M_plane_integrated_combining_methods_MMD}
\end{figure}

Figure~\ref{fig:completeness_matrices_z-M_plane_integrated_combining_methods_MMD} shows the completeness distribution in the $z$--$M$ plane obtained following the MMD procedure for the same four combinations displayed in Fig.~\ref{fig:completeness_matrices_z-M_plane_integrated_combining_methods_GM}, being the
MMD results in the four subareas consistent with the GM results. 
However, we noticed that the MMD procedure reaches greater completeness in the lower left subarea (up to 20.3\%) compared to the GM procedure (up to 15.1\%). 
The cumulative completeness values of these combinations in the full $z$--$M$ plane are reported in Table~\ref{tab:summary_performance_methods_challenge_90_percent_purity}. Considering only haloes detected by at least 2/3 of the detection methods (top left panel), the overall completeness is 6.3\%, slightly higher than with the GM procedure (6.1\%). The gap between the completeness values provided by the MMD and GM procedures increases for the other combinations. When the haloes from AMICO-WL, DoG, and W234 are combined (top right panel), the overall completeness is 19\% (16.4\% with GM). Adding O21 detections to this combination (bottom left panel) improves completeness to 19.4\% (17.2\% with GM). Finally, combining the haloes detected by all methods (bottom-right panel) yields a cumulative completeness of 23\% (18.5\% with GM). We observe that, across the full $z$--$M$ plane, the completeness is higher for the MMD approach. However, in the top-left region of the $z$--$M$ plane, the GM approach achieves about 2\% higher completeness. For the union of all methods, this difference drops to 0.4\%, which is within our uncertainty limits. 

\begin{figure}[!ht]
\begin{subfigure}{0.5\linewidth}
  \centering
  \includegraphics[width=0.805\linewidth]{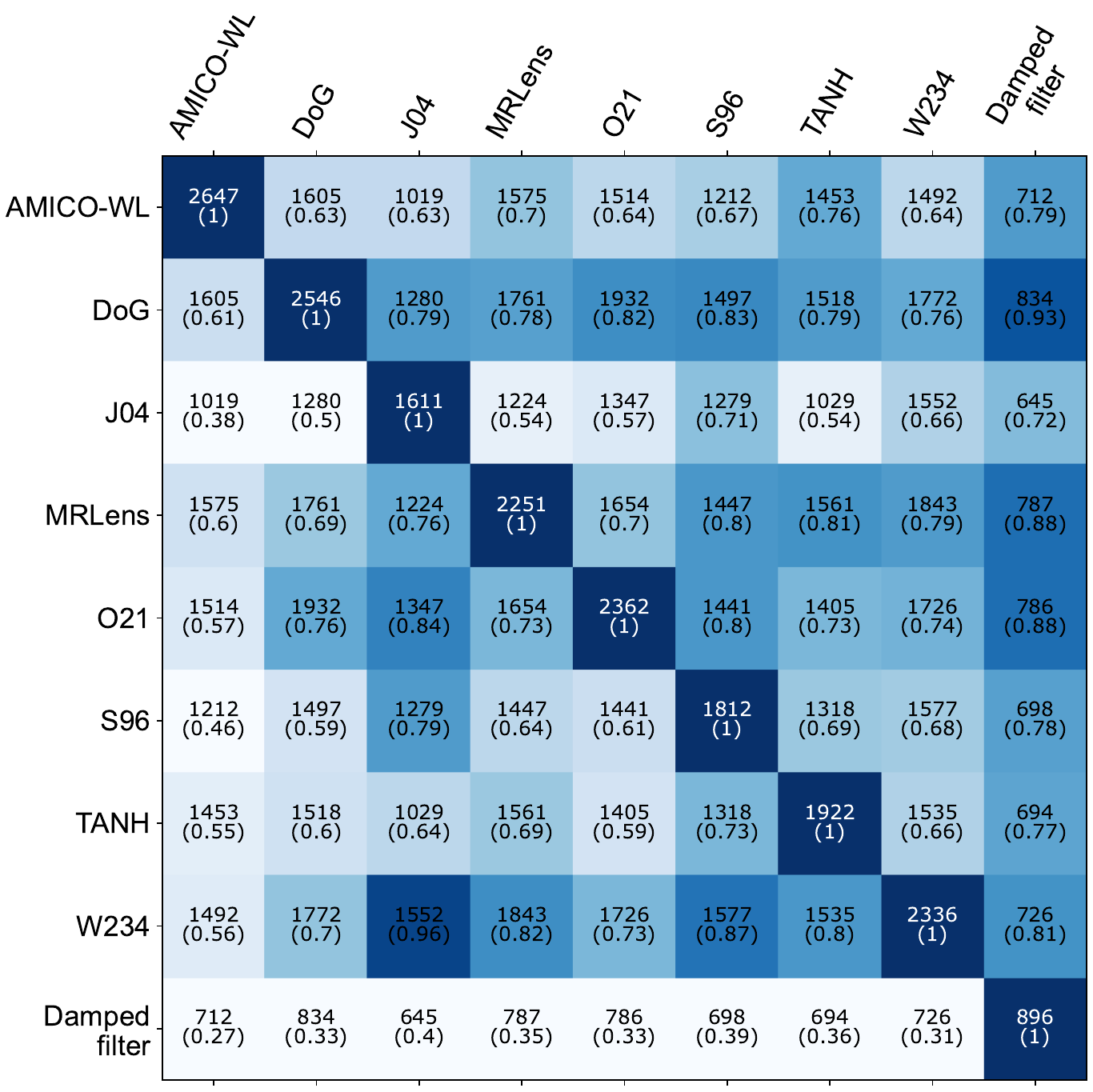}
  \caption{Integrated approach (GM).}
  \label{fig:matrix_comparing_int_methods_GM}
\end{subfigure}\hfil
\begin{subfigure}{0.5\linewidth}
  \centering
  \includegraphics[width=0.9\linewidth]{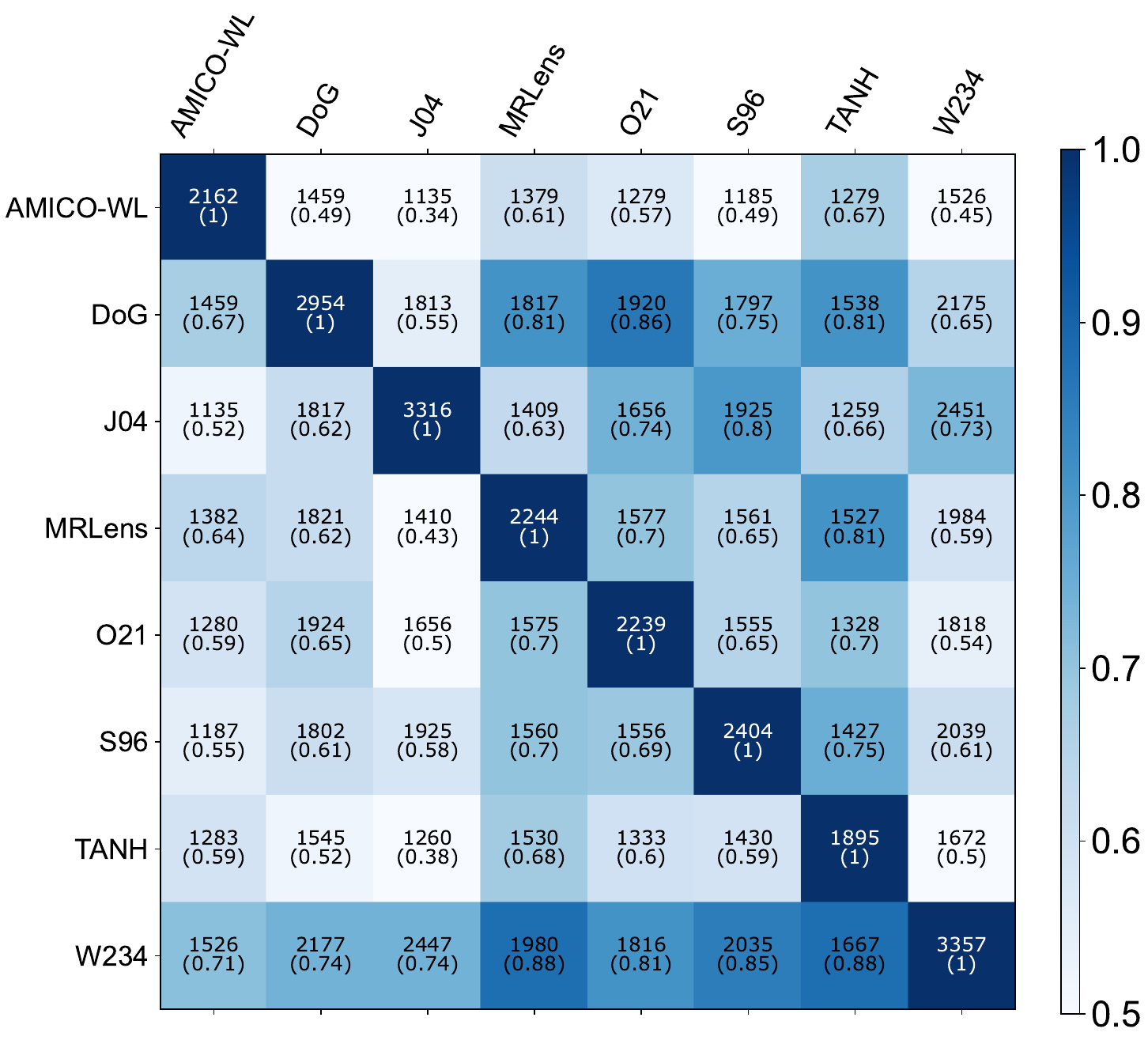}
  \caption{Integrated approach (MMD).}
  \label{fig:matrix_comparing_int_methods_MMD}
\end{subfigure}
\captionsetup{font=small}
\caption{Matrices showing the number of haloes found in common between the integrated approaches of the different WL detection methods. These results are computed at 90\% purity for both the GM  (\emph{left} panel) and  MMD (\emph{right} panel) matching procedures. The cells in each column (row) indicate the number of haloes that the method heading the column (row) has in common with the other methods. In parentheses, we show a measure of the agreement between the two methods being compared, relative to the total number of haloes in the method heading the column. The closer this value is to 1, the greater the overlap between the methods, whereas the closer it is to 0, the less the two methods have in common. In the diagonal, which corresponds to the total number of haloes found by each specific method, the overlap score is hence equal to 1.}
\label{fig:matrices_of_common_matched_haloes_GM}
\end{figure}

Combining the most complete methods in decreasing order is the most effective way to achieve maximum completeness with two, three, and four detection methods. This makes intuitive sense: by combining the two most complete methods, one includes the largest number of unique detection. However, this is not necessarily always the case. For instance, the second most complete method might rely on the same detecting principle as the most complete one but with lower efficiency, thus failing to contribute new detections. In such cases, even the combination of the most complete method with the least complete one would outperform combining the two most complete methods. For this reason, it is important to study the statistical complentarity between the haloes detected by any pair of detection methods. In Fig.~\ref{fig:matrices_of_common_matched_haloes_GM}, we present matrices showing the number of haloes, detected at 90\% purity, that two detection methods following an integrated approach have in common, that is, they tell us the degree of overlap between them. If the overlap between two methods is very large, they do not complement each other well in detecting different clusters. The number of rows and columns matches the number of detection methods, so the cells in each column indicate the number of haloes that the method heading that column has in common with the others. The quantity between parentheses is a measure (ranging from 0 to 1) of the correlation between the method heading the column and the method heading the row, with $0$ denoting no overlap and $1$ indicating full overlap. This quantity is estimated by dividing the number of haloes in the given cell by the total number of haloes for the method heading that column. By construction, the diagonal entries correspond to the total number of haloes found by each method, with an overlap score equal to 1. The results obtained using the GM approach are displayed in Fig.~\ref{fig:matrix_comparing_int_methods_GM}, while those from the MMD procedure are shown in Fig.~\ref{fig:matrix_comparing_int_methods_MMD}. We find that J04 and W234 have a high degree of overlap with the GM procedure (0.96), as 1552 of the 1611 haloes detected by J04 are also detected by W234. This result is expected given the similarity between the J04 filter and the W234 method’s W4 scale.

Finally, Table~\ref{tab:summary_forecasts_methods_challenge_90_percent_purity} presents the number of clusters per square degree detected by each method, as well as by their combinations shown in Figs.~\ref{fig:completeness_matrices_z-M_plane_integrated_combining_methods_GM} and \ref{fig:completeness_matrices_z-M_plane_integrated_combining_methods_MMD}, based on the results of this work. These outcomes allow us to forecast how many clusters per square degree would be detected in the upcoming \Euclid data releases, assuming noise and mask characteristics similar to those of the simulations considered here.

% More extended Table for the forecasts
% -------------------------------------------------------
\begin{table}[!ht]
  \centering 
  \captionsetup{font=small}
  \caption{Forecasts of the number of galaxy clusters detectable per square degree in the EWS, based on the performance at 90\% purity of the WL detection algorithms tested in this work.}
  \smallskip
\label{tab:summary_forecasts_methods_challenge_90_percent_purity}
  \smallskip
  \renewcommand{\arraystretch}{1.2}
  \begin{tabular}{lcccc}
    \hline\hline
    \noalign{\vskip 2pt}
    & \multicolumn{2}{c}{Integrated approach} & \multicolumn{2}{c}{Masked approach} \\
    \cline{2-5}
    \noalign{\vskip 2pt}
    Method & GM & MMD & GM & MMD  \\
    \noalign{\vskip 2pt}
    \hline
    \noalign{\vskip 2pt}
    \hline
    \noalign{\vskip 2pt}
    AMICO-WL & $2.21 \pm 0.03$ & $1.80 \pm 0.03$ & $1.47 \pm 0.04$ & $1.21 \pm 0.03$ \\
    Damped filter & $0.75 \pm 0.12$ & $\dots$ & $0.34 \pm 0.05$ & $\dots$ \\
    DoG & $2.12 \pm 0.03$ & $2.46 \pm 0.02$ & $1.22 \pm 0.02$ & $1.48 \pm 0.03$ \\
    J04 & $1.34 \pm 0.03$ & $2.76 \pm 0.03$ & $0.71 \pm 0.02$ & $1.55 \pm 0.03$ \\
    S96 & $1.51 \pm 0.03$ & $2.00 \pm 0.04$ & $0.88 \pm 0.02$ & $1.28 \pm 0.03$ \\
    TANH & $1.60 \pm 0.04$ & $1.58 \pm 0.04$ & $1.01 \pm 0.03$ & $1.05 \pm 0.04$ \\
    O21 & $1.97 \pm 0.04$ & $1.87 \pm 0.05$ & $1.00 \pm 0.03$ & $1.30 \pm 0.02$ \\
    MRLens & $1.88 \pm 0.04$ & $1.87 \pm 0.05$ & $1.09 \pm 0.04$ & $1.08 \pm 0.02$ \\
    W234 & $1.95 \pm 0.04$ & $2.79 \pm 0.04$ & $0.96 \pm 0.03$ & $1.72 \pm 0.04$ \\
    \hline
    AMICO-WL $\cup$ DoG $\cup$ W234 & $3.31 \pm 0.04$ & $3.82 \pm 0.04$ & $1.96 \pm 0.04$ & $2.38 \pm 0.04$ \\
    AMICO-WL $\cup$ DoG $\cup$ O21 $\cup$ W234 &  $3.46 \pm 0.04$ & $3.91 \pm 0.05$ & $2.04 \pm 0.04$ & $2.49 \pm 0.04$ \\
    Union of all the methods & $3.72 \pm 0.05$ & $4.63 \pm 0.05$ & $2.25 \pm 0.05$ & $2.95 \pm 0.04$ \\
    Intersection of $\sfrac{2}{3}$ of the methods & $1.24 \pm 0.05$ & $1.27 \pm 0.05$ & $0.69 \pm 0.05$ & $0.79 \pm 0.04$ \\
    \hline
  \end{tabular} \\
  \vspace{0.1cm}
  \begin{justify}
  {\small \textbf{Notes.} Same as in Table~\ref{tab:summary_performance_methods_challenge_90_percent_purity}, but here we show the predictions per square degree for the different detection methods.}
  \end{justify}
\end{table}

\clearpage

% --------------------------------------------------
% Part of the Appendix on Tomography results
% --------------------------------------------------
\section{Results following a tomographic approach}

\label{app:tomography_results}

\subsection{Different tomographic approaches}

We present here a comparison of the performance of the detection methods when following a tomographic approach, splitting the galaxies from the shear catalogue into redshift bins. Not all methods followed the same binning criteria; for instance, O21 followed two different approaches: O21 (3 bins) and O21 (4 bins). We therefore distinguish four main strategies. On the one hand, AMICO-WL, DoG, and O21 (3 bins) used the three non-overlapping redshift bins suggested in the challenge: $z\in[0, 0.65]$, $z\in[0.65, 1.05]$, and $z\in[1.05, 3]$. On the other hand, J04, MRLens, S96, TANH, and W234 adopted three overlapping redshift bins, shifting the minimum redshift to $z_{\rm min} > 0$, $z_{\rm min}>0.65$, and $z_{\rm min} >1.05$. O21 (4 bins) splits the shear measurements into four overlapping redshift bins with minimum redshifts of $z_{\rm min} > 0.2$, $z_{\rm min}>0.3$, $z_{\rm min} >0.5$, and $z_{\rm min} >0.7$. Finally, the damped filter divides the shear measurements into 100 linearly-spaced, equal-width thin redshift slices.

\subsection{Comparison of the tomographic approaches}

\begin{figure*}[!ht]
\minipage{0.5\textwidth}
  \includegraphics[width=\linewidth]{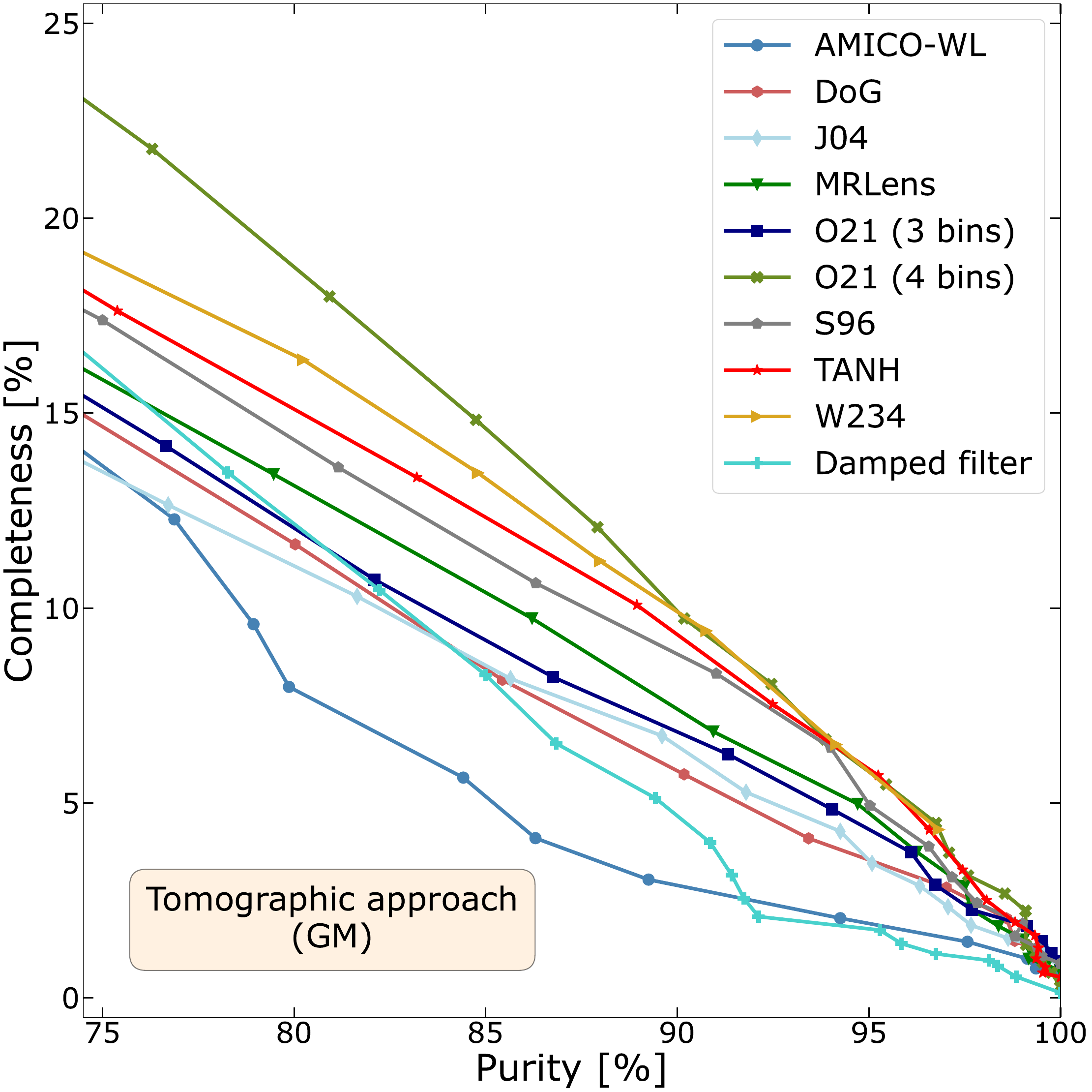}
\endminipage\hfill
\minipage{0.5\textwidth}
  \includegraphics[width=\linewidth]{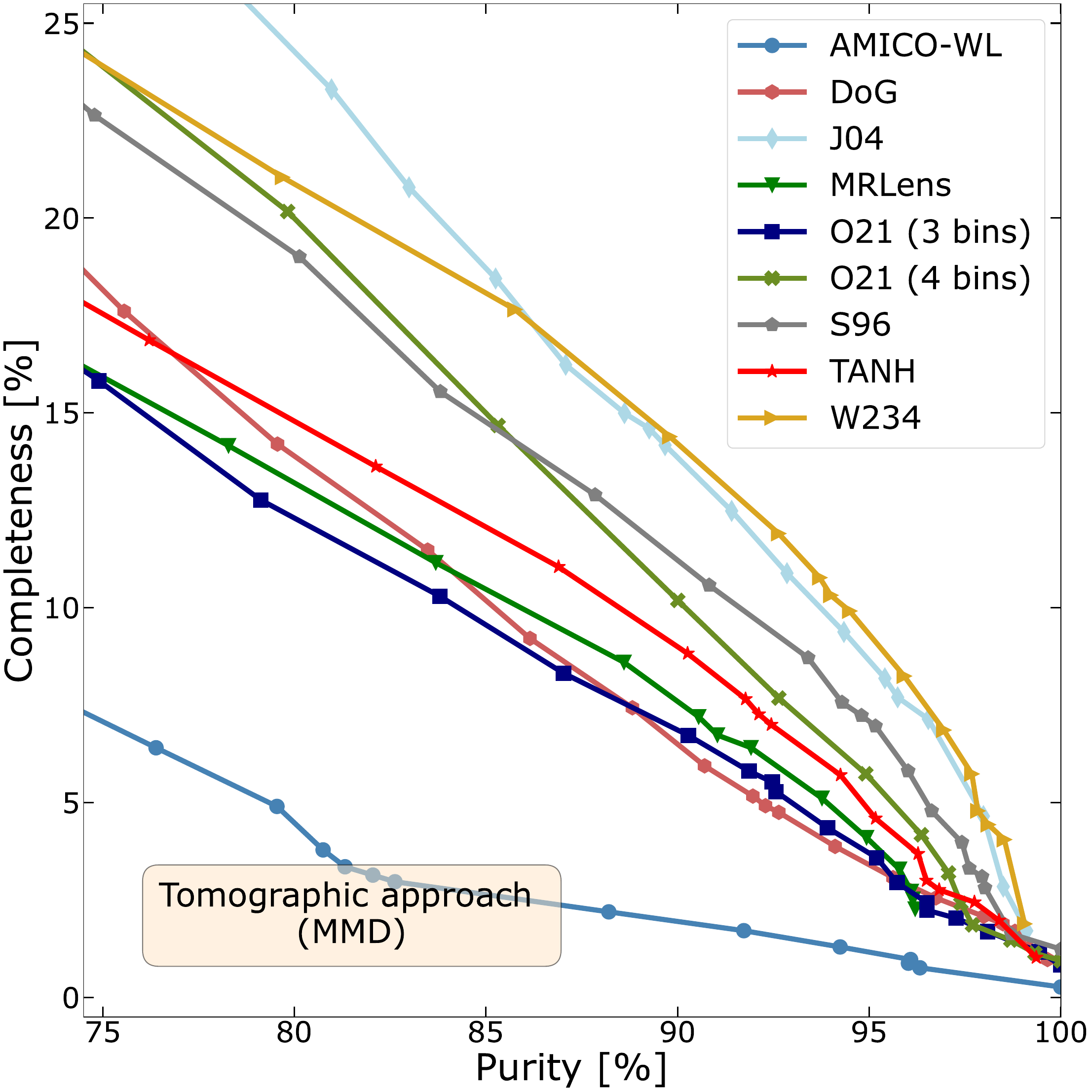}
\endminipage\hfill
\captionsetup{justification=raggedright,singlelinecheck=false}
\captionsetup{font=small}
\caption{As Fig.~\ref{fig:completeness_purity_integrated}, but for the tomographic approaches considered.} 
\label{fig:completeness_purity_tomographic}
\end{figure*}

The evolution curves of completeness as a function of purity for the tomographic approaches followed by the methods listed in Table~\ref{tab:summary_methods_challenge} are presented in Fig.~\ref{fig:completeness_purity_tomographic}. The curves corresponding to the GM procedure are displayed in the left panel, while those from the MMD procedure are shown in the right panel. The average performance of these tomographic approaches over the 12 realisations considered at 90\% purity are reported in Table~\ref{tab:tomographic_approach_90_percent_purity}. We also include a measure of the percentage relative gain or loss in the number of matched haloes, $\mathrm{\Delta_{\rm TG}}$, achieved by the tomographic approach followed by each method, compared to the integrated approach, computed according to the definition in Eq. (\ref{eq:percentage_tomographic_gain}). For the GM procedure, O21 (4 bins; 9.9\%) is the best-performing method across all purity levels. W234 (9.8\%) and TANH (9.6\%) follow in second and third place, being the only other methods to exceed the 9\% completeness ceiling. While they are close to O21 (4 bins) at 90\% purity, this method outperforms them as purity decreases. The next best-performing methods are S96 (8.9\%) and MRLens (8.1\%). For the MMD procedure, W234 (14.2\%), J04 (13.8\%), and S96 (11.2\%) clearly stand out over the entire purity range, achieving higher completeness than any of the methods with the GM procedure. W234 leads from the high-purity end until around 86\%, where it is overtaken by J04. The following best performance is achieved by TANH (9.2\%) and MRLens (7.4\%). As we have seen so far, most methods that perform well with the GM matching procedure (W234, TANH, S96, MRLens), similarly perform well with the MMD procedure. However, there are two exceptions. First, O21 (4 bins), which is the most complete method with the GM procedure (9.9\%) falls to fourth position when the MMD procedure is applied (9.8\%). The second exception is J04, which performs poorly with the GM procedure, but improves with the MMD procedure and becomes the second most complete method at 90\% purity. 

\begin{table*}[!ht]
  \centering 
\captionsetup{justification=raggedright,singlelinecheck=false}
  \captionsetup{font=small}
  \caption{As Table~\ref{tab:integrated_approach_90_percent_purity}, but for the tomographic approaches. The tomographic gain, $\Delta_{\rm TG}$, is measured as described in Sect.~\ref{subsection_performance_metrics}. } 
  \smallskip
  \label{tab:tomographic_approach_90_percent_purity}
  \smallskip
  \resizebox{\textwidth}{!}{\begin{tabular}{lcccccccccc}
    \hline\hline
    \noalign{\vskip 2pt}
    & \multicolumn{5}{c}{GM}  & \multicolumn{5}{c}{MMD} \\
    \hline
    \noalign{\vskip 3pt}
    \multicolumn{1}{c}{Method} & $N_{\mathrm{det}}$ & $N_{\mathrm{match}}$ & $C$ (\%) & $\Delta_{\rm TG}$ (\%) & $\tau_{\rm FAR}$ (\%) & $N_{\mathrm{det}}$ & $N_{\mathrm{match}}$ & $C$ (\%) & $\Delta_{\rm TG}$ (\%) & $\tau_{\rm FAR}$ (\%) \\
    \noalign{\vskip 3pt}
    \hline
    \noalign{\vskip 3pt}
    
    AMICO-WL & $\phantom{0}60 \pm \phantom{0}3$ & $\phantom{0}54 \pm \phantom{0}3$ & $2.7 \pm 0.1$ & $-76 \pm \phantom{0}2$ & $15.4 \pm 1.1$  & $\phantom{0}44\pm2$ & $\phantom{0}40\pm1$ & $\phantom{0}2.0\pm0.1$ & $-78 \pm 1$ & $\phantom{0}7.2\pm1.0$\\

    Damped filter & $106 \pm 18$ & $\phantom{0}92 \pm 14$ & $4.6 \pm 0.7$ & $\phantom{-}23 \pm 39$ & $14.9 \pm 1.2$ & \dots & \dots & \dots & \dots & \dots \\
        
    DoG & $132 \pm \phantom{0}2$ & $118 \pm \phantom{0}2$ & $5.9 \pm 0.1$ & $-44 \pm \phantom{0}2$ & $14.5 \pm 0.9$ & $139\pm2$ & $125\pm2$ & $\phantom{0}6.2\pm0.2$ & $-49 \pm 1$ & $20.4\pm1.4$ \\

    J04 & $148 \pm \phantom{0}4$ & $133 \pm \phantom{0}3$ & $6.6 \pm 0.1$ & $-\phantom{0}1 \pm \phantom{0}4$ & $15.2 \pm 0.8$  & $309\pm5$ & $278\pm5$ & $13.8\pm0.4$ & $\phantom{-0}1 \pm 3$ & $36.2\pm0.8$ \\

    S96 & $200 \pm \phantom{0}4$ & $180 \pm \phantom{0}3$ & $8.9 \pm 0.1$ & $\phantom{-}19 \pm \phantom{0}4$ & $14.6 \pm 0.7$ & $251\pm5$ & $226\pm4$ & $11.2\pm0.4$ & $\phantom{-}13 \pm 4$ & $25.4\pm0.8$ \\
        
    TANH & $215 \pm \phantom{0}3$ & $193 \pm \phantom{0}3$ & $9.6 \pm 0.2$ & $\phantom{-}21 \pm \phantom{0}5$ & $14.8 \pm 0.8$ & $205\pm3$ & $185\pm3$ & $\phantom{0}9.2\pm0.2$ & $\phantom{-}17 \pm 5$ & $11.6\pm0.7$ \\

    O21 (3 bins) & $153 \pm \phantom{0}4$ & $138 \pm \phantom{0}3$ & $6.8 \pm 0.2$ & $-30 \pm \phantom{0}3$ & $14.7 \pm 0.7$ & $154\pm3$ & $138\pm3$ & $\phantom{0}6.8\pm0.1$ & $-26 \pm 4$ & $14.5\pm0.5$ \\

    O21 (4 bins) & $221 \pm \phantom{0}6$ & $199 \pm \phantom{0}5$ & $9.9 \pm 0.2$ & $\phantom{-0}1 \pm \phantom{0}5$ & $14.5 \pm 0.6$ & $216\pm4$ & $198\pm4$ & $\phantom{0}9.8\pm0.4$ & $\phantom{-0}6 \pm 5$ & $14.5\pm0.5$ \\
        
    MRLens & $180 \pm \phantom{0}5$ & $162 \pm \phantom{0}5$ & $8.1 \pm 0.2$ & $-14 \pm \phantom{0}4$ & $14.8 \pm 0.7$ & $167\pm2$ & $150\pm2$ & $\phantom{0}7.4\pm0.1$ & $-20 \pm 3$ & $13.2\pm0.5$ \\
        
    W234 & $220 \pm \phantom{0}5$ & $198 \pm \phantom{0}4$ & $9.8 \pm 0.2$ & $\phantom{-0}2 \pm \phantom{0}4$ & $14.8 \pm 0.3$ & $319\pm6$ & $286\pm6$ & $14.2\pm0.3$ & $\phantom{-0}3 \pm 4$ & $24.0\pm0.6$ \\
    
    \hline
  \end{tabular}} \\
\end{table*}

According to the $\Delta_{\rm TG}$ scores of the tomographic approaches followed by each detection method (Table~\ref{tab:tomographic_approach_90_percent_purity}), we observe that four methods improve both the GM and the MMD procedures: O21 (4 bins), S96, TANH, and W234. The damped filter is the method that experiences the greatest improvement with the GM procedure, but it could not be analysed with the MMD approach. Setting that method aside, the two methods that improve the most under each matching procedure are the same. First, TANH, with $\Delta_{\rm TG} \approx 21\%$ using the GM procedure and $\Delta_{\rm TG} \approx 17\%$ with the MMD procedure. S96 is the other method, with $\Delta_{\rm TG} \approx 19\%$ using the GM procedure and $\Delta_{\rm TG} \approx 13\%$ with the MMD procedure. At the opposite end, the largest decreases under both procedures are observed, in order, for AMICO-WL, DoG, O21 (3 bins), and MRLens. Between these extremes, J04 exhibits roughly the same performance for its tomographic approach as for its integrated version under both matching procedures.

\subsection{Conclusions on the tomographic approaches}

The results on tomography presented in this study constitute a first attempt to further enhance cluster detection based on their WL signal. The different strategies implemented during this challenge allow us to draw some conclusions regarding their efficiency. The three methods that followed the tomography strategy suggested in the challenge (AMICO-WL, DoG, and O21 – 3 bins) performed worse than when using an integrated approach under both the MMD and GM procedures, indicating that using non-overlapping redshift bins is a poor choice. Positive results are obtained with overlapping tomography. Similar or better results than the integrated approach are achieved when using three (J04, S96, TANH, and W234) or four (O21) overlapping redshift bins, leading to average increases of 11\% and 8\% in the number of matched detections with the GM and MMD procedures, respectively. We do not have a clear answer as to why MRLens performance worsens and differs so much from those of the other methods that use the same three overlapping redshift bins, but it may be due to the way the detections made in each bin were combined. Although the damped filter's unique tomographic approach entails the greatest improvement over its integrated approach with the GM procedure, it is still far behind the completeness achieved by the rest of the methods. It remains unclear to what extent using more and thinner redshift slices would help, given that no other method has tested this strategy. For the GM approach, none of the methods achieves the completeness reached with the best integrated approach, while with the MMD procedure, only W234 achieves a slight improvement over its integrated-approach result. In summary, we find that none of the methods that performed best with the integrated approach shows significant improvement when using a tomographic strategy, consistent with the results presented in \cite{chappuis2026}. Therefore, the tomographic approach requires further investigation to carefully define and combine the different redshift bins in preparation for its application to \Euclid data.

\end{document}